\documentclass[12pt, a4paper]{article}
\usepackage[a4paper,left=25mm,right=25mm,top=25mm,bottom=25mm]{geometry}
\usepackage[utf8]{inputenc}
\usepackage{grffile}  
\usepackage{amsmath}
\usepackage{amsfonts}
\usepackage{amssymb}
\usepackage{graphicx}
\usepackage{natbib}
\usepackage{hyperref}  

\renewcommand{\v}[1]{\mathbf{#1}}		

\title{Turbulent geodynamo simulations: a leap towards Earth's core}

\author{N. Schaeffer$^1$, D. Jault $^1$, H.-C. Nataf $^1$, A. Fournier$^2$\\[0.5cm]
 \small $^1$ Univ. Grenoble Alpes, CNRS, ISTerre, \emph{F-38000} Grenoble, France \\
 \small $^2$ Institut de Physique du Globe de Paris, Sorbonne Paris Cit\'e,\\ \small Univ. Paris Diderot, CNRS, 1 rue Jussieu, F-75005 Paris, France.}

\begin{document}

\maketitle

\begin{abstract}
We present an attempt to reach realistic turbulent regime in direct numerical simulations of the geodynamo.
We rely on a sequence of three convection-driven simulations in a rapidly rotating spherical shell.
The most extreme case reaches towards the Earth's core regime by lowering viscosity (magnetic Prandtl number $Pm = 0.1$) while maintaining vigorous convection (magnetic Reynolds number $Rm > 500$) and rapid rotation (Ekman number $E=10^{-7}$), at the limit of what is feasible on today's supercomputers.
A detailed and comprehensive analysis highlights several key features matching geomagnetic observations or dynamo theory predictions -- all present together in the same simulation -- but it also unveils interesting insights relevant for Earth's core dynamics.

In this strong-field, dipole-dominated dynamo simulation, the magnetic energy is one order of magnitude larger than the kinetic energy.
The spatial distribution of magnetic intensity is highly heterogeneous, and a stark dynamical contrast exists between the interior and the exterior of the tangent cylinder (the cylinder parallel to the axis of rotation that circumscribes the inner core).

In the interior, the magnetic field is strongest, and is associated with a vigorous twisted polar vortex, whose dynamics may occasionally lead to the formation of a reverse polar flux patch at the surface of the shell.
Furthermore, the strong magnetic field also allows accumulation of light material within the tangent cylinder, leading to stable stratification there.
Torsional Alfvén waves are frequently triggered in the vicinity of the tangent cylinder and propagate towards the equator.

Outside the tangent cylinder, the magnetic field inhibits the growth of zonal winds and the kinetic energy is mostly non-zonal.
Spatio-temporal analysis indicates that the low-frequency, non-zonal flow is quite geostrophic (columnar) and predominantly large-scale: an m=1 eddy spontaneously emerges in our most extreme simulations, without any heterogeneous boundary forcing.

Our spatio-temporal analysis further reveals that
(i) the low-frequency, large-scale flow is governed by a balance between Coriolis and buoyancy forces -- magnetic field and flow tend to align, minimizing the Lorentz force;
(ii) the high-frequency flow obeys a balance between magnetic and Coriolis forces;
(iii) the convective plumes mostly live at an intermediate scale, whose dynamics is driven by a 3-term MAC balance -- involving Coriolis, Lorentz and buoyancy forces.
However, small-scale ($\simeq E^{1/3}$) quasi-geostrophic convection is still observed in the regions of low magnetic intensity.

\end{abstract}

\tableofcontents
\newpage

\section{Introduction}

Earth's magnetic field is generated by a turbulent flow of liquid metal in the core.
The pioneering work of \citet{glatzmaier1995} launched the first generation of numerical simulations of the geodynamo.
For the first time, a self-sustained magnetic field was produced by a self-consistent convective geodynamo model.
It exhibited several key features of the Earth's magnetic field: a mostly dipolar field aligned with the rotation axis, and polarity reversals.
These findings were somewhat surprising, considering the huge gap between
the parameters used in the simulations and their expected values in the
Earth's core.

Second generation numerical simulations of the geodynamo explored the parameter space and derived scaling laws \citep{christensen2006}.
As the mean properties of the simulations could be cast in laws that excluded diffusion properties (viscosity, thermal and magnetic diffusivities), it was argued that the simulations had reached an asymptotic regime, which allowed extrapolation to the Earth \citep{christensen2010}, planets \citep{christensen2010b}, and some stars \citep{christensen2009}.

Rapidly, several teams questioned that these simulations really reached an asymptotic regime, and that it was the right asymptotic regime for the Earth.
Indeed, \citet{soderlund2012} highlighted the important role of viscosity in their simulation set, and that magnetic forces did not seem to play a major role, contrary to what is expected for the Earth's core.
Reanalyzing the large suite of simulations of \citet{christensen2006}, \citet{king2013} confirmed that viscous dissipation was far from negligible in these simulations;
\citet{cheng2016} showed that the diffusionless scaling laws were hiding an actual dependency upon viscosity;
\citet{oruba2014} derived alternative scaling laws in which the magnetic field intensity depends upon viscosity and rotation rate.

In the mean time, analyses of geomagnetic observations suggested a quasi-geostrophic (invariant along the rotation axis) description of the flow \citep{gillet2011}.
Using this assumption, the large-scale flow at the top of the core inferred from geomagnetic observations can be continued throughout the core, leading to a root-mean-square velocity of about 10~km/yr, that is $\bar{U}_{rms} \simeq 3 \times 10^{-4}$~m/s, encompassing large-scale flow, and consistent with previous estimates of the flow speed at the top of the core.
In addition, several remarkable features were found \citep{pais2015,gillet2015}:
the analysis of \citet{pais2008} yielded a large-scale off-centered quasi-geostrophic gyre around the inner core \citep[also see:][]{schaeffer2011};
\citet{finlay2003} showed evidence for equatorial magnetic waves propagating at decadal time-scales;
\citet{gillet2010} detected torsional oscillations and, from their propagation velocity, inferred the magnetic field inside the core.
Indeed, the propagation speed of torsional waves scale with the intensity $B_{rms}$ of the magnetic field perpendicular to the rotation axis.
Assuming isotropy, \citet{gillet2010,gillet2015} found $B_{rms} = 4$~mT, a value that includes the contribution of all length scales.

These discoveries prompted modelers to look for these signatures in their simulations, but also highlighted that magnetic energy $E_m$ in the Earth's core is much larger than kinetic energy $E_k$.
Indeed, estimating the former using $B_{rms}$ and the latter assuming only the large-scale flow $\bar{U}_{rms}$ leads to $E_m/E_k \simeq 10^4$.
Even allowing for hidden and energetic small scales of about $U_{rms} \simeq 3\,\bar{U}_{rms}$ leads to $E_m/E_k \simeq 1000$.

\begin{figure}
\includegraphics[width=1\textwidth]{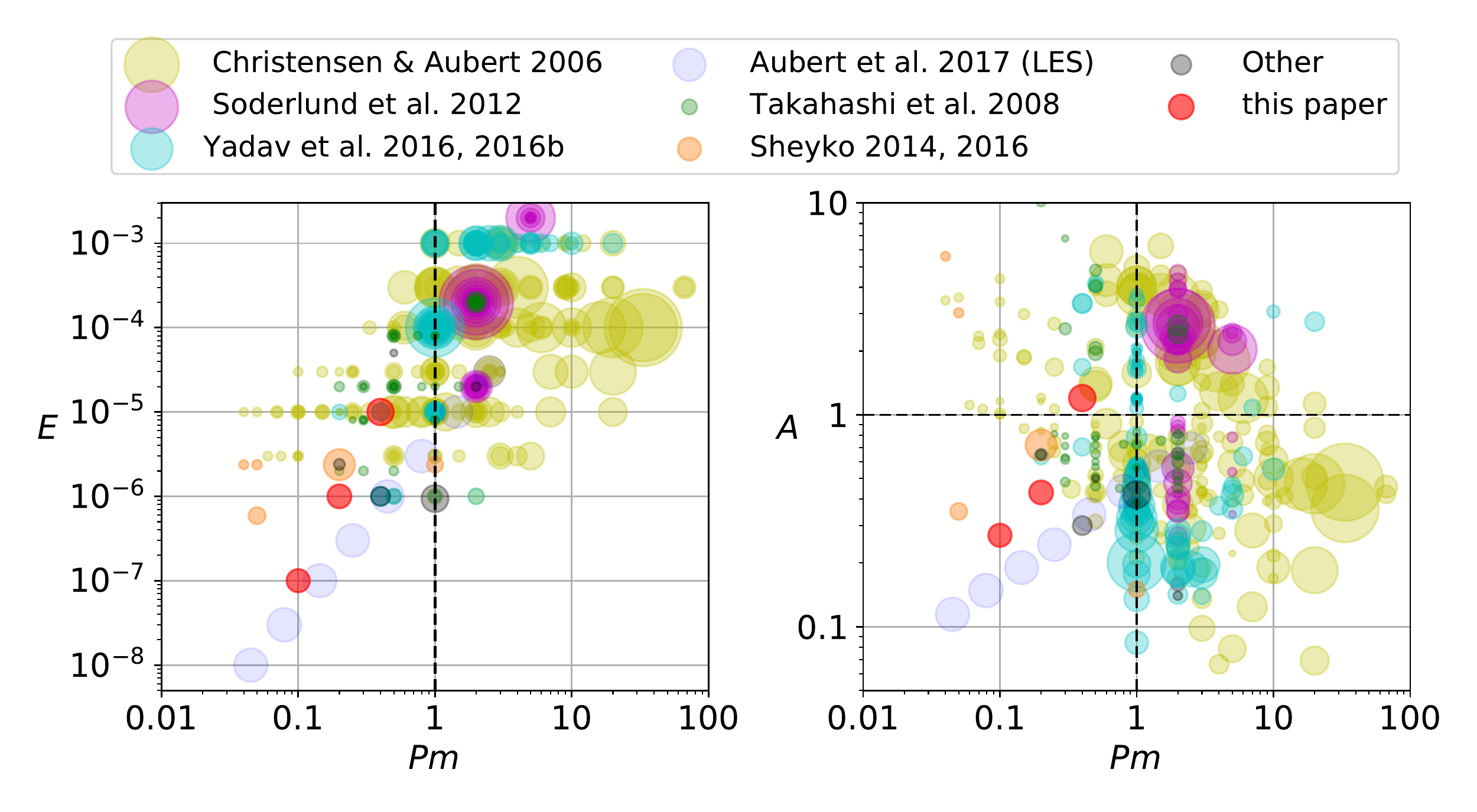}
\caption{Some numerical simulations of the geodynamo represented in parameter space ($E$, $Pm$, $Rm$, $A$), with the surface of the discs being proportional to $Rm$.
$E$ is the Ekman number, $Pm$ the magnetic Prandtl number, $Rm$ the magnetic Reynolds number and $A$ the Alfvén number (see expressions in section \ref{sec:model}).
Our three simulations are shown in red.
In each plot, the Earth's core lies far away beyond the bottom left corner (see table \ref{tab:simus}), out of reach from current models.
The datasets include the work of \citet{christensen2006}, \citet{takahashi2008}, \citet{soderlund2012}, \citet{yadav2016a,yadav2016b}, \citet{sheyko2014phd}, \citet{sheyko2016},
the simulations with hyperviscosity and $Rm \simeq 1000$ of \citet{aubert2017}, and other simulations displayed in table \ref{tab:simus}.
}
\label{fig:Pm_vs_E}
\end{figure}

This strong magnetic field revived the search for inviscid magnetostrophic dynamos -- where Pressure, Coriolis and Lorentz forces form the main balance \citep{livermore2013, wu2015}.
Taking an opposite approach, \citet{dormy2016} obtained strong-field dynamos for large magnetic Prandtl numbers $Pm$ -- the ratio of kinematic viscosity to magnetic diffusivity.
Increased computational power also made it possible to push the parameters towards more realistic values \citep{kageyama2008,sakuraba2009, sheyko2014phd, yadav2016b}.
Figure \ref{fig:Pm_vs_E} displays the current parameter-space coverage that has been achieved by several groups, in terms of Ekman number $E$ (ratio of viscous to Coriolis forces), magnetic Prandtl number $Pm$, magnetic Reynolds number $Rm$ (ratio of induction over ohmic dissipation), and Alfvén number $A$ (ratio of fluid velocity over Alfvén wave velocity, or equivalently the square root of kinetic over magnetic energies).
It shows that having a large $Rm$ is difficult to achieve at low viscosity (low $E$ and low $Pm$, as in the Earth's core), because the flow becomes turbulent, implying that smaller and smaller scales need to be resolved, leading to tremendous computing costs.
Similarly, few simulations lie in the low $Pm$, low $A$ region where the Earth's core is found.
In contrast to direct simulations that are committed to resolve all dynamical scales, \citet{aubert2017} have used a form of eddy viscosity to suppress the smallest scales while keeping the largest ones unaffected.
This allowed them to reach very low values of the large-scale viscosity and argue for a continuous path connecting today's simulations with the Earth's core.

Here, we present an attempt to reach turbulent regime in direct numerical simulations of the geodynamo.
By not artificially damping the smallest flow scales, we ensure unbiased dynamics at all scales.
We have been able to reach strongly forced dynamos at low viscosity, where the magnetic energy becomes one order of magnitude larger than the kinetic energy.
In this way, all obvious time-scales have the same ordering than what is inferred from observations.
From the shortest to the longest time: the rotation period $2\pi/\Omega$, the Alfvén time $\tau_b=D\sqrt{\mu_0\rho}/B$ (typical time for an Alfvén wave to propagate across the core), the turnover time $\tau_u=D/U$ (time for a fluid parcel to travel across the core), the magnetic diffusion time ($D^2/\eta$), and the viscous damping time ($D^2/\nu$).
Fast and slow variations of the magnetic field can be observed, while the flow exhibits tall and thin structures under the effect of the strong global rotation, with a wide range of excited scales.
We also observe torsional waves, which appear to be forced near the tangent cylinder.
Using intensive post-processing, we analyze the fields and forces as a function of time- and length-scale in the two dynamically distinct regions separated by the imaginary cylinder aligned with rotation axis and tangent to the inner-core.

The rest of this paper is organized as follows.
In the next section, we describe our numerical model, as well as the simulation targets.
Then we describe and analyze our set of simulations, focusing on both averaged fields and rapid dynamics.
We conclude the paper with a discussion.

\section{Model equations, target, and numerical methods}

\subsection{Model} \label{sec:model}

In order to simulate the liquid core of the Earth, confined between the solid inner core and mantle, we consider a spherical shell with an inner boundary at radius $r=r_i$ and an outer boundary at $r=r_o$.
The aspect ratio is fixed to $r_i/r_o = 0.35$, as for the present Earth.
The solid boundaries are at rest in the reference frame rotating at a constant rate $\Omega$ along the $z$-axis.
The fluid has an electrical conductivity $\sigma$ and a kinematic viscosity $\nu$, both being constants (in space and time).
Its magnetic permeability $\mu_0$ is that of empty space.

The widely used Boussinesq approximation, coupled with the induction equation via the Lorentz force, is solved numerically from an initial condition.
We use the codensity formulation \citep{braginsky1995}, which combines buoyancy effects due to temperature variations and chemical species concentration into one scalar codensity field $C$, with associated diffusivity $\kappa$.
The acceleration of gravity is in the radial direction and proportional to the radius $r$.
At the outer boundary ($r=r_o$) the radial gravity is $g$ and the codensity gradient is $\beta = -\partial_r C_0|_{r_o}$.
We choose as length-scale the shell thickness $D=r_o-r_i$, and as time-scale the viscous time-scale $D^2/\nu$.
The resulting non-dimensional equations that will be time-stepped numerically read:
\begin{align}
  \partial_t \v{u} + \left(\frac{2}{E}\,\v{e_z} + \nabla \times \v{u}\right) \times \v{u} &= -\v{\nabla} p + \Delta \v{u} + (\nabla \times \v{b}) \times \v{b} - \frac{Ra}{\beta}\,C\,\frac{D}{r_o}\vec{r}   \label{eq:NS} \\
  \partial_t \v{b} &= \nabla \times (\v{u} \times \v{b}) + \frac{1}{Pm} \Delta \v{b} \\
  \partial_t C + \v{u}.\v{\nabla} (C + C_0) &= \frac{1}{Pr} \Delta C \label{eq:codensity} \\
  \nabla . \v{u} = 0 & \quad \quad \quad \quad \quad \nabla . \v{b} = 0
\end{align}
with the Ekman number $E = \nu / D^2 \Omega$, the Rayleigh number $Ra = \beta g D^4/\kappa\nu$, the magnetic Prandtl number $Pm = \nu \mu_0 \sigma$, and the Prandtl number fixed to $Pr = \nu / \kappa = 1$.
$\v{u}$ is the velocity field and $\v{b}$ the magnetic field in (Alfv\'en) velocity units (i.e. it has been scaled by $\sqrt{\mu_0\rho}$, where $\rho$ is the homogeneous core density).

The conductive codensity profile $C_0(r)$ is obtained using the thermochemical model of \citet{aubert2009}, as detailed in appendix \ref{sec:codensity}.

At both inner and outer boundaries, the codensity deviation from $C_0$ is imposed to have a zero gradient $\partial_r C = 0$  (effectively fixing the constant uniform heat flux set by $C_0$), the velocity is zero (no-slip), and the magnetic field is matched to a potential field outside the fluid domain (the inner-core and mantle are both modeled as electrically insulating).
Note that fixing the codensity flux also enforces uniform crystallization of the inner-core.

\subsection{Target features for our simulations}

The Earth's core operates at low viscosity (as measured by $E \simeq 10^{-15}$ and $Pm \simeq 10^{-6}$), so we should try to lower $E$ and $Pm$ as much as possible.
Figure \ref{fig:Pm_vs_E} (left) illustrates the difficulty to lower $E$ and $Pm$: the magnetic Reynolds number $Rm$ (represented by the size of the circles) decreases and at some point it is not enough to maintain a magnetic field against ohmic dissipation.

Another outstanding feature of Earth's dynamo is the small ratio of kinetic to magnetic energy, which is the squared Alfvén number $A^2 \simeq 10^{-4}$.
It is apparent in Figure \ref{fig:Pm_vs_E} (right) that low $A$ are difficult to obtain for $Pm<1$.
Low Alfv\'en number dynamos are readily obtained with $Pm \geq 1$ \citep[e.g.][]{kageyama2008, dormy2016} and just above the convection threshold \citep[e.g.][]{takahashi2012}.
\citet{sakuraba2009} argued that imposing the heat flux rather than the temperature at the boundaries also helps to get a strong magnetic field (i.e. a low $A$).
However, the convective power was not the same for their fixed flux and fixed temperature cases.
Indeed, for fixed flux the convective power is proportional to $Ra$, whereas for fixed temperature the convective power is proportional to $(Nu-1)Ra$, leading to weaker driving power in the fixed flux case for the same value of super-criticality $Ra/Ra_c$ \citep[Fig. 1]{aubert2017}.
According to the scaling laws of \citet{christensen2006}, the Alfvén number $A$ increases when the convective power increases, and the trend is in broad agreement with the field strength obtained in our simulations.

Our goal is to obtain vigorous convection (far above its onset) and low $Pm$ dynamo with a strong magnetic field ($A < 1$).

When trying to compute geodynamo models as close as possible to the parameters of the Earth's core, the computation cost increases not only because of the higher and higher spatial resolution required, but also because the time-step is smaller and smaller compared to the magnetic diffusion time.
Hence, in order to reach a statistically stationary dynamo regime, the time needed for a simulation to run increases prohibitively \footnote{for our series of simulations, the computing resources needed to span a magnetic diffusion time increase by a factor of about 30 to 100 when the Ekman number is divided by 10}.
To reach extreme parameters in our simulations, we prepare the initial conditions close to the expected statistical equilibrium state, in order to reduce the duration of transients.
Those initial conditions are obtained by applying previously established scaling laws \citep{christensen2006} to the output of a lower resolution simulation at parameters further from the Earth's core.
This procedure can be repeated to achieve simulations that are closer and closer to the conditions of the Earth's core.

\subsection{Numerical implementation}

We use the XSHELLS code, available as free software\footnote{\label{xsweb}\url{https://www.bitbucket.org/nschaeff/xshells/}}.
It passes the dynamo benchmark of \citet{christensen2001} as reported by \citet{matsui2016}.
It uses second order finite differences in radius and pseudo-spectral spherical harmonic expansion.
The time-stepping scheme is second order in time, and treats the diffusive terms implicitly, while the non-linear and Coriolis terms are handled explicitly.
We have carefully optimized the code for speed.
The SHTns library \citep{schaeffer2013} performs all the spherical harmonic transforms and, thanks to its low memory requirement and high performance, we can reach high resolutions (up to harmonic degree $\ell_{max}=1000$ in this study).
A domain decomposition in the radial direction allows efficient parallel execution using multiple processes (using the MPI standard).
In addition, within each process, multiple threads (using the OpenMP standard) are used for an added level of parallelism.
The most demanding simulation presented below has been run routinely on up to 8192 cores with a good scaling of the run time with the number of cores.
More details about performance can be found in \citet{matsui2016} and in the user manual\footnotemark[2].

\begin{table}
 \footnotesize
\begin{tabular}{c|ccccc|ccc|c}
     &  \textit{K08} & \textit{S09} & \textit{S14} & \textit{Y16} & \textit{A17} & \textbf{S0} & \textbf{S1} &  \textbf{S2} & Earth\\
\hline
$N_r$      & 511  & 160 & 528 & 181 & 624 & 256  & 512 &  1280 &  \\
$N_\theta$ & 1024 & 384 & 384 & 640 & 200 & 448  & 720  & 1504 &  \\
$N_\phi$   & 2048 & 768 & 768 & 1280 & 400 & 720 & 1440  & 2688 & \\
time & 0.016 & & 0.027 & 0.05 & 0.12 & 2.2 & 0.51 & 0.052 &  \\
\hline
$E$ &  $9.4\,10^{-7}$ & $2.4 \, 10^{-6}$ & $5.9 \, 10^{-7}$ & $10^{-6}$ & $10^{-8}$ & $10^{-5}$ & $10^{-6}$ & $10^{-7}$ & $3\,10^{-15}$ \\
$Ra/Ra_c$ & $\sim 300$ & 144 & 470 & 400$^*$ & & 4879 & 5770 & 6310 & $10^{6}$ ?\\ 
$Pm$ & 1 & 0.2 & 0.05 & 0.4 & 0.045 & \textbf{0.4} & \textbf{0.2} & \textbf{0.1} & $2\, 10^{-6}$\\
$Pr$ & 1 & 1 & 1 & 1 & 1 & 1 & 1 & 1 & 0.1 - 10 \\
\hline
$Rm$ &  700 & 120 & 274 & 346 & 1082 & \textbf{671} & \textbf{546} & \textbf{514}  & 2000 \\
$A$  &  0.42 & 0.46 & 0.35 &  0.3 & 0.11 & \textbf{1.2} & \textbf{0.43} & \textbf{0.27} & 0.01\\ 
$Re$ &  700 & 600 & 5480 & 865 & 24000 & 1680 & 2730 & 5140 & $10^9$\\
$Ro$ & $7.6\,10^{-4}$ & $1.4\,10^{-3}$ & $3.2\,10^{-3}$ & $8.7\,10^{-4}$ & $2.4\,10^{-4}$ & 0.017  & $2.7\,10^{-3}$ & $5.1\,10^{-4}$ & $3\, 10^{-6}$ \\
$Le$ &  $1.8\,10^{-3}$ & $2.2\,10^{-3}$ &$9.2\,10^{-3}$ & $2.9\,10^{-3}$ & $2.1\,10^{-3}$ & 0.014 & $6.4\,10^{-3}$ & $1.9\,10^{-3}$ & $10^{-4}$\\
$\Lambda$ & 3 & 0.8 & 3.6 & 3.4 & 20 & 8.0 & 8.2 & 3.7 & $\gtrsim 10$
\end{tabular}
\caption{Various input and output parameters of our dynamo simulations (S0, S1 and S2) compared to the Earth's core and the simulation of \citet{kageyama2008} labeled \textit{K08}, the UHFM case of \citet{sakuraba2009} labeled \textit{S09}, and the least viscous cases of \citet{sheyko2014phd}, \citet{yadav2016b} and \citet{aubert2017} labeled \textit{S14}, \textit{Y16} and \textit{A17} respectively.
We use the following definitions: $E=\nu/D^2\Omega$, $Pm = \nu\mu_0\sigma$, $Pr = \nu/\kappa$, $Rm=UD\mu_0\sigma$, $A=\sqrt{\mu_0\rho}U/B$, $Re=UD/\nu$, $Ro=U/D\Omega$, $Le=B/\sqrt{\mu_0\rho}D\Omega$, $\Lambda=B^2\sigma/\rho\Omega$.
The numbers of \textit{K08}, \textit{S09}, \textit{Y16} and \textit{A17} have been cast to our definitions, where $D$ is the shell thickness, $U$ the rms velocity and $B$ the rms magnetic field averaged over the whole fluid domain.
The number of discretization points in the radial, latitudinal and longitudinal directions are denoted by $N_r$, $N_\theta$ and $N_\phi$ respectively.
The \textit{time} refers to the simulated time, normalized by the magnetic diffusion time $D^2\mu_0\sigma$.
$Ra/Ra_c$ is the Rayleigh number divided by its value $Ra_c$ at the onset of convection for the same $Pr$ and $E$.
For \textit{S09, S14,} S0, S1 and S2, $Ra_c$ has been computed precisely using the SINGE eigenmode solver \citep{vidal2015}.
$^*$ Note that $Ra/Ra_c$ for \textit{Y16} has been multiplied by their Nusselt number to account for the different thermal boundary conditions (fixed temperature for $Y16$).
}
\label{tab:simus}
\end{table}

\section{The simulations}

\subsection{Overview}

\begin{figure}
\includegraphics[width=0.99\textwidth]{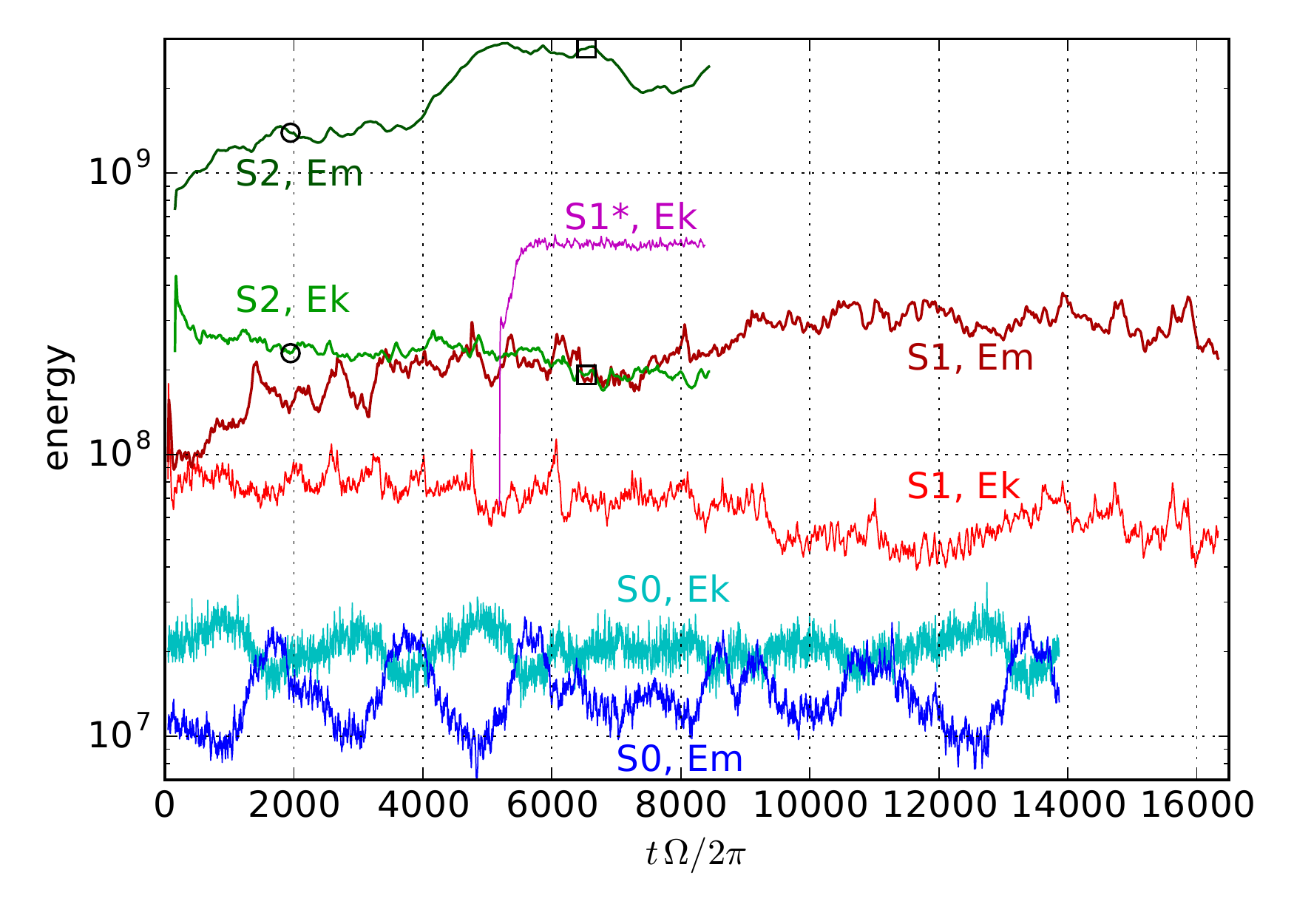}
\caption{Kinetic (Ek) and magnetic (Em) energies as a function of time (normalized by the rotation period $2\pi/\Omega$) for our four cases (S0, S1, S1* and S2).
Simulations S0, S1 and S2 span respectively 2.2, 0.51 and 0.052 magnetic diffusion times ($D^2/\eta$).
For S1 and S2, magnetic energy is much larger than kinetic energy.
The circles and squares mark the instants that are displayed in section \ref{sec:instant}.
}
\label{fig:nrj}
\end{figure}


We have run three dynamo simulations (S0, S1, S2) that have decreasing Ekman numbers, $E=10^{-5}$, $10^{-6}$ and $10^{-7}$ respectively.
They all have a magnetic Reynolds number $Rm$ between 500 and 700.
This is achieved by lowering the magnetic Prandtl number $Pm$ and increasing the Rayleigh number to preserve high supercriticality ($Ra/Ra_c > 4870$ is almost constant, where $Ra_c$ is the Rayleigh number for the onset of non-magnetic convection).
In order to avoid transients, initial fields for simulations S1 and S2 are produced from S0 and S1 respectively, by multiplying each field by a scaling factor to reach the energy levels expected from the scaling laws of \citet{christensen2006}.
Note that the S0 simulation is a continuation of the simulation presented by \citet{fournier2012egu}, and that parts of both S0 and S1 were used by \citet{bouligand2016} with the same name.

In order to quantify the influence of the magnetic field, we have run a fourth simulation, S1*, which has the same parameters as S1, but without magnetic field.
S1* is started from velocity and codensity fields of S1 at a given time, without any rescaling.

Table \ref{tab:simus} summarizes the input parameters of our simulations, and compares it to the simulations of \citet{kageyama2008}, \citet{sakuraba2009}, \cite{sheyko2014phd}, \citet{yadav2016b} and \citet{aubert2017}.
The table also includes some output parameters computed from averaged quantities.
No magnetic polarity reversal occurred in our simulations.

Figure \ref{fig:nrj} shows both kinetic and magnetic energies as a function of time, for our four simulations.
It is clear that the ratio of magnetic over kinetic energy increases significantly from S0 to S2.
In addition, the kinetic energy of S1* is about 10 times larger than that of S1, showing the strong influence of the magnetic field.
In S1 and S2, the magnetic energy has fluctuations of larger amplitudes but lower frequencies than the kinetic energy.
Furthermore, all the dynamo simulations exhibit much larger variations of their energy levels compared to the non-magnetic case S1*, with much longer correlation times.
This seems to be an inherent property of turbulent dynamo simulations.
We note that the magnetic energy in S1 and S2 increases above the level expected by the scaling laws of \citet{christensen2006}, but not by more than a factor 3, which corresponds also to the fluctuations observed in S0.
However, the magnetic energy of all three simulations correspond to the trend exhibited by the \citet{christensen2006} dataset, which also displays a scatter by a factor 3 to 10 (not shown).

We have checked the spatial convergence of S2 by multiplying the number of radial shells by 1.5, and the maximum degree of spherical harmonics by 1.12.
No noticeable difference was found in the energy levels.
Furthermore, the highest degrees of the spherical harmonic spectra have two orders of magnitude lower energy than the most energetic degree, at each radius.
As a supplementary convergence check, the time-averaged total energy dissipation rate is equal to the time-averaged power of the buoyancy force within less than 1\% in all our dynamo simulations (values listed in table \ref{tab:mysimus}).

\begin{table}
\begin{center}
\begin{tabular}{ccccc}
     &  \textbf{S0} & \textbf{S1} & \textbf{S1*} & \textbf{S2} \\
\hline
$E$ &  $10^{-5}$ & $10^{-6}$ & $10^{-6}$ & $10^{-7}$  \\
$Ra$ & $6.34\,10^{9}$ & $1.27\,10^{11}$ & $1.27\,10^{11}$ &  $2.54\,10^{12}$ \\
$Ra/Ra_c$ & 4879 & 5770 & 5770 & 6310 \\
$Pm$ & 0.4 & 0.2 & 0 & 0.1 \\
\hline
$Rm$ &  671 & 546 & 0  & 514 \\
$A$  &  1.2 & 0.43 & & 0.27 \\ 
$Re$ &  1680 & 2730 & 8760 & 5140 \\
$Ro$ & 0.017 & $2.7\,10^{-3}$ & $8.8\,10^{-3}$  & $5.1\,10^{-4}$ \\
$Le$ & 0.014 & $6.4\,10^{-3}$ & 0  & $1.9\,10^{-3}$\\
$\Lambda$ & 8.0 & 8.2 & 0 & 3.7 \\
\hline
$Nu$ & 31 & 45 & 42 & 59 \\
$P_{conv}$ & $2.643\, 10^{11}$ & $5.579 \, 10^{12}$ & $5.54 \, 10^{12}$ & $10.3 \, 10^{13}$\\
$D_\eta$ & $1.540\, 10^{11}$ & $4.451 \, 10^{12}$ & 0 & $8.71 \, 10^{13}$ \\
$D_\nu$ & $1.085\, 10^{11}$ & $1.113 \, 10^{12}$ & $5.48 \, 10^{12}$ & $1.44 \, 10^{13}$\\
$f_{ohm}=D_\eta/(D_\nu+D_\eta)$ & 59\% & 80\% & 0\%   & 86\%  \\
$L_u$ & 0.020 & 0.011 & 0.014 &  $5.2\,10^{-3}$ \\
$L_b$ & 0.022 & 0.025 & & 0.023 \\
$m_c$ & 15 & 32 & 32 & 67 \\
$\bar{\ell}$ & 26 & 46 & 24 & 86 \\
\hline
$Ek(m=0)/Ek$ & 0.09 & 0.12 & 0.50 & 0.06 \\
$f_{dip}$ & 0.49 & 0.72 & & 0.67 \\
$B_{surf}(\ell=1) / B_{rms}$ & 0.11 & 0.16 &   & 0.10  \\
\hline
$\langle|\mathcal{T}(s,t)|\rangle_{s,t}$    & 0.447 & 0.161 & & 0.119  \\
$\langle|\langle\mathcal{T}(s,t)\rangle_t|\rangle_s$  & 0.257 & 0.0247 & & 0.0219 \\
\hline
$P_{conv}/(D_\nu+D_\eta)$ & 1.007 & 1.003 & 1.011 & 1.010 \\
$N_r$     & 256 & 512 & 512 & 1280 \\
$\ell_{max}$ & 297 & 479 & 479 & 1000 \\
$m_{max}$ & 238 & 479 & 479 & 893 \\
million core.hours & $\simeq 0.4$ & $\simeq 1.4$ & $\simeq 0.4$ & $\simeq 10$
\end{tabular}
\end{center}
\caption{
The Nusselt number $Nu$ is computed from the averaged codensity profile $\bar{C}$ by $Nu = (1+\Delta\bar{C}/\Delta C_0)^{-1}$, with $\Delta C = C(r_o)-C(r_i)$ the codensity drop across the shell.
$L_u = u_{rms}/\omega_{rms}$ and $L_b=b_{rms}/j_{rms}$ are velocity and magnetic field length-scales.
$m_c$ is the critical azimuthal wavenumber at the onset of (non-magnetic) convection, computed using SINGE \citep{vidal2015}.
$\bar{\ell}$ is the mean degree computed from kinetic energy spectrum (see \S\ref{sec:spec}).
Values given for S2 are obtained using time-averages over the last half of the time series (with strongest field).
For definition of $f_{dip}$ and $\mathcal{T}$ see section \ref{sec:br_surf} and \ref{sec:taylor} respectively.
The time-averaged power of the buoyancy force $P_{conv}$, viscous dissipation rate $D_\nu$ and ohmic dissipation rate $D_\eta$ are computed (including boundary layers).
The value $P_{conv}/(D_\nu+D_\eta)$ should be close to one and is listed as a convergence check.
The approximate cost of our simulations is given in millions of core.hours on an \textit{Intel Sandy Bridge} machine.
}
\label{tab:mysimus}
\end{table}

\subsection{Mean fields}

\begin{figure}
\includegraphics[width=0.24\linewidth]{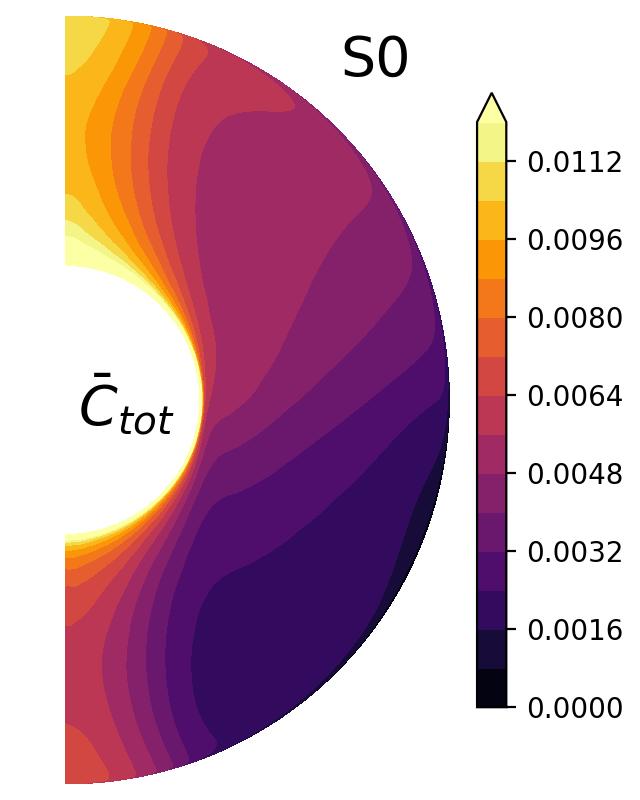}
\includegraphics[width=0.24\linewidth]{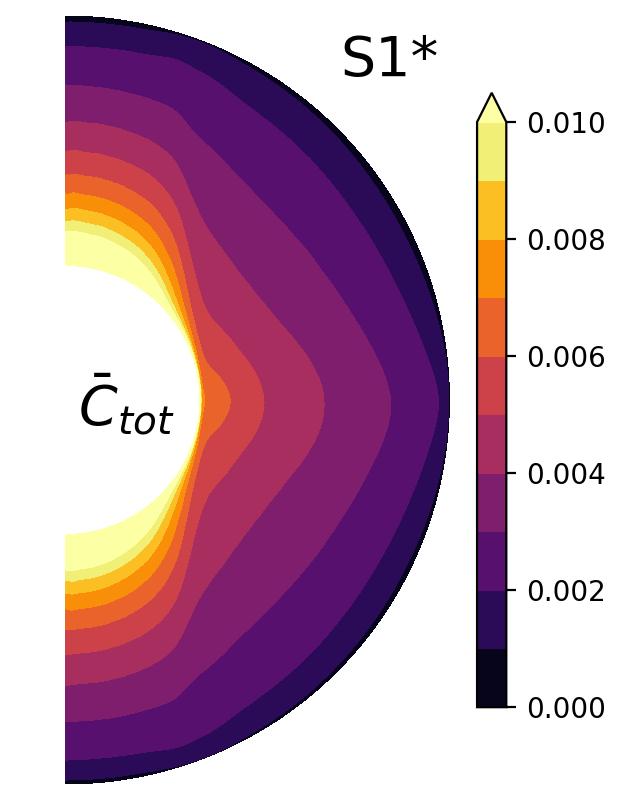}
\includegraphics[width=0.24\linewidth]{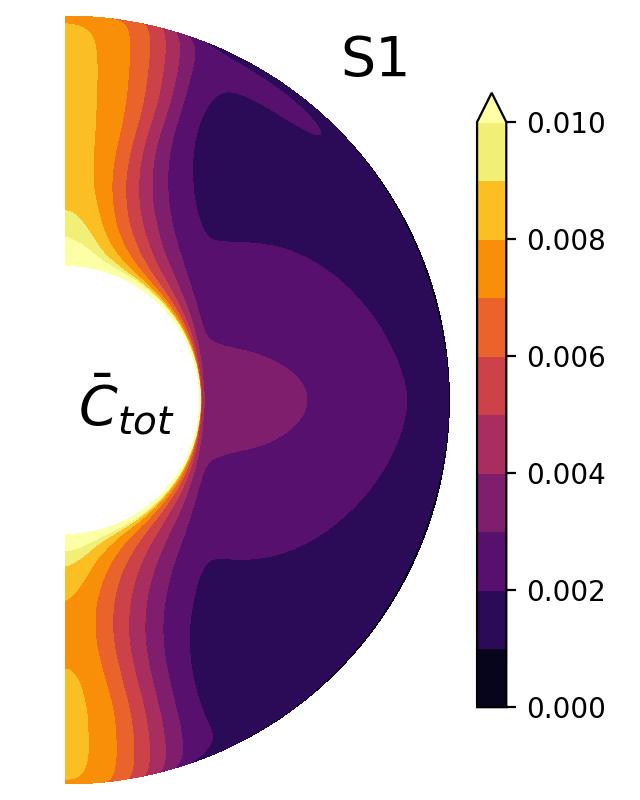}
\includegraphics[width=0.24\linewidth]{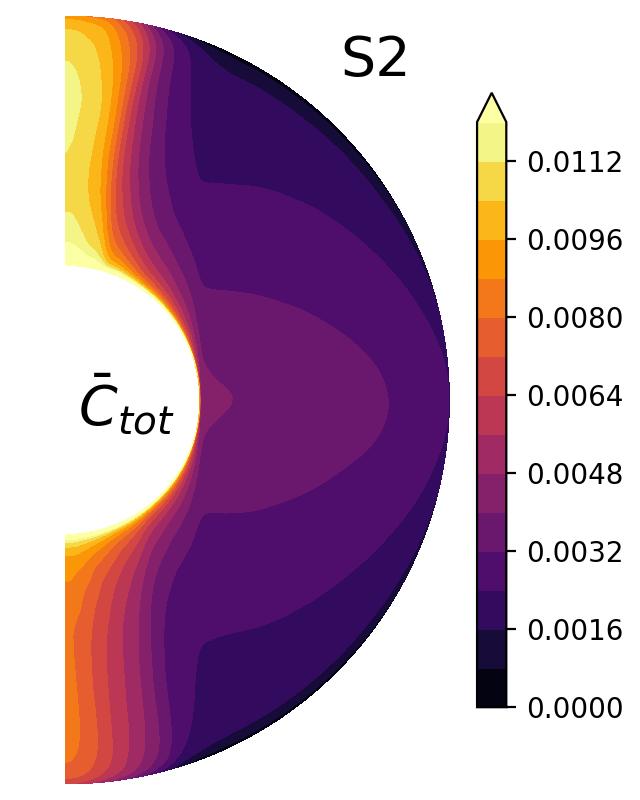} \\
\includegraphics[width=0.24\linewidth]{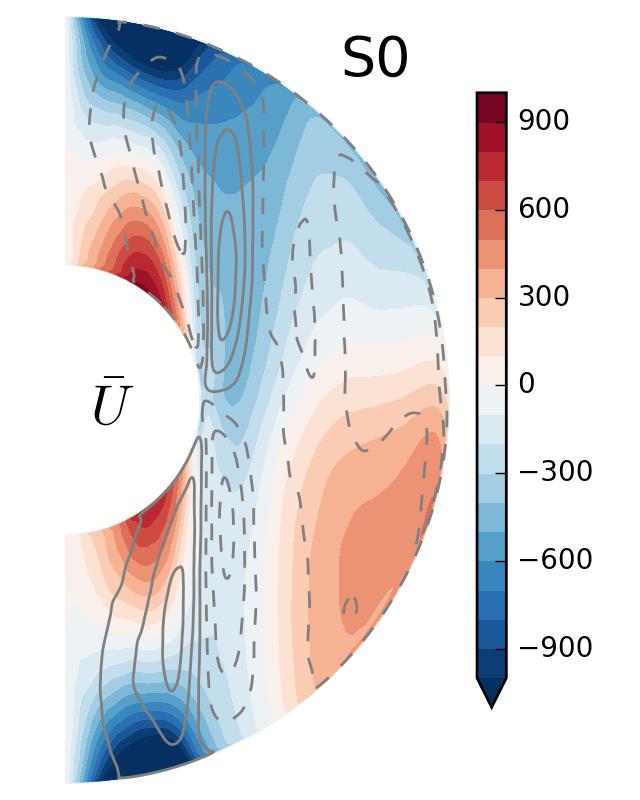}
\includegraphics[width=0.24\linewidth]{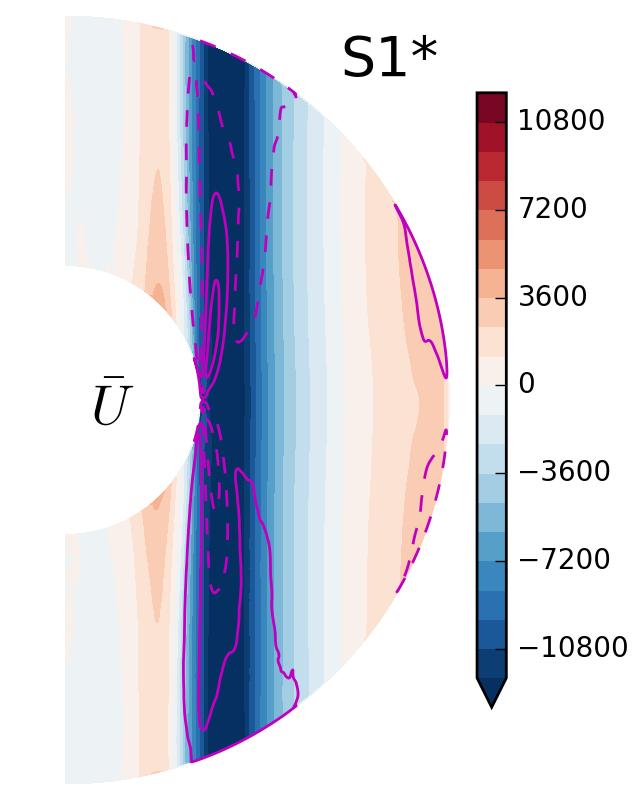} 
\includegraphics[width=0.24\linewidth]{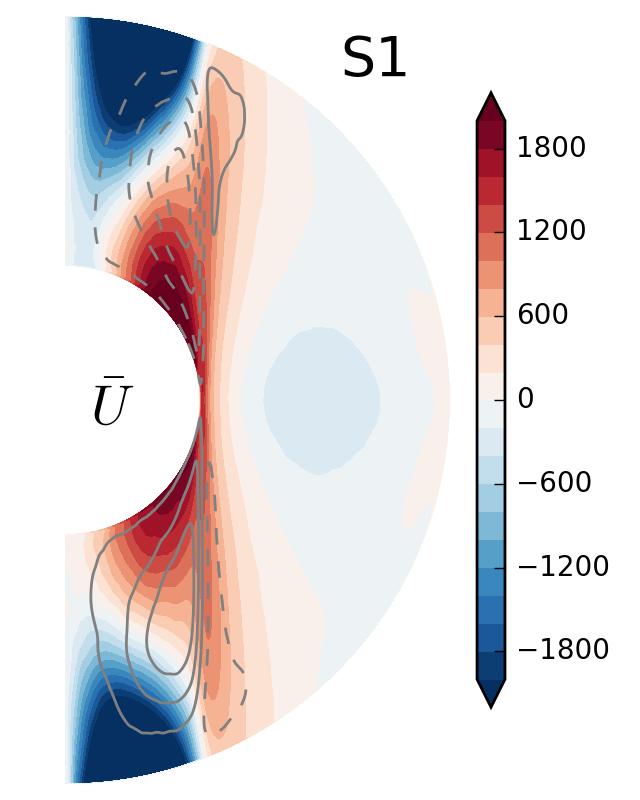} 
\includegraphics[width=0.24\linewidth]{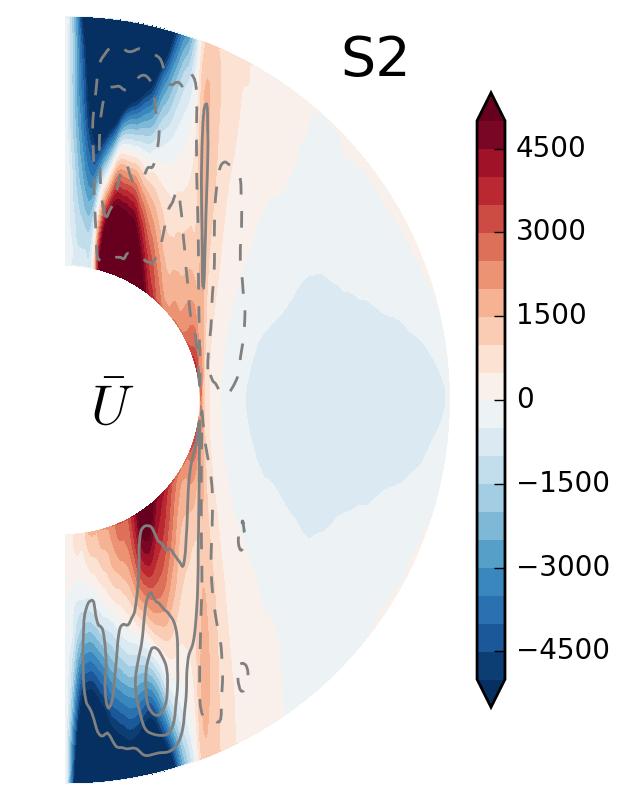} \\
\includegraphics[width=0.24\linewidth]{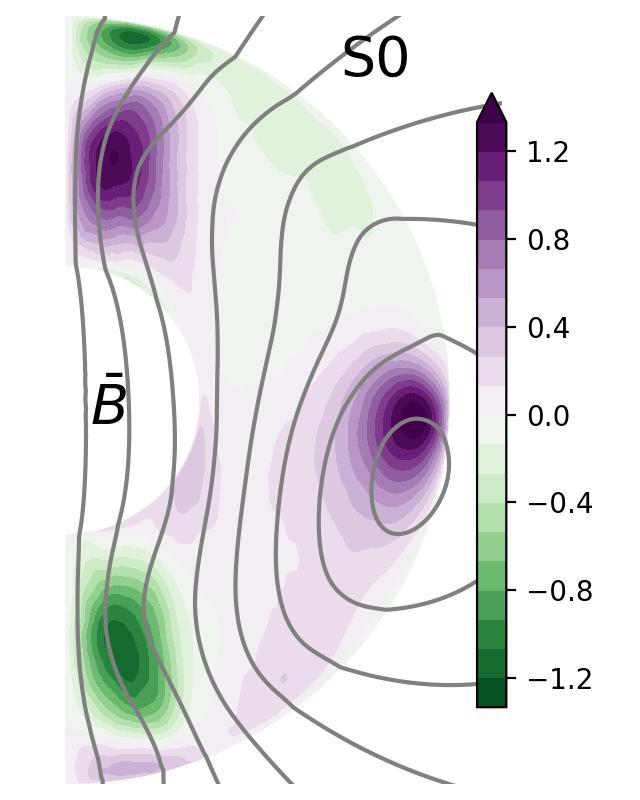}
\hspace{0.24\linewidth}
\includegraphics[width=0.24\linewidth]{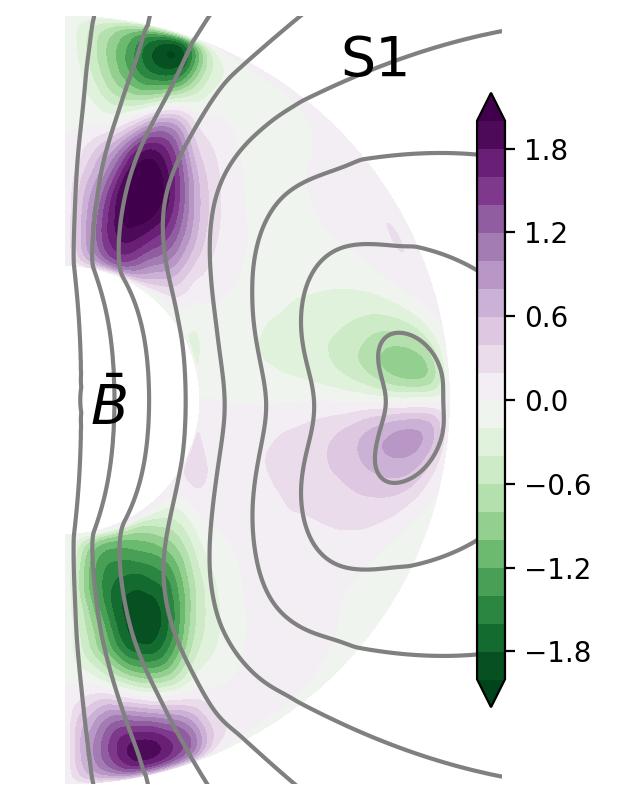}
\includegraphics[width=0.24\linewidth]{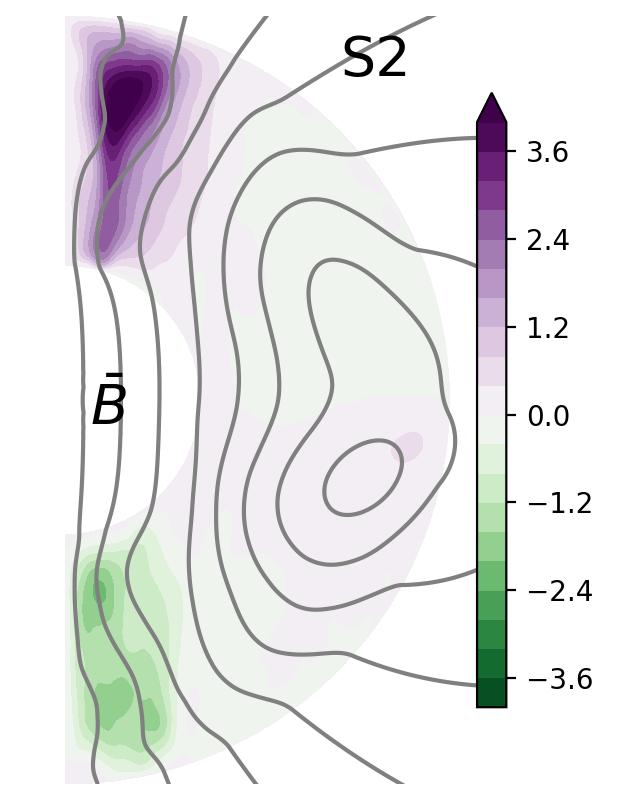} \\
\includegraphics[width=0.24\linewidth]{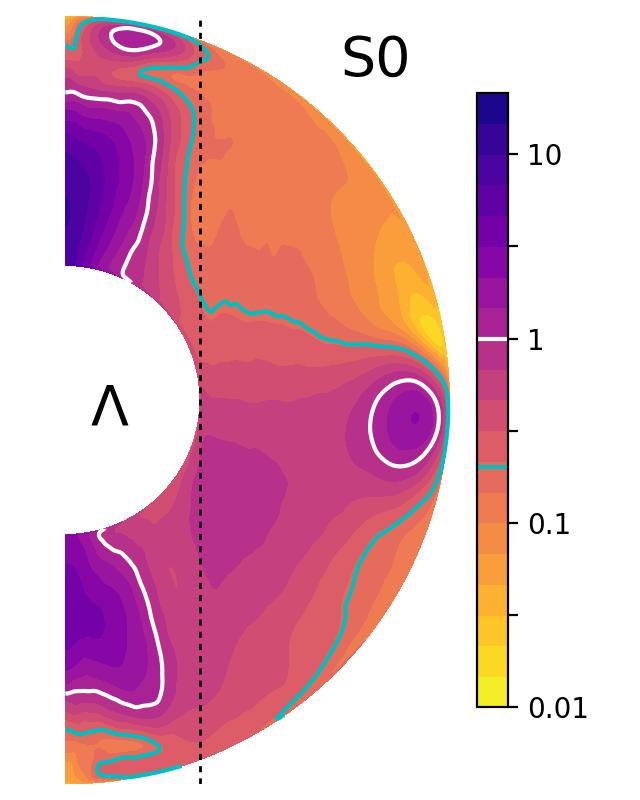}
\hspace{0.24\linewidth}
\includegraphics[width=0.24\linewidth]{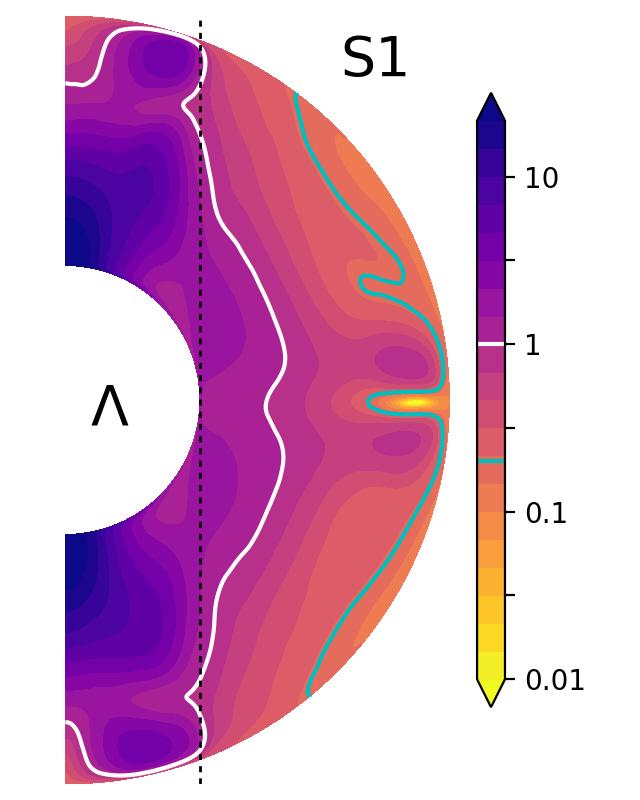}
\includegraphics[width=0.24\linewidth]{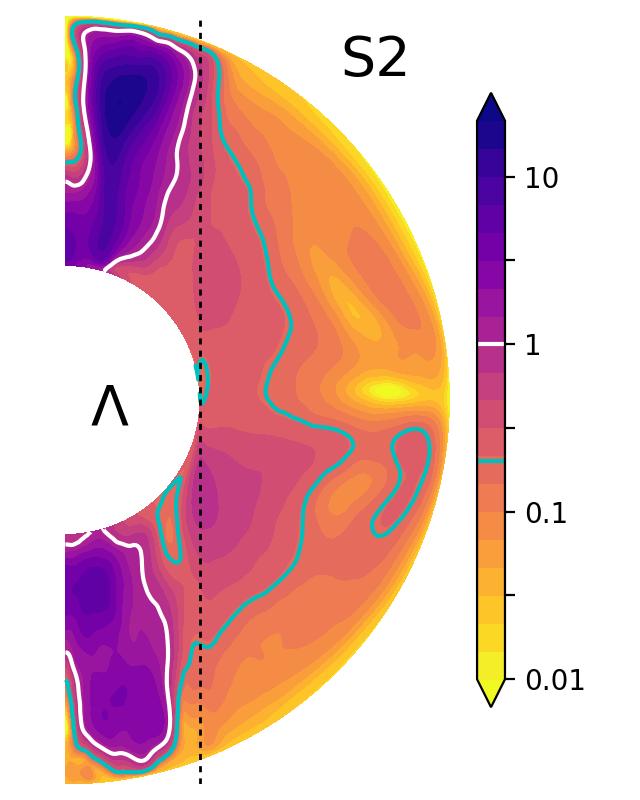}
\caption{Time and longitude averaged codensity (top row, $\bar{C}_{tot}=\bar{C}+C_0$), velocity (second row) and magnetic (third row, in Elsasser units) fields in our simulations.
The bottom row is the Elsasser number $\Lambda = |\bar{B}|^2$ computed from the mean magnetic field, with the white and cyan contours for $\Lambda=1$ and $\Lambda=0.2$ respectively.
For vector fields, the color map shows the azimuthal (toroidal) field ($U_\phi$ or $B_\phi$), while the contours represent the meridional (poloidal) field lines.
The magnetic energy contained in the poloidal field is about 0.9 the one contained in the toroidal field.
The parameters of simulations S0, S1, S1* and S2 are given in table \ref{tab:mysimus}.
}
\label{fig:avg}
\end{figure}

Time and longitude averaged fields are represented in figure \ref{fig:avg}, for the three dynamo simulations S0, S1 and S2, and for the non-magnetic convection simulation S1*.

The root-mean-square (rms) poloidal magnetic field is comparable to the toroidal one, with a poloidal over toroidal ratio of $1.73$, $0.95$ and $0.83$ for simulations S0 to S2.
In all three cases, the amplitude of meridional velocity is weak, about 4 to 6 \% of the zonal one.

A North-South asymmetry remains in S0 and S2 after averaging over the time $T_{avail}$ when the required data is available (in terms of turn-over time $\tau_u = D/U$: $T_{avail}/\tau_u = 1390$ for S0, $190$ for S1, and $8$ for S2).
Note that a similar asymmetry is present in snapshots of S1 (not shown).

The imaginary cylinder tangent to the inner-core and aligned with the rotation axis (hereafter named tangent cylinder or TC) separates two regions of different dynamics.
The lower the Ekman number, the sharper the transition between these two regions.
This dichotomy is visible on both velocity components as well as on the toroidal magnetic field, for S1 ($E=10^{-6}$) and S2 ($E=10^{-7}$).
A sharp shear layer, associated with a meridional circulation materializes the tangent cylinder, and is reminiscent of Stewartson layers \citep{stewartson1966}.
The zonal mean flow inside the tangent cylinder has amplitudes about 10 times larger than outside.
This contrasts with the findings of \citet{aubert2005} who argued for iso-rotation along magnetic field lines and had comparable flow amplitudes inside and outside the tangent cylinder.
However, we checked that the scaling of the zonal flow proposed by \citet{aubert2005} seems to hold.

Interestingly, the total codensity shows that lighter fluid is trapped within the tangent cylinder (see also Fig \ref{fig:3D}), with strong lateral gradients.
This feature is much less pronounced in the non-magnetic case S1*. 
A more quantitative view is given by the codensity profiles of \ref{fig:Cprof} in the Appendix.
It is noteworthy that all dynamo simulations present an adverse codensity gradient within the tangent cylinder.
This emergence of a stable stratification within TC is most likely an effect of the strong magnetic field, as the stratification remains unstable everywhere in the non-magnetic S1*.
Note that in S0, light material is not well contained and appears to leak (especially in the northern hemisphere, see fig.~\ref{fig:avg} top).
In S1, the leak is barely visible and the confinement seems even better in S2.

\subsubsection{Outside the tangent cylinder}
Outside the tangent cylinder, the averaged flow is weak in both S1 and S2.
The zonal wind amplitude is $Ro \simeq 5 \times 10^{-5}$ in S2, about 10 times smaller than the overall rms averaged Rossby number (see table \ref{tab:mysimus}).
In particular, the zonal wind is 10 times smaller outside than inside the tangent cylinder, and 30 times smaller in S1 than in the corresponding non-magnetic thermal convection of S1* (see figure \ref{fig:avg}), showing the strong effect of the magnetic field on the mean flow.
Note also that the weak mean flow next to the equator of the core-mantle boundary (CMB) is rather eastward (prograde) in S1, but westward (retrograde) in S2.
The mean toroidal magnetic field is also weak in S1 (corresponding Elsasser number $\Lambda \simeq 0.05$) and even more so in S2 ($\Lambda \simeq 0.01$), but the poloidal field is rather strong, especially near the inner-core (where the associated Elsasser number reaches 0.25 in S2, see fig.~\ref{fig:avg} bottom).
The much lower values of Elsasser numbers here compared to the global root-mean-square Elsasser given in table \ref{tab:simus} emphasize the weak average toroidal magnetic field outside TC.

\subsubsection{Inside the tangent cylinder}
Inside the tangent cylinder, a strong azimuthal flow takes place, in the prograde direction close to the inner-core, and in the retrograde direction close to the mantle, with an associated meridional circulation (one cell).
The zonal flow amplitude is $Ro \simeq 5 \times 10^{-4}$ in S2, similar to the overall rms Rossby number (see table \ref{tab:mysimus}).
They form what has been coined as the polar vortex \citep{olson1999,aubert2005}.
This twisted polar vortex is anticyclonic next to the CMB.
It has been associated with a low influence of inertia and a strong magnetic field \citep{sreenivasan2006}.
This flow resembles the Taylor vortices described in rotating thermal convection \citep[e.g.][]{grooms2010}, or the Von-Karman flow generated by two impellers co-rotating at different speeds (in order to obtain one meridional circulation cell).
In this respect, the VKS experiment \citep{monchaux2009} may be not so far from the flow inside one hemisphere of the tangent cylinder.
This contrasts with the non-magnetic case S1*, where the zonal flow is almost invariant along the rotation axis.
Conversely, codensity shows little variations along the rotation axis in the dynamo simulations, but important ones in S1*.
The twisted zonal flow within the tangent cylinder also reaches higher speeds in S1 than in S1*, suggesting that the strong toroidal field there allows the flow to break the Taylor-Proudman constraint imposed by the global rotation.

The toroidal magnetic field is concentrated here, suggesting that an omega effect is associated with the strong twisted vortices that dominate the mean flow.
The Elsasser number associated with this strong toroidal field is close to unity in S0, S1 and S2.
At Elsasser close to one, convection onsets more easily and its length-scale increases dramatically from order $E^{1/3}$ to order 1 \citep{chandrasekhar1961,busse2011,aujogue2015}.
All the conditions for such an effect are met within the tangent cylinder of S1 and S2.
Note also that the toroidal magnetic field in S2 seems to have switched to another topology (only one sign in each hemisphere of S2 instead of a change of sign in S1), with minor impact on the mean poloidal field (but see Fig. \ref{fig:Br}).

\begin{figure}
\includegraphics[width=0.49\linewidth]{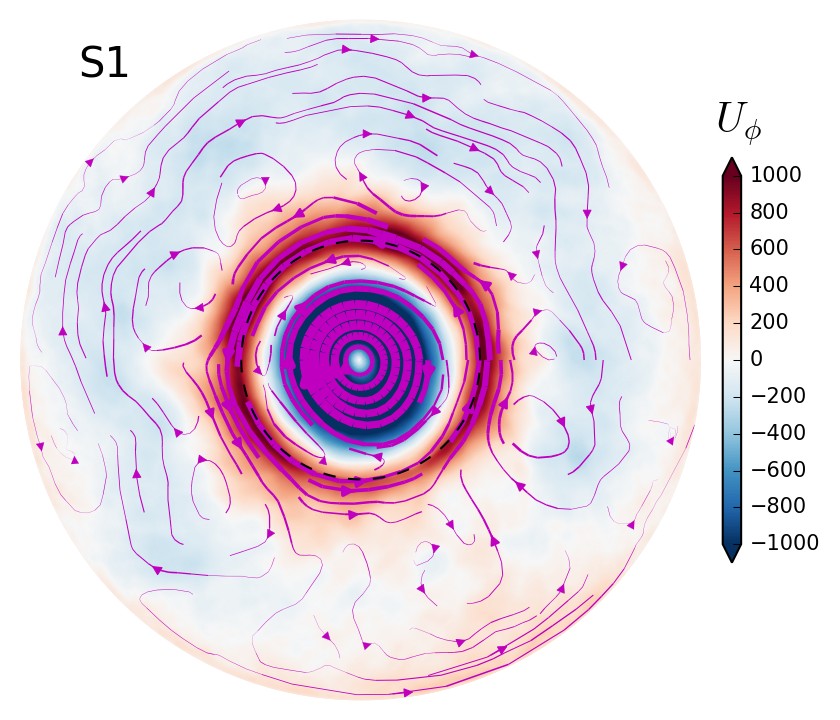}
\includegraphics[width=0.49\linewidth]{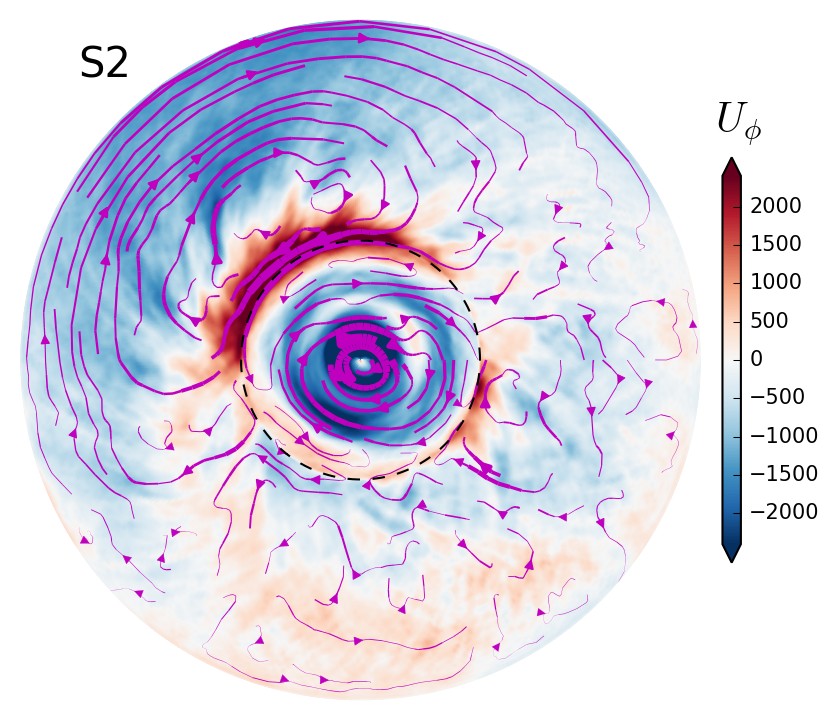} \\
\includegraphics[width=0.49\linewidth]{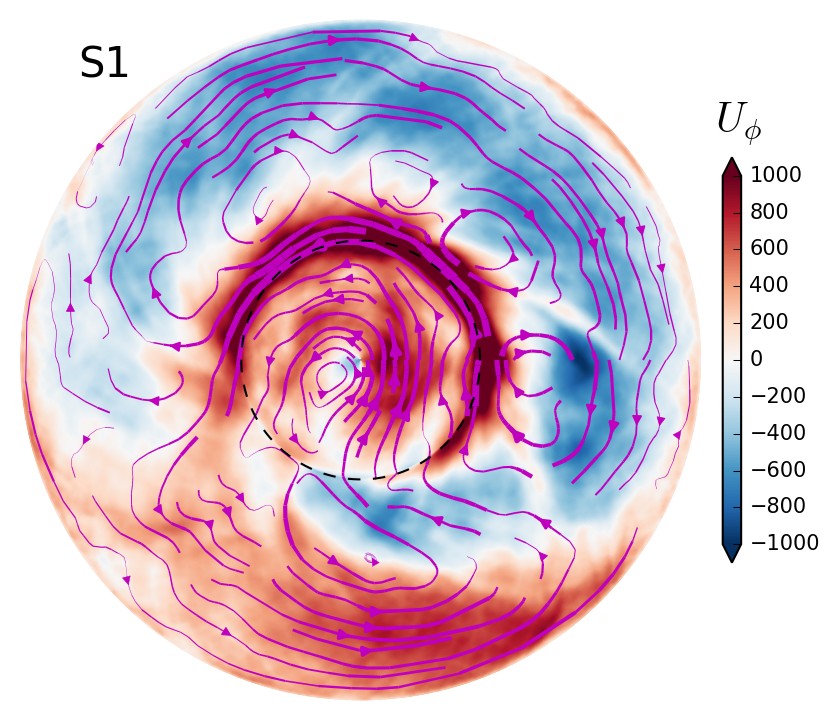}
\caption{Mean flow (averaged in time and along the rotation axis) in simulations S1 (left) and S2 (right).
The streamlines materialize the velocity field, while the color map highlights the azimuthal component.
In these views from the north-pole, the black dashed circle indicates the location of the tangent cylinder.
Top row: the time-average spans 0.35 magnetic diffusion time or 190 turn-over times for S1, and 0.016 magnetic diffusion time or 8 turn-over times for S2.
Bottom row:  the time-average spans 0.044 magnetic diffusion time or 24 turn-over times for S1.
}
\label{fig:Uavg_zavg}
\end{figure}

\subsubsection{Non-zonal mean flows}
Non-zonal mean flows appear to be a prominent feature of S2.
Figure \ref{fig:Uavg_zavg} shows velocity field averaged in time and along the rotation axis ($z$-axis -- note that the average along $z$ spans only one hemisphere within the tangent cylinder).
In S1, the $z$-averaged mean flow is dominated by a prograde jet at the tangent cylinder and a retrograde circulation within.
A weak large-scale non-zonal flow is still visible outside the TC in S1.
When averaging over a shorter time-span (24 turn-over times instead of 190), the non-zonal flow dominates.
The least viscous and strongly magnetized simulation S2 is dominated by a large non-zonal gyre outside the TC.
Although arguably limited, the time span available for averaging the S2 flow is significant and would translate to about thousand years if the turn-over time is used for scaling to Earth values.
Similarly, the non-zonal $z$-invariant flow seems to last about 24 turn-over times in S1, corresponding to more than 5000-10000 years.
One outstanding difference between S1 and S2 is that the anti-cyclonic gyre leads to strong westward velocities at the equator in S2, whereas it is weak and mostly eastward in S1.

\subsection{Instantaneous fields}	\label{sec:instant}

\begin{figure}
\includegraphics[width=0.49\linewidth]{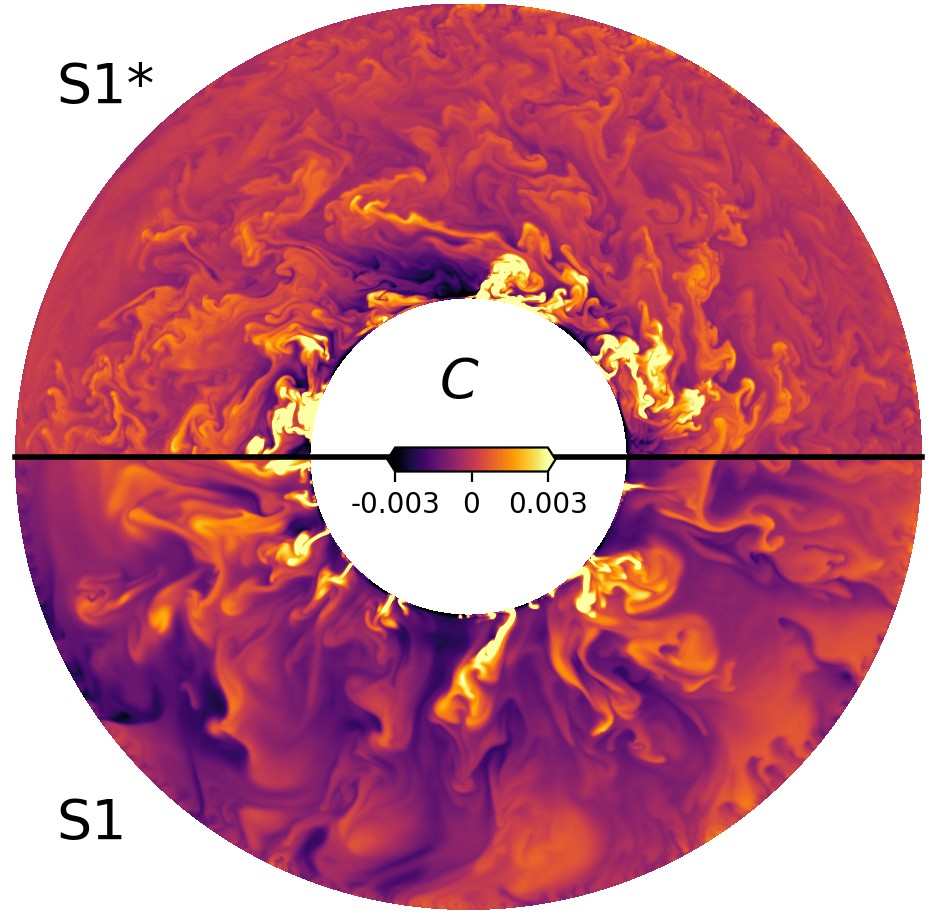}
\includegraphics[width=0.49\linewidth]{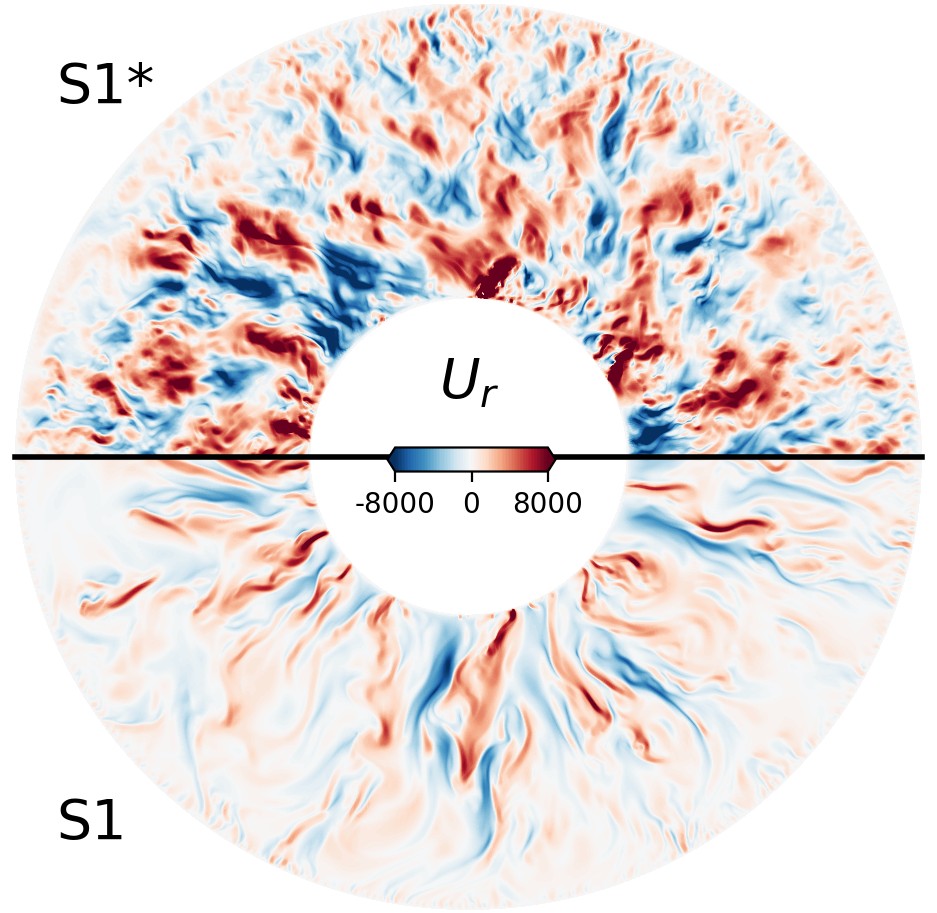} \\[2mm]
\includegraphics[width=0.49\linewidth]{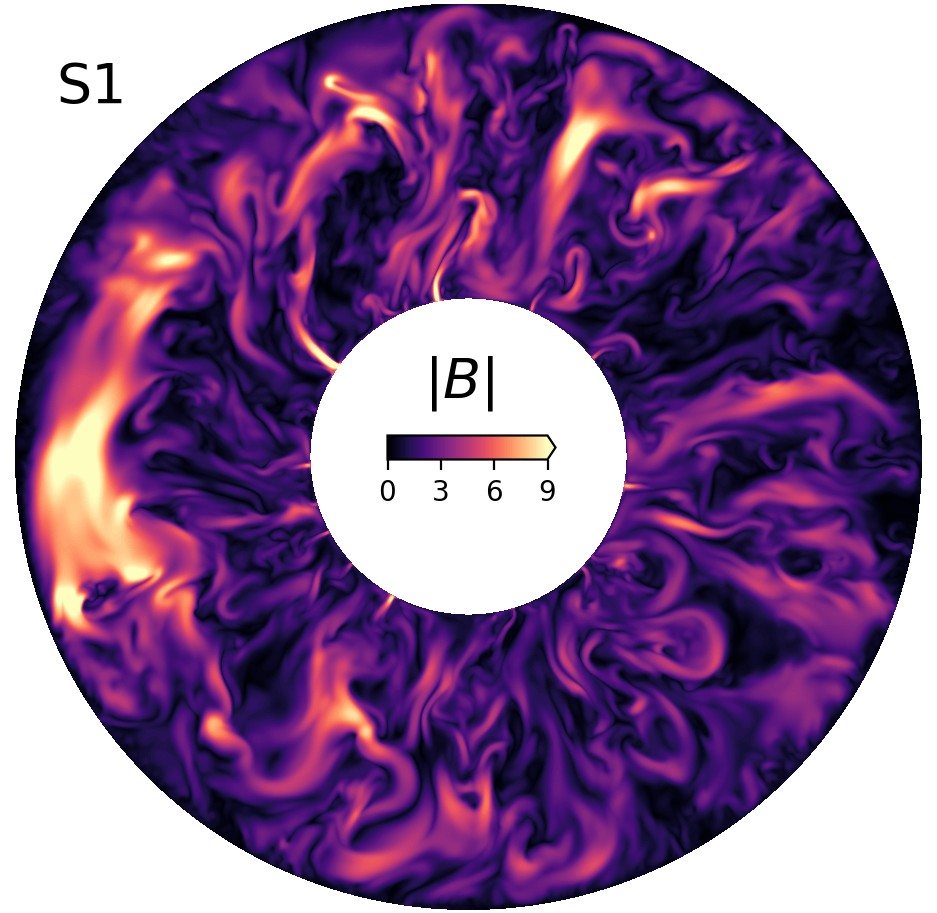}
\includegraphics[width=0.49\linewidth]{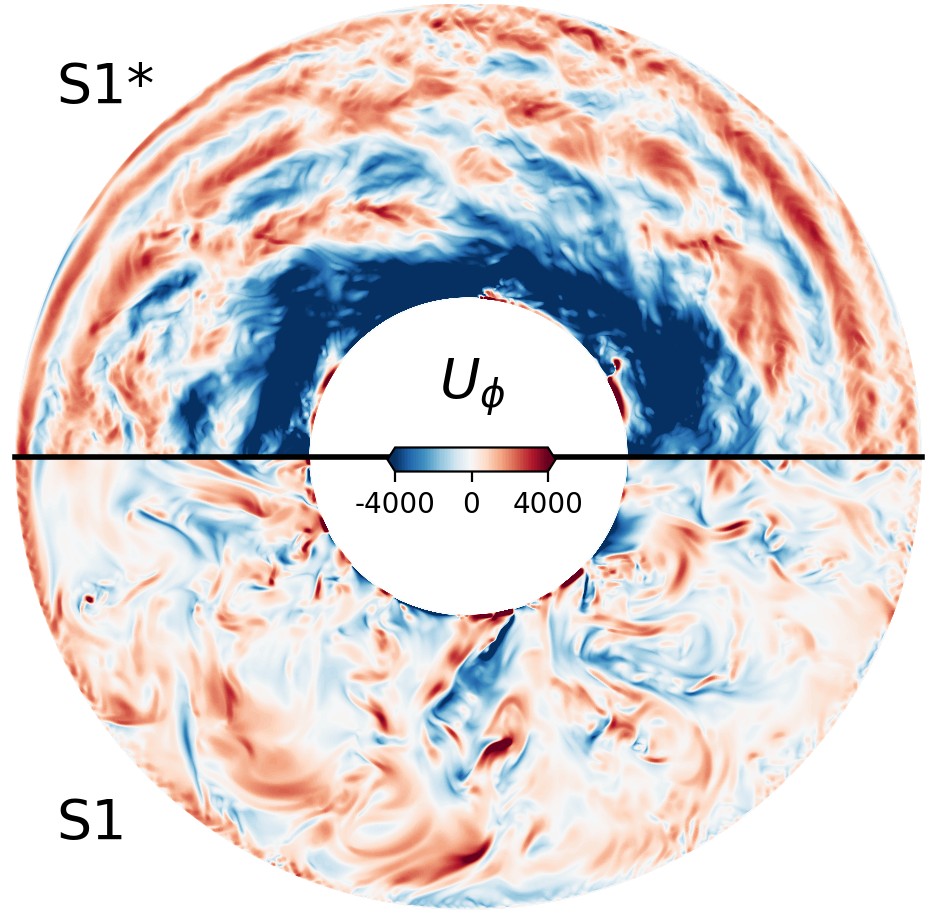}
\caption{Snapshot of fields in the equatorial plane of simulation S1 ($E=10^{-6}$, $Pm=0.2$) and S1* ($E=10^{-6}$, without magnetic field).
The codensity (upper left panel) is represented after removing the mean value at each radius in the plane.
The azimuthal velocity $U_\phi$ has been divided by 3 in S1* to fit the same color scale as S1.
The magnetic field is in Elsasser units (such that $\Lambda = |B|^2$).
The snapshots are taken at the end of S1 and S1* simulations.
}
\label{fig:equat_S1}
\end{figure}

We now turn to instant snapshots of the fields, which are represented in the equatorial plane for simulations S1 and S1* in Figure \ref{fig:equat_S1}, contrasting the differences between dynamo (S1) and non-magnetic convection (S1*).
The codensity field $C$ exhibits very small scales near the inner-core where the plumes originate in both S1 and S1*.
Further away from the inner-core, the codensity field exhibits much larger structures in S1 than in S1*.
The plumes also reach further out in S1 whereas they seem to be stopped by the zonal winds in S1* (see $U_r$ in Fig \ref{fig:equat_S1}).
However, the overall state is better mixed in S1* (lower contrasts in variations of $C$).
This illustrates the effect of the magnetic field on the convection.


\begin{figure}
\begin{center}
\includegraphics[width=0.9\linewidth]{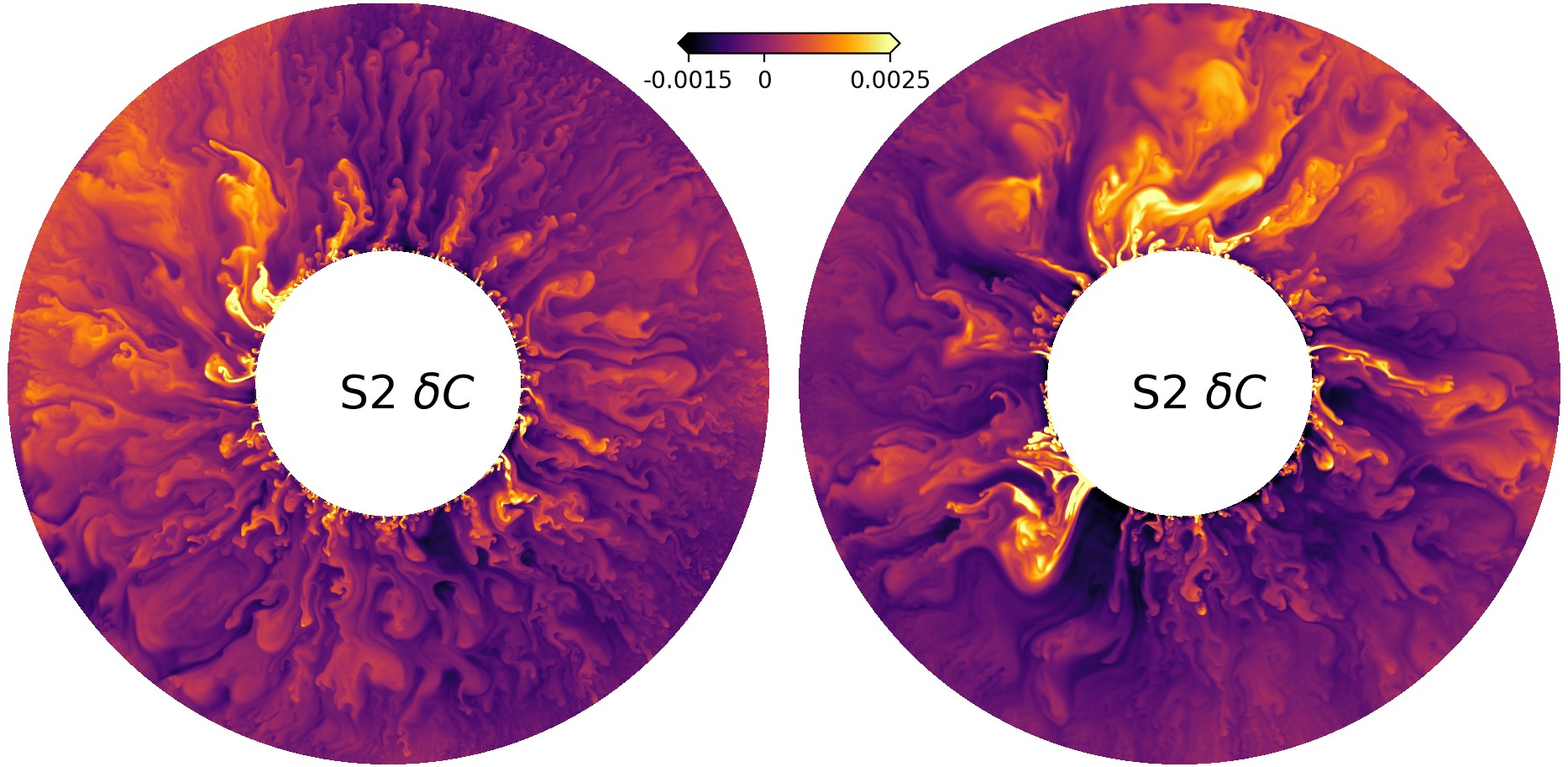} \\[2mm]
\includegraphics[width=0.9\linewidth]{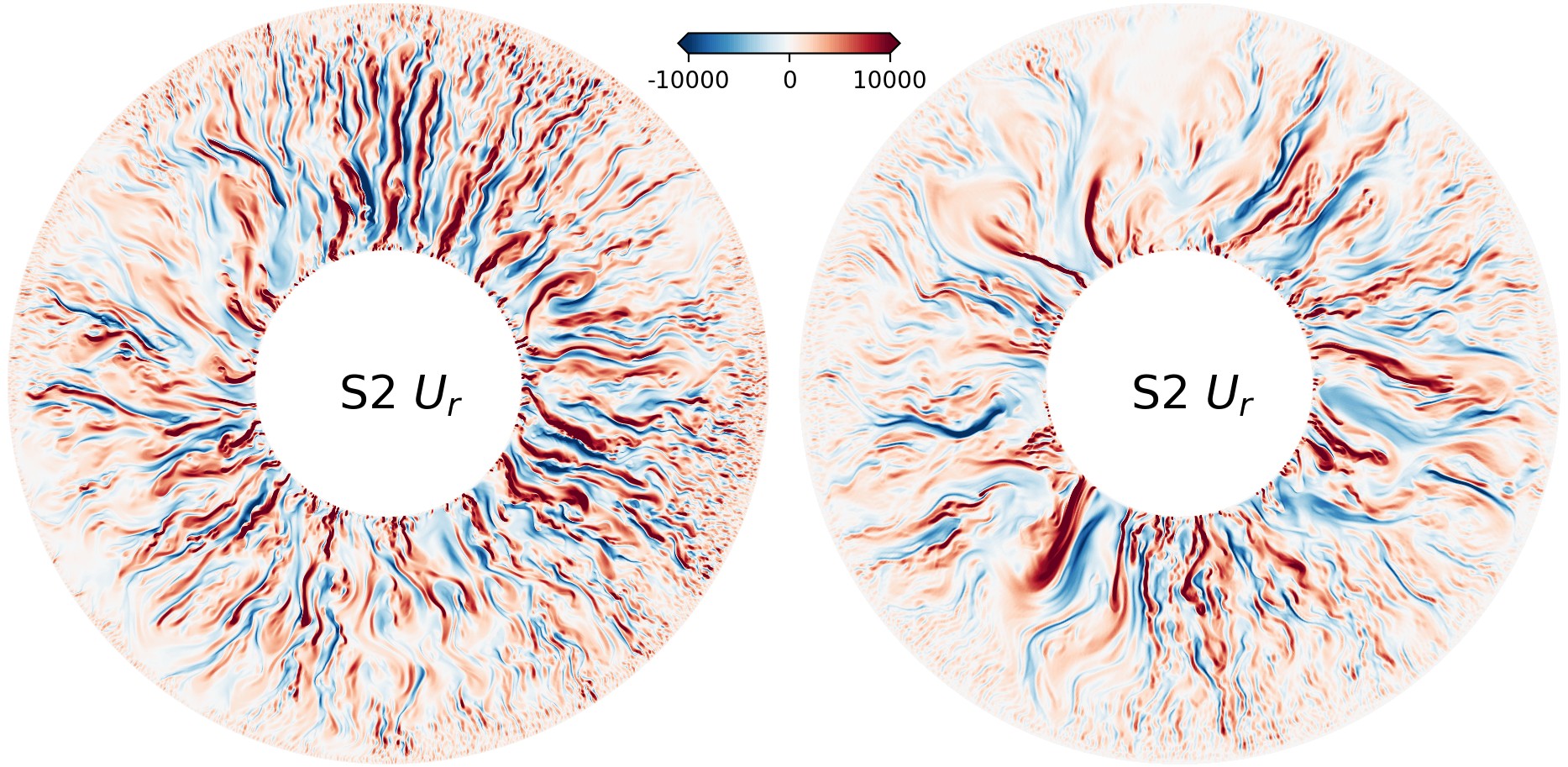} \\[2mm]
\includegraphics[width=0.9\linewidth]{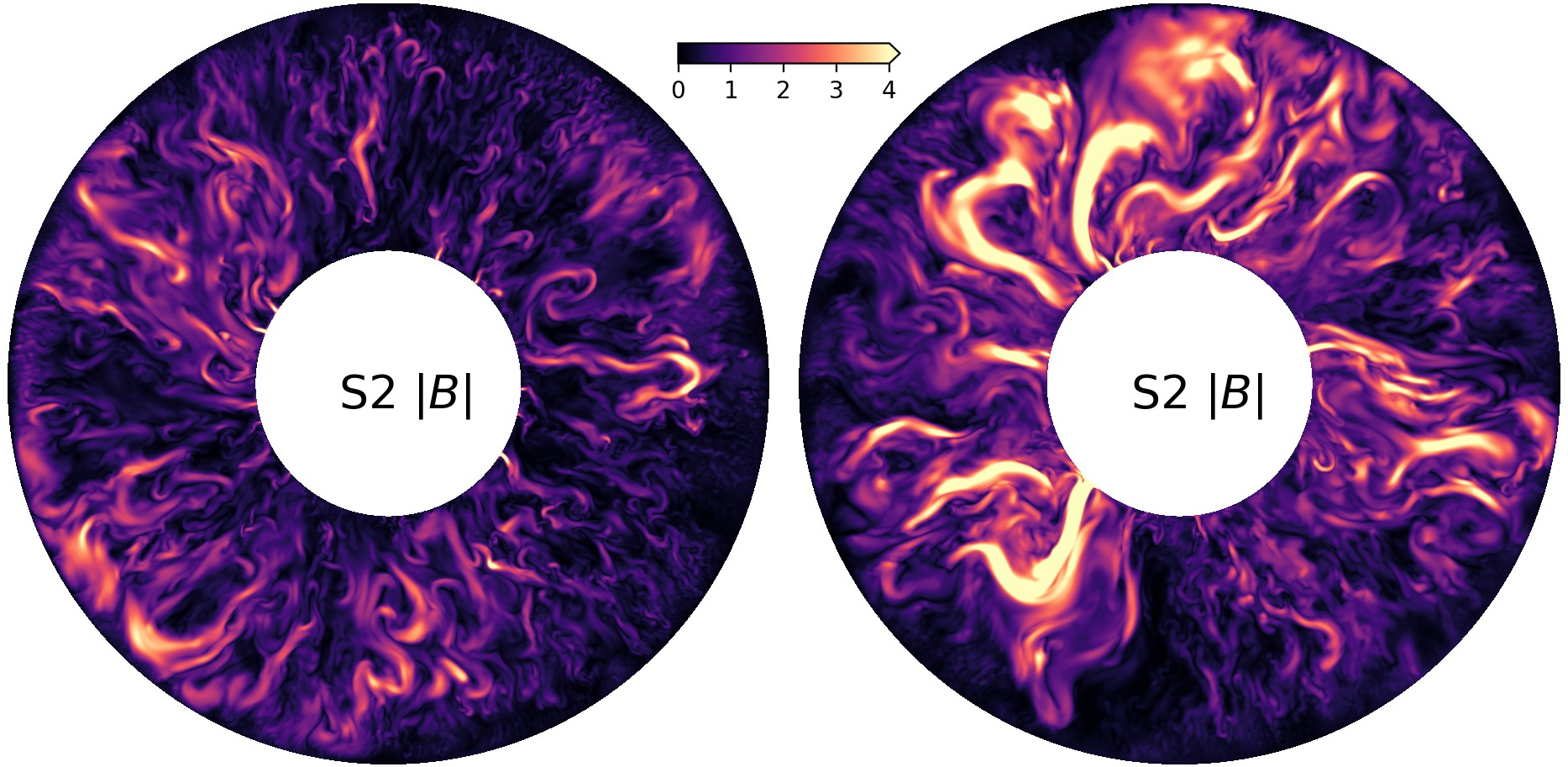}
\caption{Snapshot of fields in the equatorial plane of simulation S2 ($E=10^{-7}$, $Pm=0.1$).
Left: at a time with low magnetic energy (marked by a circle in Fig. \ref{fig:nrj}). Right: high magnetic energy (marked by a square in Fig. \ref{fig:nrj}).
Top row: codensity (after removing the mean value at each radius in the plane). Middle row: radial component of the velocity field; Bottom row: intensity of the magnetic field.
}
\label{fig:equat_S2}
\end{center}
\end{figure}

\begin{figure}
\includegraphics[width=0.75\textwidth]{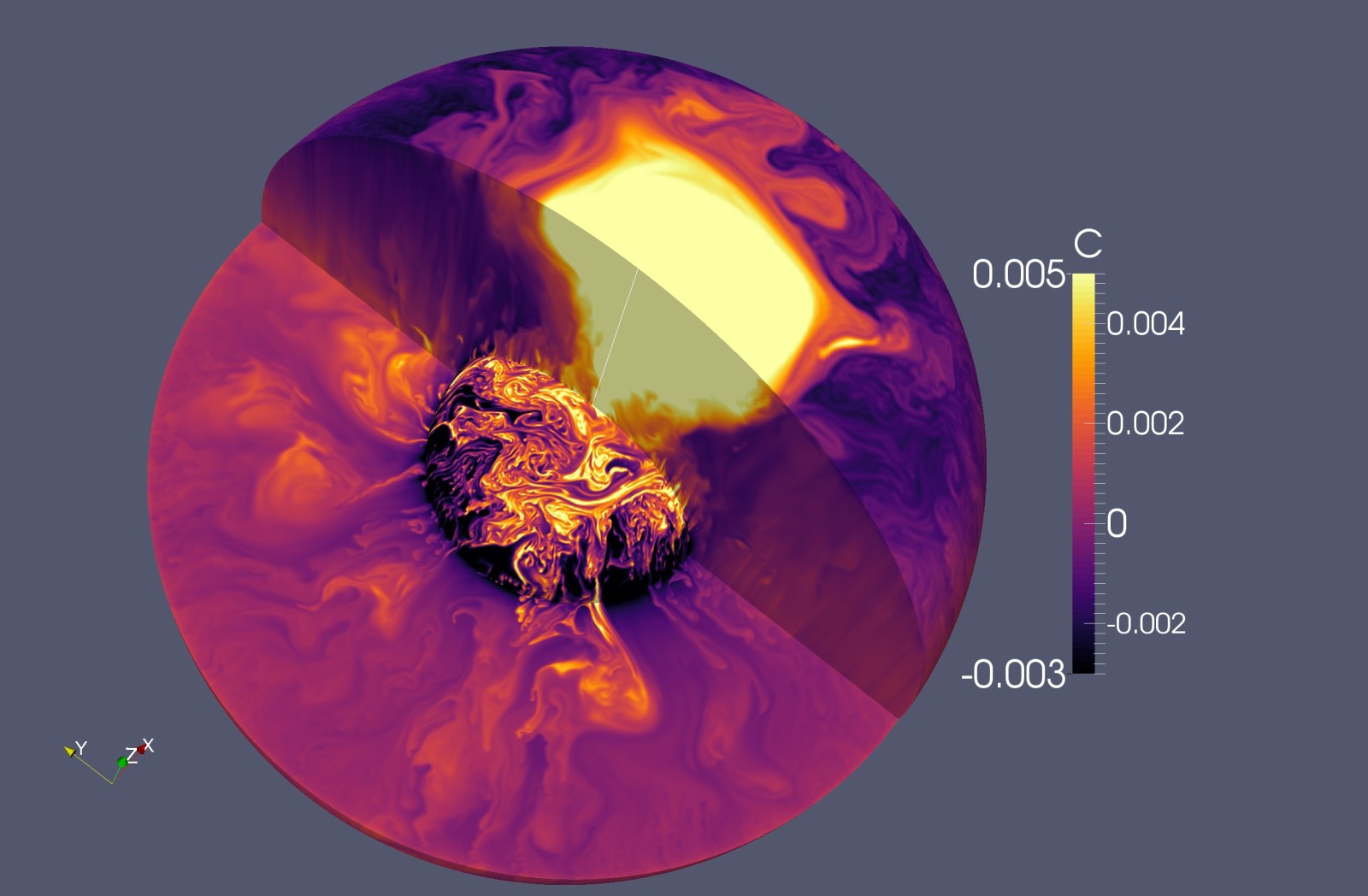} \\
\includegraphics[width=0.75\textwidth]{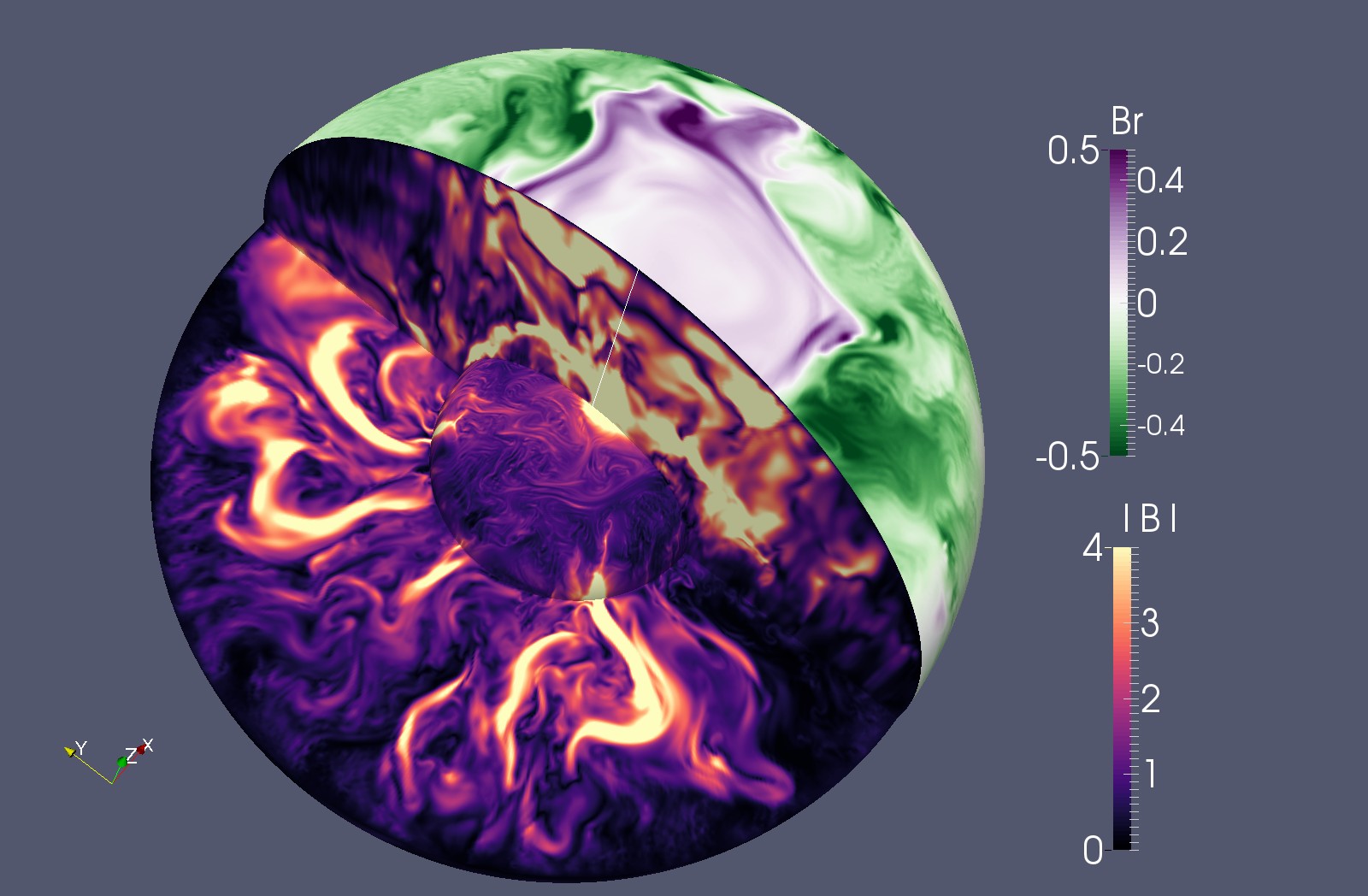} \\
\includegraphics[width=0.75\textwidth]{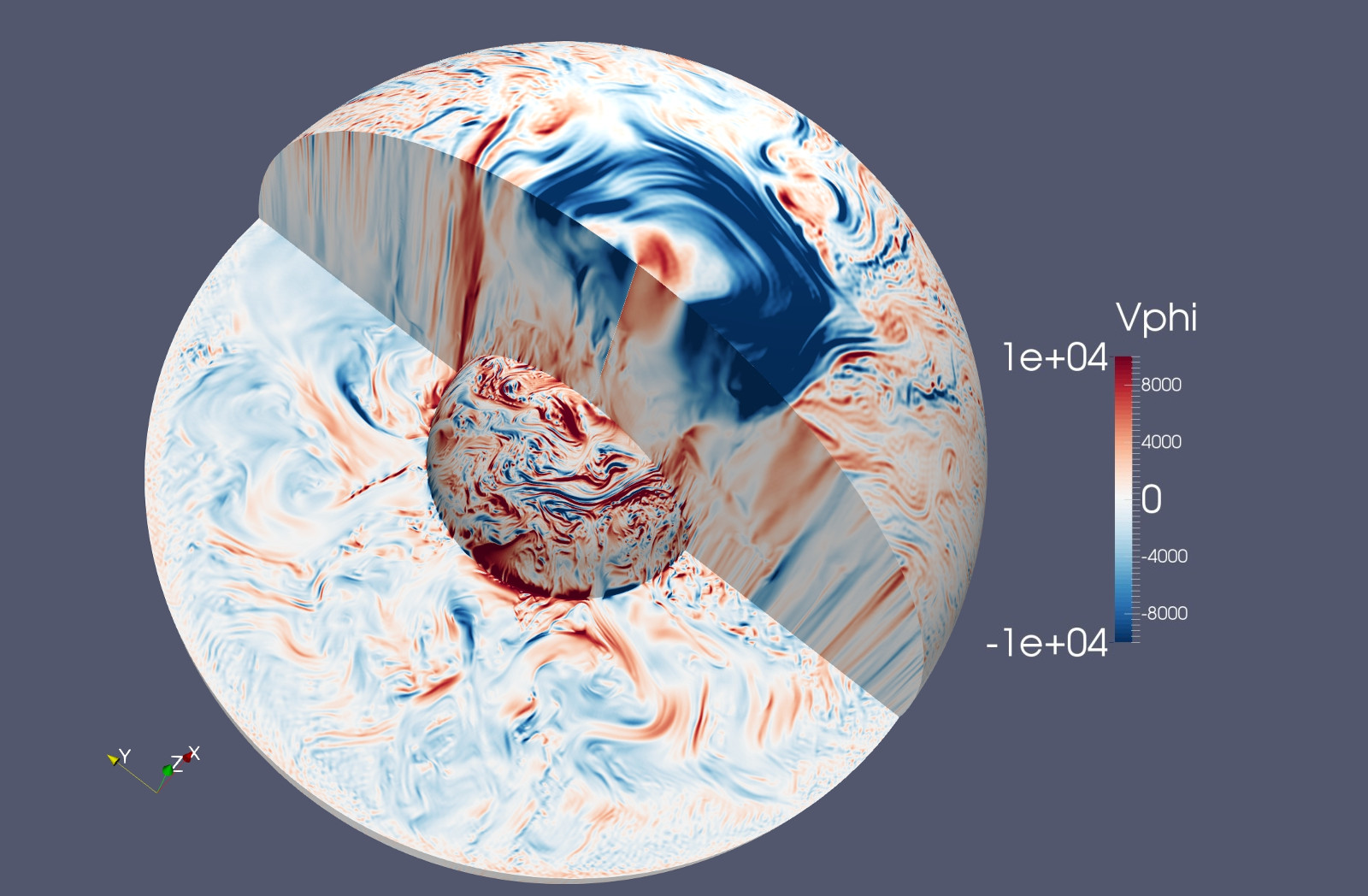}
\caption{Three-dimensional renderings of the fields in the S2 simulation, at the instant marked by a square in figure \ref{fig:nrj}.
The faint white line is the rotation axis.
}
\label{fig:3D}
\end{figure}

Figure \ref{fig:equat_S2} shows similar views for S2 at two different times characterized by moderate and strong magnetic fields (respectively marked by circle and square in Figure \ref{fig:nrj}).
Three-dimensional renderings of the strongest field situation are shown in Figure \ref{fig:3D}, displaying also meridional cuts\footnote{Note also that supplementary meridional slices of the instantaneous velocity field can be seen at \url{https://doi.org/10.6084/m9.figshare.5002199}}.
Very small-scale buoyant plumes originate near the inner-core for the moderate and strong field snapshots, but further away the scales appear larger where the strongest field reigns.
Radial velocities are also weaker in these regions.
There are noticeable velocity field patterns with azimuthal wavenumbers much smaller than $m=67$ -- the critical wavenumber at onset -- especially in the strong field regions.
These large scales coexist with smaller scales close to the inner-core, but also in regions of weaker magnetic field.

This enlargement of convection scale 
is in broad agreement with the observations made by \citet{matsui2014} and \citet{yadav2016b} at higher Ekman numbers and lower forcing \citep[see also][]{hori2012}.
The mechanism proposed by \citet{matsui2014} does also fit our simulations.
Namely, the presence of important variations of codensity over large regions (Fig. \ref{fig:equat_S2} top) produces non-axisymmetric thermal winds that convert the poloidal field into azimuthal field.
The field inhibits motions, preventing the codensity anomalies to mix, thus sustaining the phenomena.
Large scale motions induce less shear and are thus favored compared to small-scale convection.

We would like to emphasize that the large scale codensity anomalies can build up only because the zonal flow is suppressed by the magnetic field.
Otherwise codensity anomalies in the azimuthal direction are quickly sheared away by the zonal jets (which effectively improve lateral mixing), as seen in the non-magnetic convection of S1* (Fig. \ref{fig:equat_S1}).
This is also supported by the codensity profiles (see fig.~\ref{fig:Cprof}) that are better mixed (flatter) in S1 than in S1*

In our simulations S1 and S2, magnetic field intensity is largely inhomogeneous (see $|B|$ in Figures \ref{fig:equat_S1}, \ref{fig:equat_S2} and \ref{fig:3D}).
This does not seem to affect the surface field which is already an order of magnitude smaller than the bulk average.
Because the strength of the magnetic field is expected to control the length-scale of convection, then convection in planetary core would span a wide range of length-scales, possibly from the viscous scale $E^{1/3}$ to the planetary scale.
It is important to take this into account when conceptualizing turbulence in planetary cores \citep[e.g.][]{nataf2015}.

We remark also that because the kinetic energy spectra are not steep, using the length-scale diagnostic introduced by \citet[and widely used afterwards]{christensen2006}  fails to capture accurately the change in length-scale, especially since the viscous length-scale may still be present in the system when the Rayleigh number is highly super-critical as in our case, in contrast to the work of \citet{matsui2014}.
This failure may also explain why the viscous length-scale was found to play a role in previous dynamo studies \citep{king2013} despite large Elsasser numbers.

\subsection{Magnetic field at the core surface}		\label{sec:br_surf}

\begin{figure}
\centering
\includegraphics[width=0.7\linewidth]{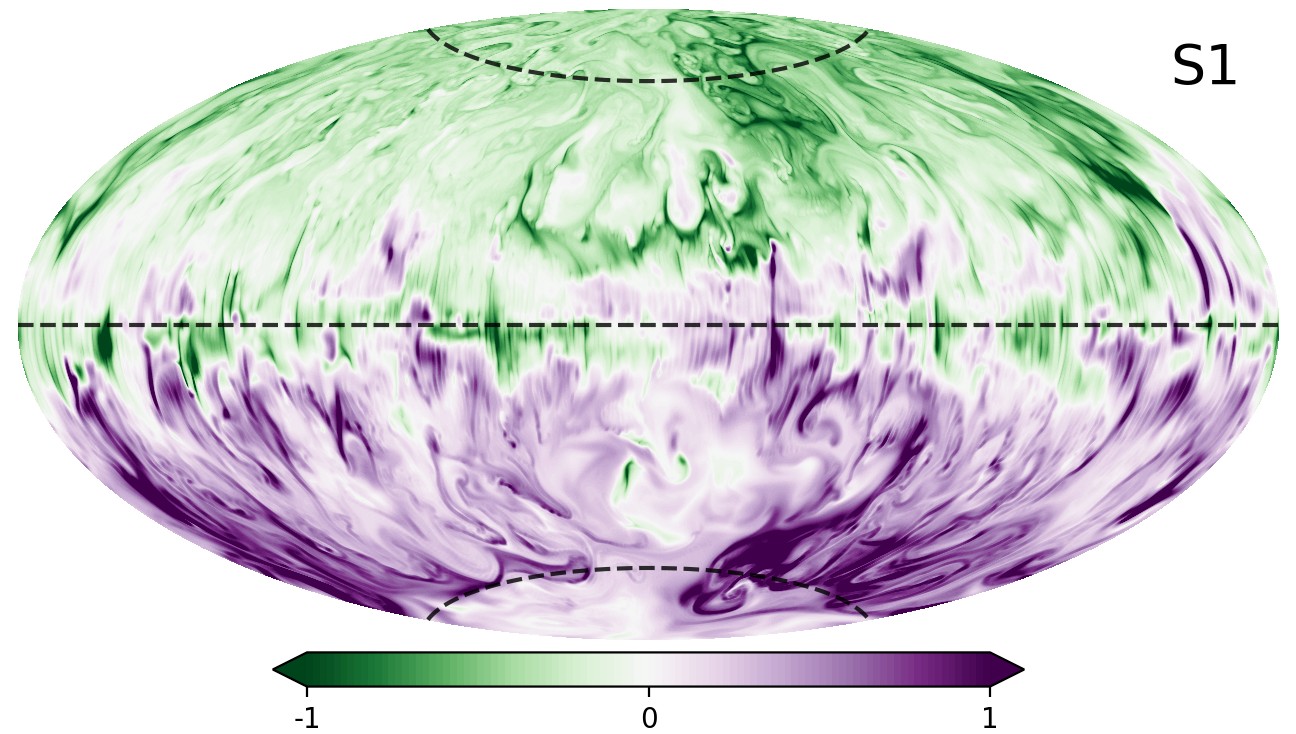}
\includegraphics[width=0.29\linewidth]{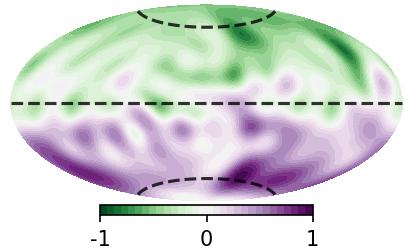} \\[1em]
\includegraphics[width=0.7\linewidth]{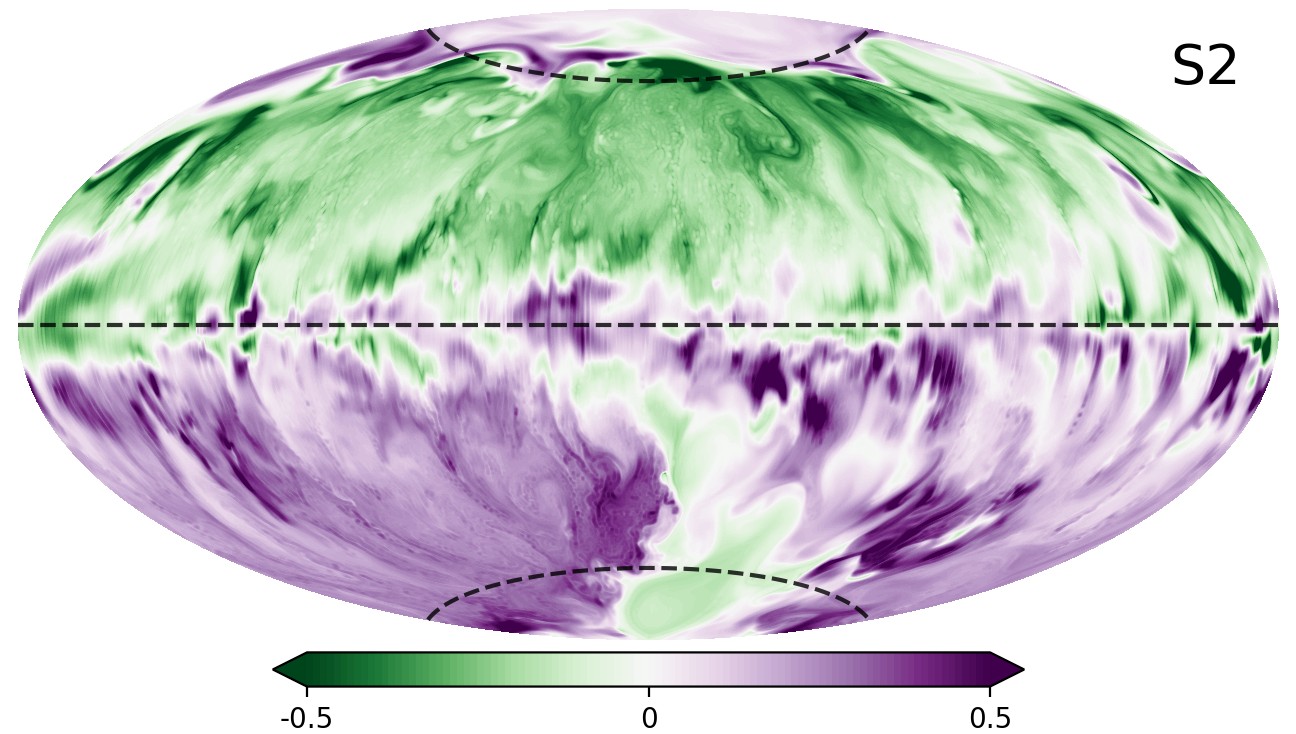}
\includegraphics[width=0.29\linewidth]{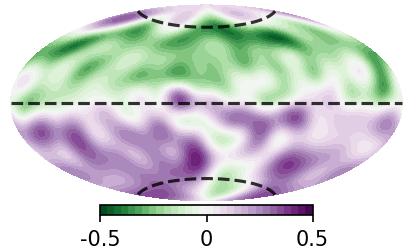}

\caption{Aitoff projection of the radial magnetic field at the core surface in full resolution (left) and truncated after harmonic degree $\ell=13$ (right).
Top row: S1 ($E=10^{-6}$, $Pm=0.2$); Bottom row: S2 ($E=10^{-7}$, $Pm=0.1$).
The dashed lines indicate the equator and the intersection with the tangent cylinder (latitude 69.5$^\circ$).
The magnetic field is in (non-dimensional) Elsasser units (scaled by $\sqrt{\rho\Omega/\sigma}$).
A movie of the S2 fields evolving with time can be found at \url{https://doi.org/10.6084/m9.figshare.4924223}.
}
\label{fig:Br}
\end{figure}

It is important to look at the magnetic field at the surface of the core, as it can be readily compared with the geomagnetic field.
Following \citep{christensen2006}, the fraction of axial dipole in the observable spectrum (up to $\ell=13$) is given by $f_{dip}$ in table \ref{tab:mysimus}, and is in reasonable agreement with the Earth's value ($f_{dip} \simeq 0.68$).
Figure \ref{fig:Br} shows the magnetic field at the surface of S1 and S2 at a given time (the same time as in fig. \ref{fig:equat_S1} and \ref{fig:3D}).
Interestingly, the surface magnetic field displays similarities with the surface codensity (see Fig. \ref{fig:3D}), in particular sharp gradients near the tangent cylinder.
We note that the snapshot of S2 in fig. \ref{fig:Br} displays a reversed flux patch at high latitude (within the tangent cylinder) in the northern hemisphere. This feature is not persistent, as can be seen in the movie (see caption of fig. \ref{fig:Br}).
We remark that magnetic field models derived from satellite measurements show an inverse flux patch (albeit smaller) within TC \citep[see e.g.][fig.~25]{hulot2015}.

\begin{figure}
\centering
\includegraphics[width=0.49\linewidth]{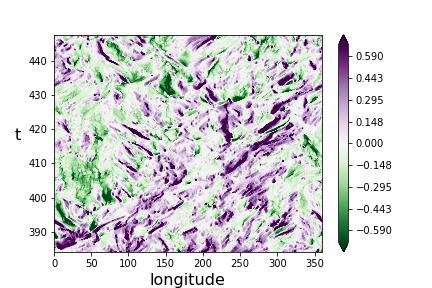}
\includegraphics[width=0.49\linewidth]{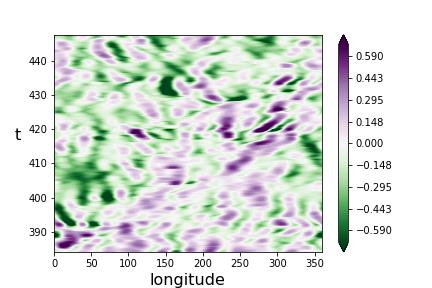} \\
\includegraphics[width=0.49\linewidth]{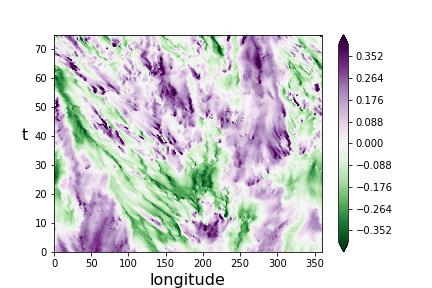}
\includegraphics[width=0.49\linewidth]{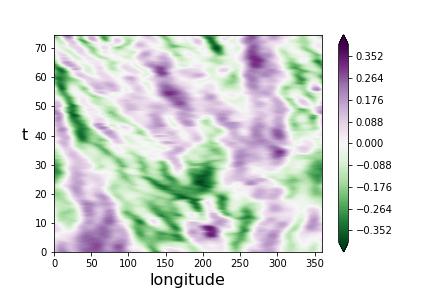}

\caption{Time-longitude representation of the radial magnetic field along the equator.
Top: simulation S1; bottom: simulation S2. Left: full field; right: truncated at $\ell=13$.
Time is in Alfvén time-scale units, and magnetic field in Elsasser units.
A movie of the S2 fields evolving with time can be found at \url{https://doi.org/10.6084/m9.figshare.4924223}.
}
\label{fig:wdrift}
\end{figure}

An important feature of the Earth's magnetic field is its secular variation, and in particular the westward drift of magnetic flux patches near the equator of the Atlantic hemisphere.
The evolution in time of the field at the equator is represented in figure \ref{fig:wdrift} for S1 and S2.
Whereas there is mostly eastward drift in S1, S2 displays an important region of westward drift (roughly between $30^\circ$ and $180^\circ$ longitude) together with a region with no net drift (around $300^\circ$).
These plots also highlight the larger longitudinal extent of patches in S2 than in S1 and their longer life-span.
There is a one-to-one correspondence between the westward drifting region in S2 seen in figure \ref{fig:wdrift} and the location of the strong westward circulation in figure \ref{fig:Uavg_zavg}.
This demonstrates that large-scale westward drift arises naturally in our less viscous simulation.
We also notice the appearance of alternating polarity magnetic field with a well determined wave-length drifting west in the upper left part of the bottom left panel of Figure \ref{fig:wdrift}, akin to the equatorial slow waves of \citet{finlay2003}.

\subsection{Fluctuations and Helicity}

\begin{figure}
\includegraphics[width=0.24\linewidth]{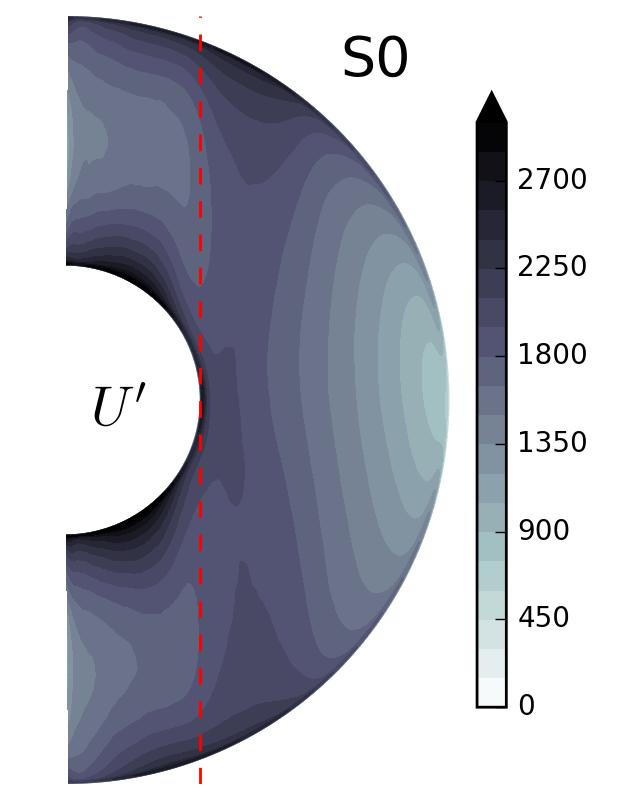}
\includegraphics[width=0.24\linewidth]{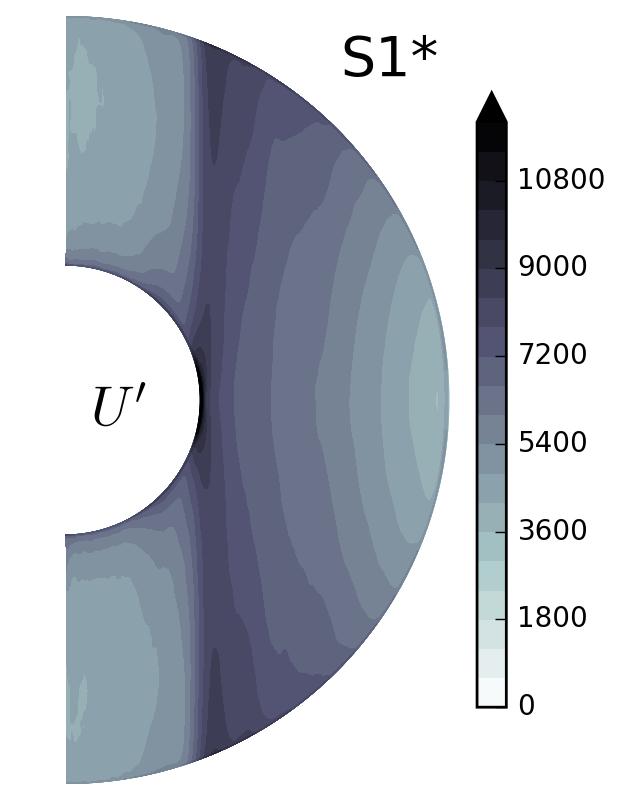}
\includegraphics[width=0.24\linewidth]{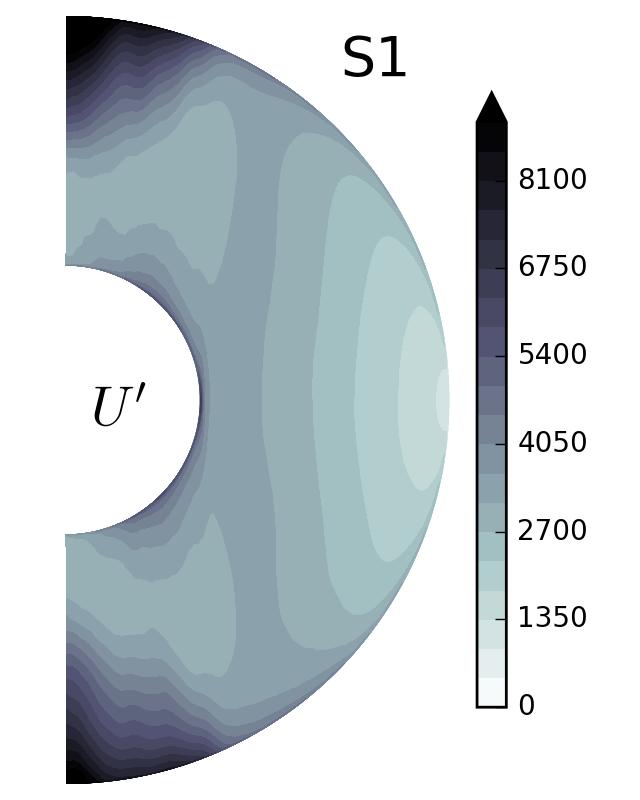}
\includegraphics[width=0.24\linewidth]{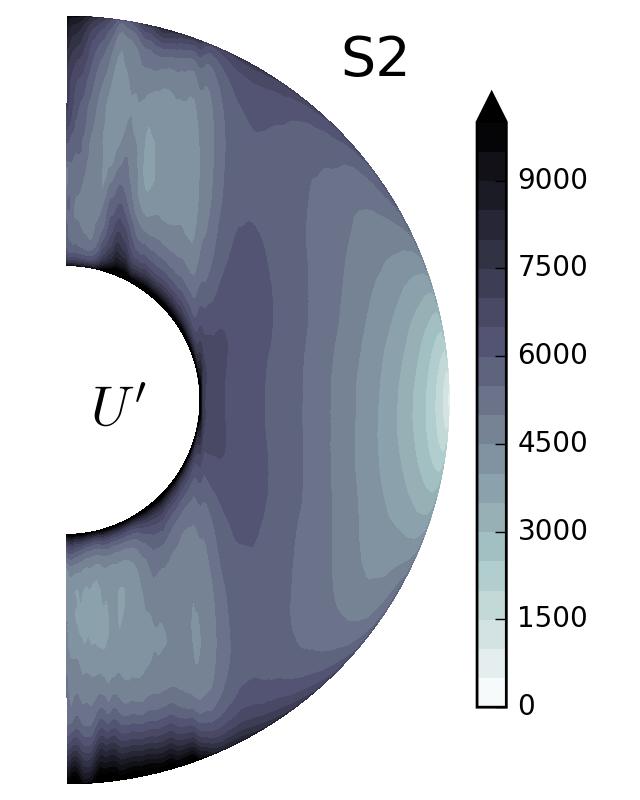} \\
\includegraphics[width=0.24\linewidth]{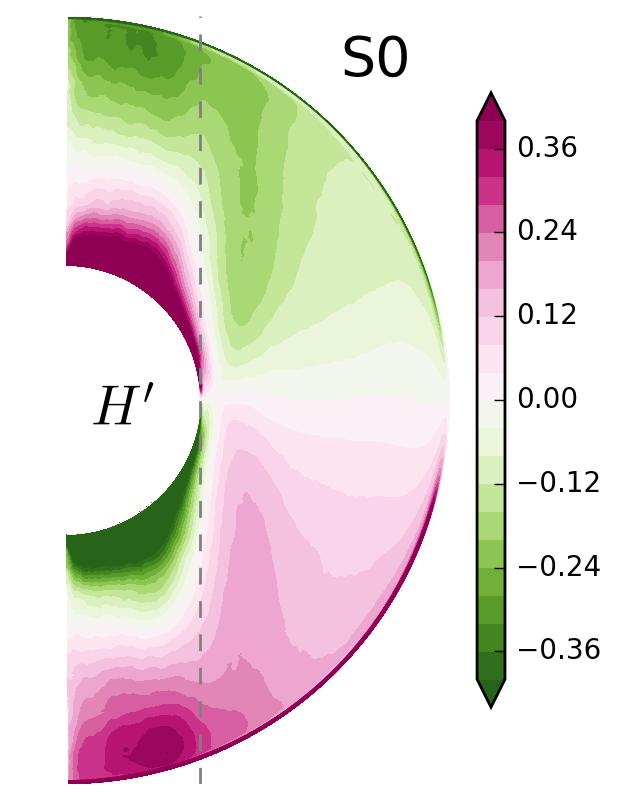}
\includegraphics[width=0.24\linewidth]{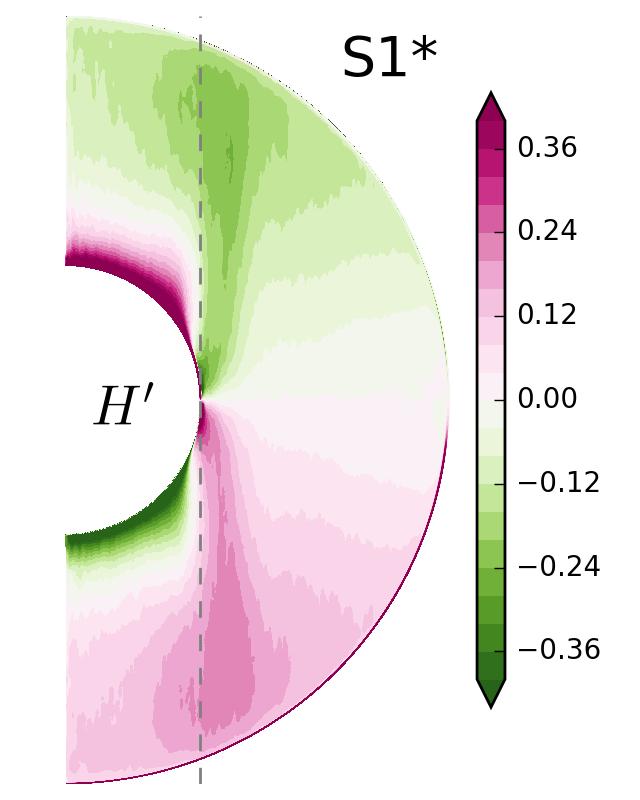}
\includegraphics[width=0.24\linewidth]{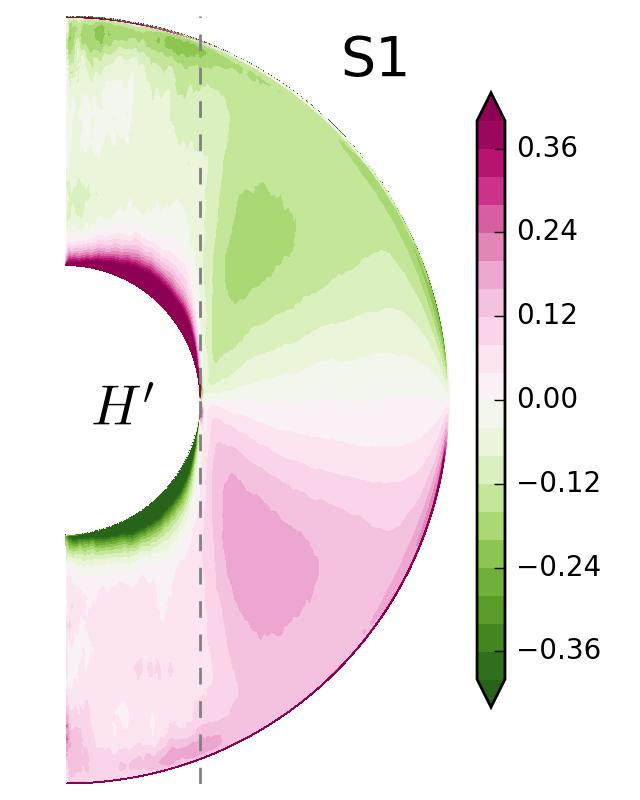}
\includegraphics[width=0.24\linewidth]{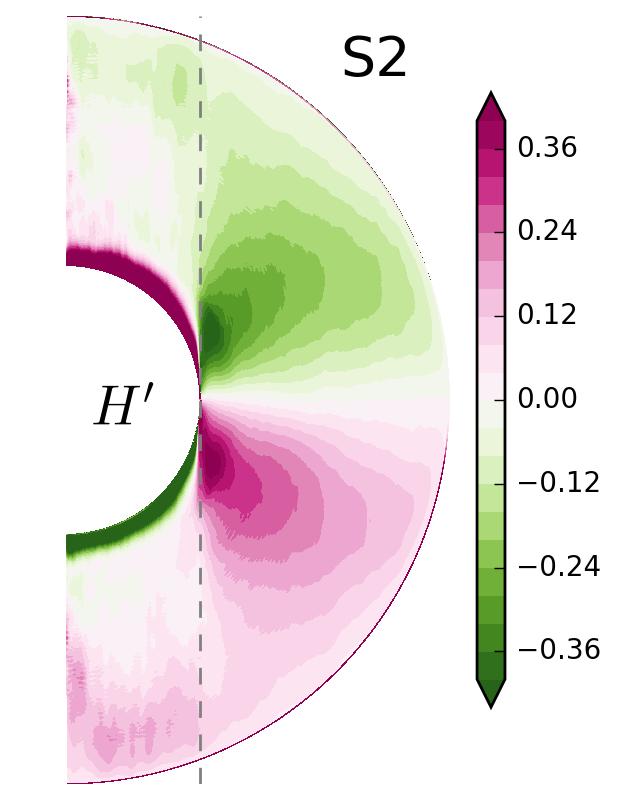} \\
\caption{Time- and longitude-averaged fluctuating velocity ($U'$) and relative helicity ($H'$). See text for definitions.}
\label{fig:avg_fluct}
\end{figure}

We compute the fluctuation maps shown in figure \ref{fig:avg_fluct} by
\begin{equation}
U'(r,\theta) = \sqrt{\langle|U(r,\theta,\phi,t) - \bar{U}(r,\theta,\phi)|^2\rangle}
\end{equation}
where $\langle.\rangle$ denotes the average over longitude $\phi$ and time $t$.
Similarly, the relative mean helicity $H'$ from fluctuating field is computed with:
\begin{align}
\mathbf{u}(r,\theta,\phi,t) &= U(r,\theta,\phi,t) - \bar{U}(r,\theta,\phi) \\
\mathbf{\omega}(r,\theta,\phi,t) &= \nabla \times \mathbf{u}  \\
H'(r,\theta) &= \frac{\langle\mathbf{u}.\mathbf{\omega}\rangle}{\sqrt{\iint_{r,\theta}\langle\mathbf{u}^2\rangle \: \iint_{r,\theta}\langle\mathbf{\omega}^2\rangle}}
\end{align}



Velocity fluctuations are large in the polar regions just below the CMB and near the inner core boundary (ICB), but rather weak in between.
This suggests that the velocity fluctuations inside TC are associated to variability of the twisted polar vortices.
Outside TC, velocity fluctuations are always small near the equator of the CMB, and increasing gradually towards the tangent cylinder, where the mean poloidal field is also concentrated.

\begin{figure}
\includegraphics[width=0.99\linewidth]{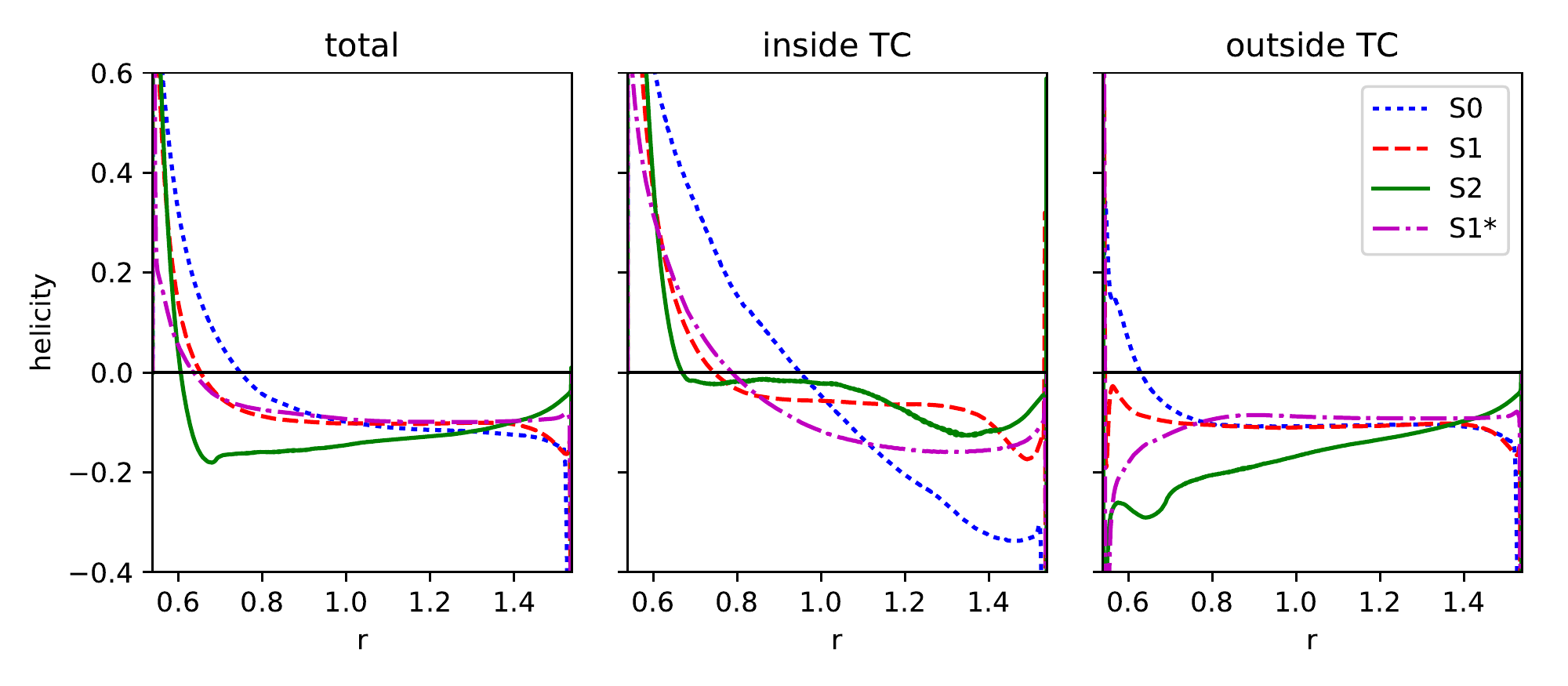}
\caption{Profiles of relative helicity ($H'$, as in fig. \ref{fig:avg_fluct}) averaged over each shell (left), within the tangent cylinder (middle) and outside the tangent cylinder (right), as a function of the radius.
Note also that both hemispheres are averaged, with a weight of $+1$ and $-1$ respectively for the northern and southern hemispheres.}
\label{fig:helprof}
\end{figure}

To link these fluctuations with a possible poloidal magnetic field generation, we turn to helicity (fig. \ref{fig:avg_fluct} bottom and fig. \ref{fig:helprof}), which is often associated with alpha effect \citep[e.g.][]{moffatt1978,jones2008lecture} whereby poloidal magnetic field is produced from toroidal field.
Helicity maps exhibit a gradual change.
In S0 helicity is maximum near the boundaries and extends towards the equator just outside the TC.
In S2 the maximum helicity in the outside region is located near the inner-core.
S1 shows both boundary bound helicity together with local maximum in the bulk (fig. \ref{fig:avg_fluct} bottom).
Outside the tangent cylinder, relative helicity is larger in S2 than in S1.
This suggests a transition from helicity mostly due to recirculation associated with the strong zonal jets just outside TC in S0 and S1* \citep[see][]{busse1975} to bulk helicity in S1 and S2, which might be a hint for the mechanism proposed by \citet[relying on magnetic pumping]{sreenivasan2011}, or the one proposed by \citet[relying on inertial waves]{davidson2014}.
Inside TC, it may be simply the contribution of the helical convective vortices \citep[see e.g.][]{aurnou2003,grooms2010}.
In any case, we can confidently say that the S2 dynamo does not rely on helicity induced by Ekman pumping.
Similarly, Ekman pumping cannot be an important source of helicity in Earth's core \citep[e.g.][]{schaeffer2006}.
In fig. \ref{fig:helprof}, strong helicity is found at the boundaries.
The outer boundary is the Ekman layer, and scales as $E^{1/2}$ (barely visible in fig. \ref{fig:helprof}).
The inner boundary is much larger and we could not find any scaling that convincingly collapse the three dynamos.
The profiles are significantly different inside and outside TC.
The effect of the magnetic field on these profiles seems weak outside TC, but seizable inside TC (compare S1 and S1* in fig. \ref{fig:helprof}).

\subsection{Spectra} \label{sec:spec}

\begin{figure}
\begin{center}
\includegraphics[width=0.99\textwidth]{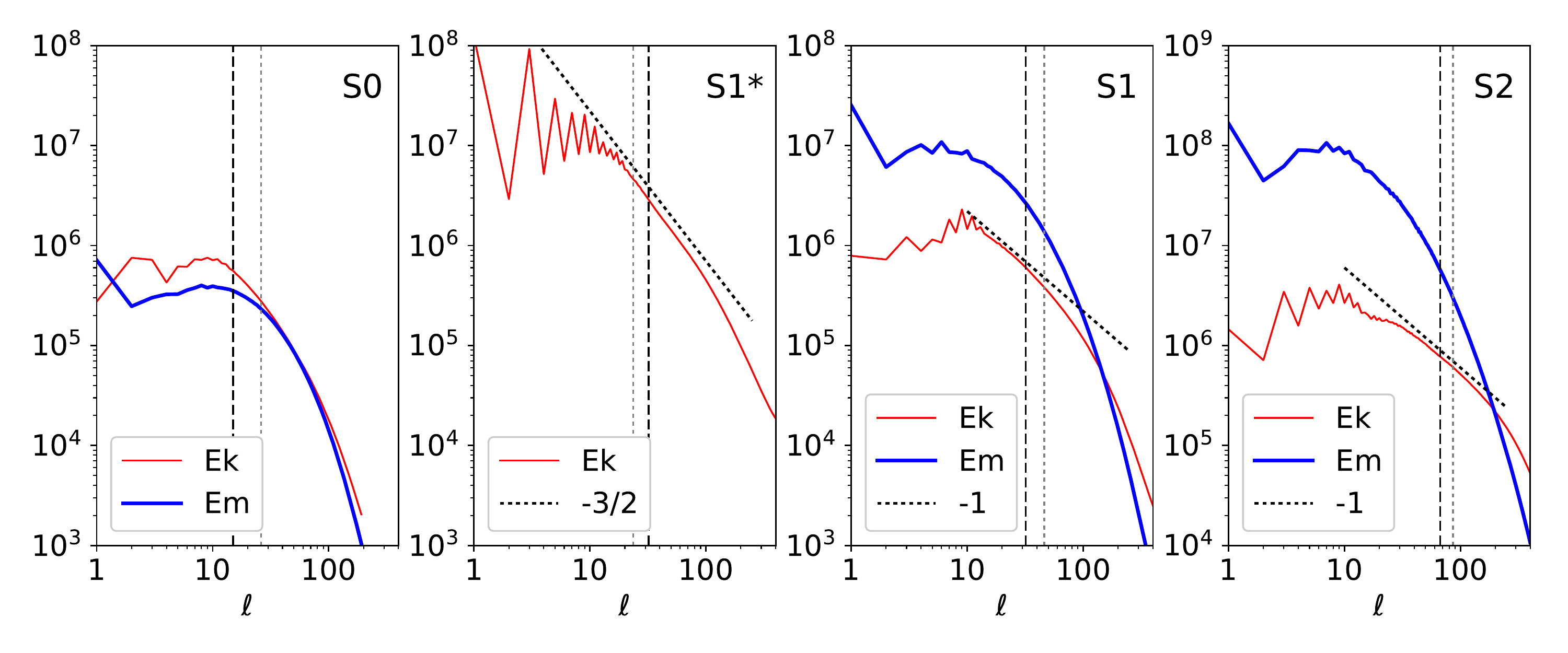}
\caption{Time- and radius- averaged energy density spectra in our simulations, as a function of harmonic degree $\ell$.
The thin red line is the kinetic energy spectrum (Ek), while the thick blue line is the magnetic energy spectrum (Em).
Dotted lines are arbitrary slopes to guide the eye.
The dashed vertical line marks the unstable wavenumber $m_c$ at the onset of (non-magnetic) convection.
The vertical dotted gray line marks the mean degree $\bar{\ell}$ of the kinetic energy spectrum (see table \ref{tab:mysimus}).
}
\label{fig:spec_avg}
\end{center}
\end{figure}

Figure \ref{fig:spec_avg} shows kinetic and magnetic energy spectra for all our simulations.
These time- and radius- averaged spectra are misleading in several ways.
First, they encompass regions with vastly different dynamics and length scale (inside and outside the tangent cylinder).
Second, even in these regions, fluctuations are largely inhomogeneous (see Fig. \ref{fig:avg_fluct}).
Finally, a given harmonic degree $\ell$ corresponds to different length-scale depending on the radius.

We remark however that the non-magnetic simulation S1* displays a significant range of kinetic energy spectrum obeying a $\ell^{-3/2}$ law.
This contrasts with the kinetic energy spectra in the dynamo simulations S1 and S2, which are much less steep, being almost flat at large scales, and then displaying more or less a $\sim \ell^{-1}$ range.
Note that the exponents quoted here are purely empirical, and that no clear relation to theoretical Cartesian spectra can be made because of the spherical geometry and the anisotropy of such rotating flows.
\citet{christensen2006} introduced an \textit{average harmonic degree} $\bar{\ell} = \sum_\ell {\ell E_\ell(u)}/\sum_\ell {E_\ell(u)}$.
With a kinetic energy spectrum $E_\ell(u) \sim \ell^{-1}$, $\bar{\ell}$ will be determined by the smallest scales, although larger scales are present (see Fig. \ref{fig:Uavg_zavg} and \ref{fig:spec_avg}).
This may explain why \citet{king2013} found that $\bar{\ell}$ seems to scale as the inverse of the viscous scale in a large database of dynamo simulations.


\begin{figure}
\begin{center}
\includegraphics[width=0.99\textwidth]{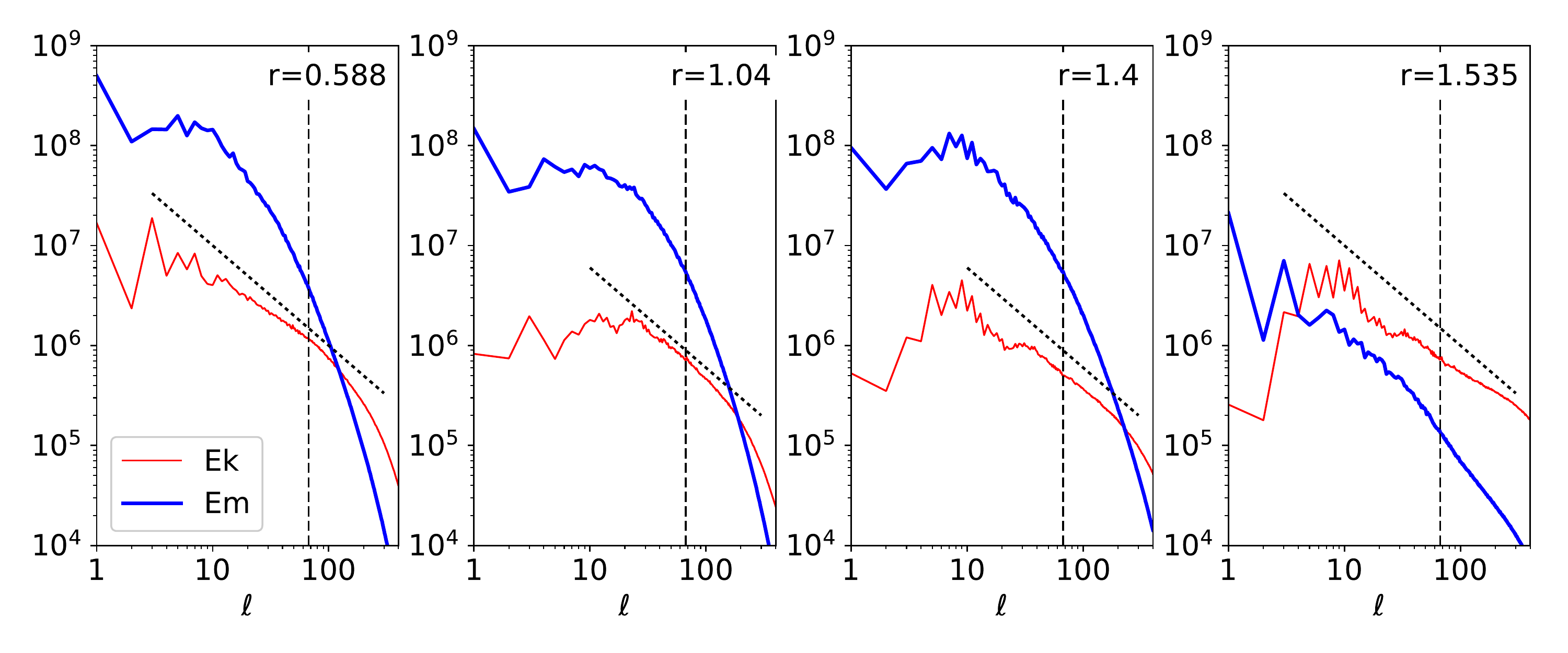}
\caption{Time-averaged energy density spectra in simulation S2 at different radii.
The dashed vertical line marks the unstable wavenumber $m_c=67$ at the onset of (non-magnetic) convection.
The dotted line is proportional to $\ell^{-1}$ to guide the eye.}
\label{fig:spec_S2}
\end{center}
\end{figure}

Figure \ref{fig:spec_S2} shows the energy spectra at 4 different radii in the S2 simulation.
Near the inner-core ($r=0.588$), the spectra are dominated by the fields inside of the tangent cylinder.
The strength of both velocity and magnetic fields at large scales ($\ell \leq 5$) is much larger than at larger radii.

In the middle of the shell ($r=1.04$), the magnetic spectrum is not clearly dominated by the dipole component but is rather flat up to degree 20 to 30, and then decays quickly beyond $\ell \sim 50$.
The kinetic energy spectrum is about 20 to 30 times smaller than the magnetic spectrum, and also peaks at degree $\ell \simeq 20$.

In the outer part ($r=1.4$) the magnetic spectrum is not yet dominated by the dipole component $\ell=1$, but peaks for degree $5 \leq \ell \leq 10$, just as the velocity spectrum.

At the surface ($r=1.535$ -- just below the Ekman layer), the magnetic spectrum is clearly dominated by the dipole.
However, magnetic energy is much weaker than deeper in the shell.
It is actually barely larger than the kinetic energy, unlike in the Earth.
If we exclude the dipole ($\ell=1$) and octupole ($\ell=3$) parts, the magnetic spectrum is rather flat up to degree $\ell=10$.
Beyond $\ell=10$, both velocity and magnetic spectra decay slower than at lower radii.

Finally, note that the peaks in the spectra are always located at wavenumber significantly smaller than those expected at the onset of thermal convection ($m_c = 67$).

\subsection{Spatio-temporal Fourier analysis}

Because of the vastly different dynamics occuring inside and outside the tangent cylinder, we need to analyze these regions separately.
This cannot be done with simple harmonic degree spectra.
Furthermore, the temporal fluctuations visible in Figure \ref{fig:nrj} also prompt us to analyze the fields and spectra beyond their time-averages.

\begin{figure}
\begin{center}
\includegraphics[width=0.35\textwidth]{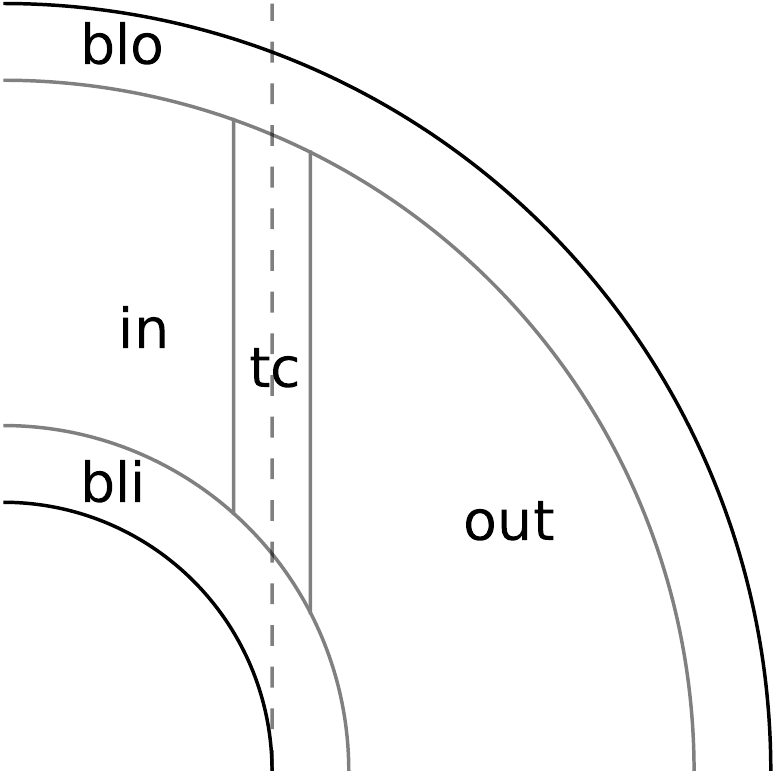}
\caption{Meridional cut showing the five different axisymmetric regions of the simulation domain that are studied separately in the spatio-temporal Fourier analysis.
\textit{bli} and \textit{blo} are boundary layer regions close to the inner-core and mantle respectively, of radial extent $0.1 \, r_o$.
\textit{tc} is a region centered around the tangent cylinder spanning cylindrical radii $0.3<s/r_o<0.4$.
\textit{in} is a region located inside the tangent cylinder and excluding all other regions.
\textit{out} is the region outside the tangent cylinder, also excluding all other regions.
 }
\label{fig:regions}
\end{center}
\end{figure}

We have recorded snapshots of the fields at regular time intervals.
A full snapshot of S2 needs 32 gigabytes of disk space, and we could not afford to save many such large snapshots.
For this reason, the many recorded snapshots are truncated at a spherical harmonic degree $\ell_{tr}$ that is smaller than the maximum $\ell_{max}$ resolved by the simulation: $\ell_{tr} = 399$ for S1 and and $\ell_{tr} = 299$ for S2.
To save even more disk space, we also use single-precision to store the snapshots.
We apply a Fourier transform in time and longitude (the two homogeneous directions) to the whole snapshots series.
Unfortunately, some snapshots have been lost.
For S1, we could effectively use a series of 523 regularly spaced snapshots sampled at period $25/\Omega$ (every $\sim 4$ rotation periods or 8000 samples per magnetic diffusion time).
For S2, only a series of 102 snapshots sampled at period $80/\Omega$ (every 12.7 rotation periods or 12500 samples per magnetic diffusion time).
We then study separately five different regions, represented in Figure \ref{fig:regions}, inspired by the behavior of mean fields described above:
an inner and outer boundary layer region, located near the inner-core and near the mantle respectively, with a radial extent of 10\% of the outer shell radius; a tangent cylinder region, excluding the previous boundary layers and spanning cylindrical radii $s$ from $0.3 r_o$ to $0.4 r_o$; an inner region for $s<0.3$ and an outer region for $s>0.4$, both also excluding the two boundary layers.
This allows us to describe quantitatively the temporal behavior of the fields in our simulations.
For each snapshot, we also compute the different terms in the evolution equations and perform the same Fourier analysis, allowing to describe the dynamical balances in our simulations in terms of spatial and temporal scales \citep{nataf2015}.

\subsubsection{Velocity and Magnetic fields}


\begin{figure}
\begin{center}
\includegraphics[width=0.9\textwidth]{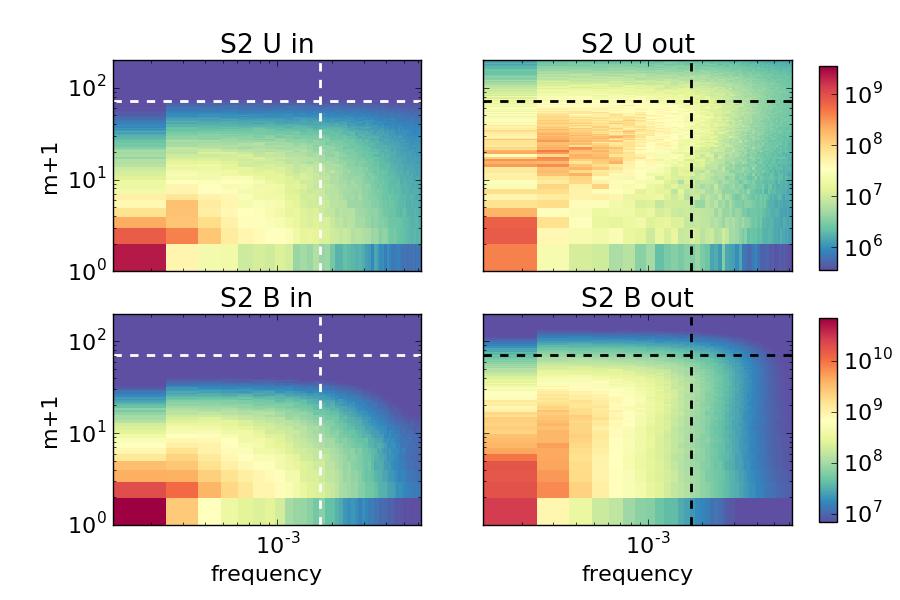}
\caption{Space-time spectra of the kinetic (top) and magnetic (bottom) energies in S2, inside (left) and outside (right) the tangent cylinder (as defined in figure \ref{fig:regions}).
The colormaps use a logarithmic scale, which is the same across regions allowing to compare or sum the energies of each region.
Frequency is in units of rotation rate $\Omega$, and has been shifted so that the leftmost column in each plot is the time-average (zero frequency).
The dashed horizontal line marks the unstable mode at onset of convection, while the dashed vertical line marks the Alfvén frequency (Lehnert number).
}
\label{fig:ub_spec2d_S2}
\end{center}
\end{figure}

\begin{figure}
\begin{center}
\includegraphics[width=0.9\textwidth]{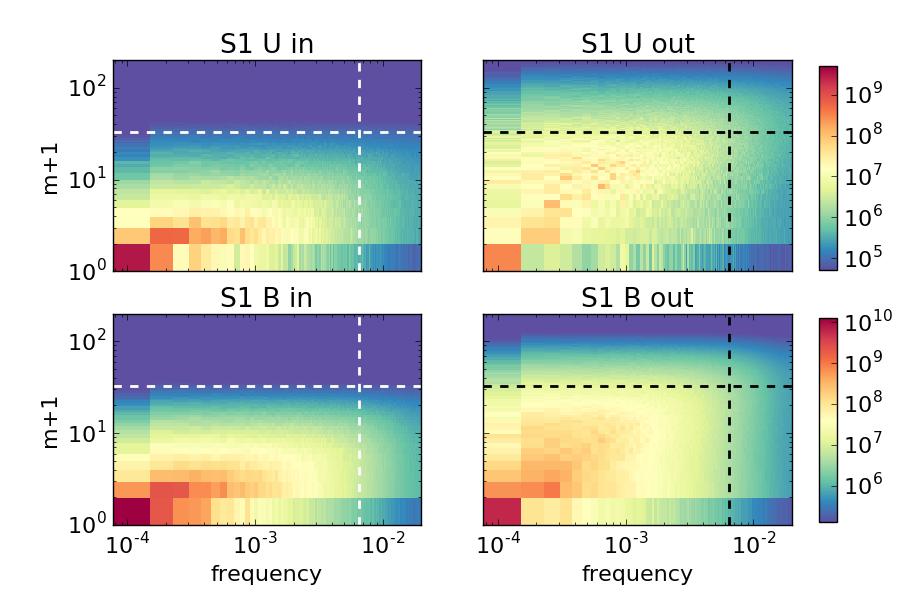}
\caption{Space-time spectra of the kinetic and magnetic energies in S1.
See caption of Fig. \ref{fig:ub_spec2d_S2}.
}
\label{fig:ub_spec2d_S1}
\end{center}
\end{figure}

\begin{figure}
\begin{center}
\includegraphics[width=0.9\textwidth]{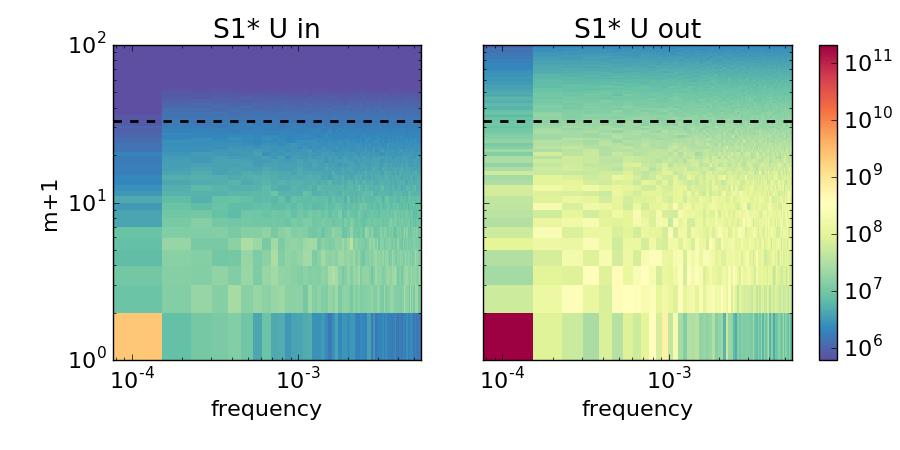}
\caption{Space-time spectra of the kinetic energy in S1* (without magnetic field).
See caption of Fig. \ref{fig:ub_spec2d_S2}.
}
\label{fig:u_spec2d_S1nob}
\end{center}
\end{figure}

The critical azimuthal order for the onset of convection is $m_c = 15$, 32 and 67 for S0, S1 and S2 respectively.
These are exact values computed by SINGE \citep{vidal2015} and are consistent with $m_c = 0.31\, E^{-1/3}$.
Figures \ref{fig:ub_spec2d_S2} and \ref{fig:ub_spec2d_S1} show that the azimuthal order of convection outside the tangent cylinder is around $m=10$ for S1 and $m=20$ for S2.
These correspond to length scales three times larger than the ones for the onset of convection, although they are still dependent on the viscosity, scaling roughly as $m \simeq 0.1 \, E^{-1/3}$.
However, in addition to this small scale convection, a non-zonal mean flow is produced in S2, and low frequency, low $m$ flows are produced outside the tangent cylinder in S1 and S2.
It is remarkable that in S2 the mean flow is dominated by $m=1$ and $2$ rather than $m=0$. This mean flow is represented in Figure \ref{fig:Uavg_zavg}.

As already noticed, the dynamics is very different within the tangent cylinder.
There, large scale motions dominate (the twisted polar vortices), and the time-variability peaks at $m=1$, meaning that this vortex keeps a rather constant amplitude but deviates from the global rotation axis.
Axisymmetric terms ($m=0$) also dominate the magnetic field spectra.
However, small-scale convection is also present where the magnetic field is weaker (as visible near the inner-core on fig.~\ref{fig:3D}, bottom).

The effect of the magnetic field can be seen by comparing figures \ref{fig:ub_spec2d_S1} and \ref{fig:u_spec2d_S1nob}.
Without magnetic field, the mean zonal flow dominates both inside and outside of the tangent cylinder, while more variability in both space and time is observed when the magnetic field is present, with important fluctuations of the $m=1$ flow inside and outside TC.

\subsubsection{Dynamical balances}

We apply a Fourier transform, in both azimuthal and time directions, to the different terms in the vorticity equation (the curl of equation \ref{eq:NS}).
Note that examining the terms in the vorticity equation effectively extracts the second order balance, after the main geostrophic balance in which most of the Coriolis and pressure force have canceled each other.
Fields were truncated to $\ell_{max}=299$ before computing the terms, so we lose the contribution of the smallest scales to the non-linear terms.
The spatial spectra shown in Figure \ref{fig:spec_avg} suggest that this is not a problem within the bulk, as the spectra start to decay well before $\ell=100$, but we might lose significant contributions in the boundary layers.
However, we have checked that truncating further at $\ell_{max}=100$ does not alter the broad picture, although it lowers the relative contribution from non-linear terms (Lorentz and advection).

\begin{figure}
\begin{center}
\includegraphics[width=0.99\textwidth]{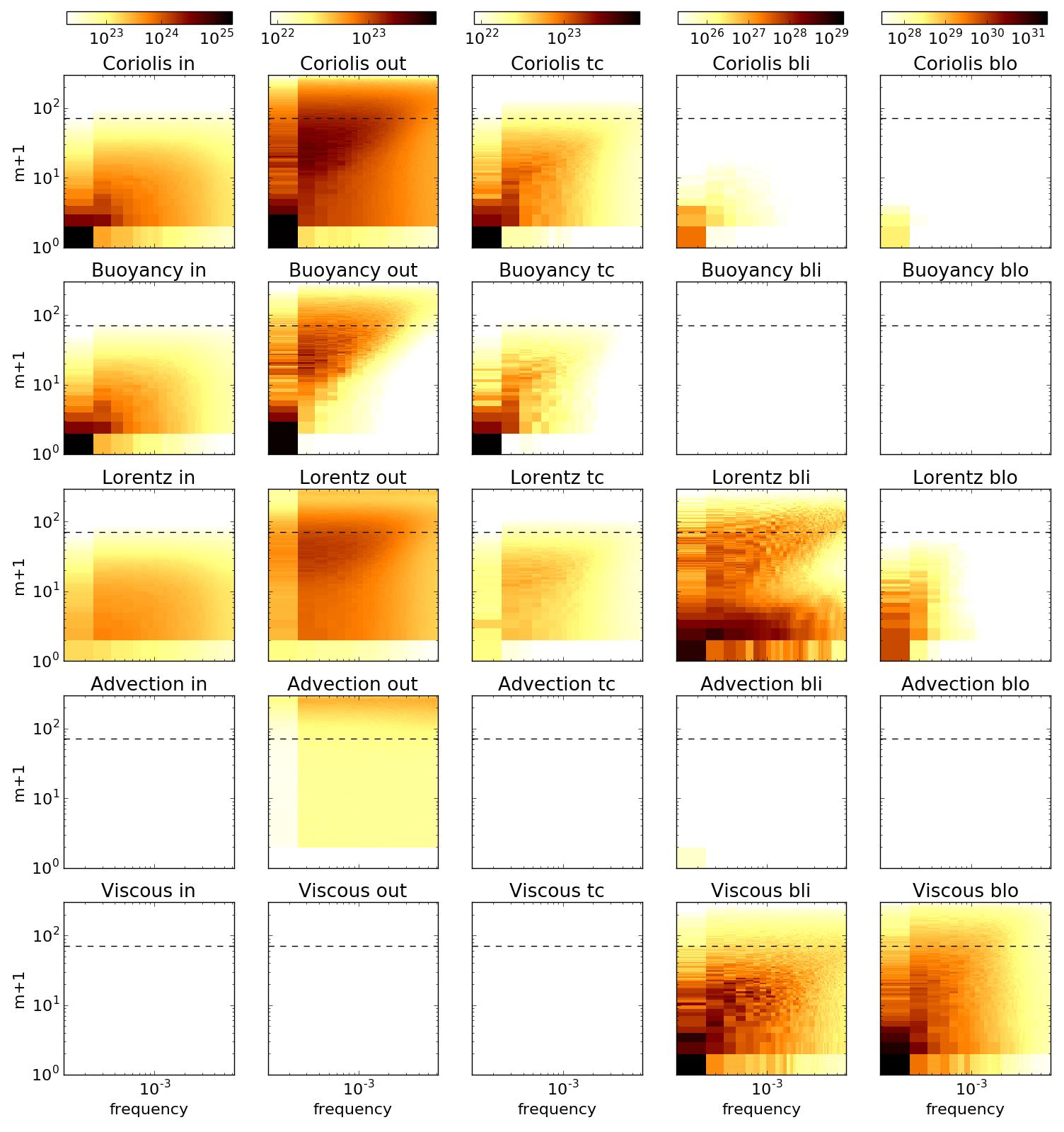}
\caption{Space-time spectra of various terms in the vorticity equation of S2, in the regions defined in figure \ref{fig:regions}.
The colormaps use a logarithmic scale.
Advection is the curl of $u\nabla u$.
Fields were truncated to $\ell_{max}=299$ before computing each terms.
The dashed horizontal line marks the unstable mode at onset of convection.
Frequency is in units of rotation rate $\Omega$, and has been shifted so that the leftmost tile in each plot is the time-average (zero frequency).
We have also computed the global time variation of vorticity ($\partial_t\, \omega$, not shown) which would appear white everywhere, except at the highest frequency in the \textit{out} region.
}
\label{fig:fb2d_S2}
\end{center}
\end{figure}

The resulting two-dimensional spectra are represented in Figure \ref{fig:fb2d_S2} for each region and each term of simulation S2. 
Clearly, the different regions have very different dynamics, and averaging over all of them will definitely blur the analysis.
Because the analyzed series spans only 1300 rotation periods for S2 (2081 for S1), we focus on the so-called rapid dynamics.

The broad picture, shown in Figure \ref{fig:fb2d_S2}, consists of a clear dominance of Coriolis, Buoyancy and Lorentz terms in the interior regions (far from boundaries).
This is consistent with the so-called MAC (Magneto-Archimedes-Coriolis) balance.

Close to the boundaries the balance involves mostly viscosity and Lorentz forces perturbed by the Coriolis term, which is compatible with the balance in a Hartmann layer.
We notice that all forces are smaller for $m=0$ than for $m>0$, except for the time-average, which stands clearly out as a thermal wind balance, even outside the tangent cylinder.
Remember however that the zonal flow is very much affected by the magnetic field (compare S1 and S1* in Figure \ref{fig:avg}) even though the Lorentz force is absent from the main thermal wind balance.
This analysis also illustrates the complexity of the spatio-temporal balances, which is completely overlooked when averaging in time.
Note that for S1*, the balances do not depend on frequency (not shown).

\begin{figure}
\begin{center}
\includegraphics[width=0.99\textwidth]{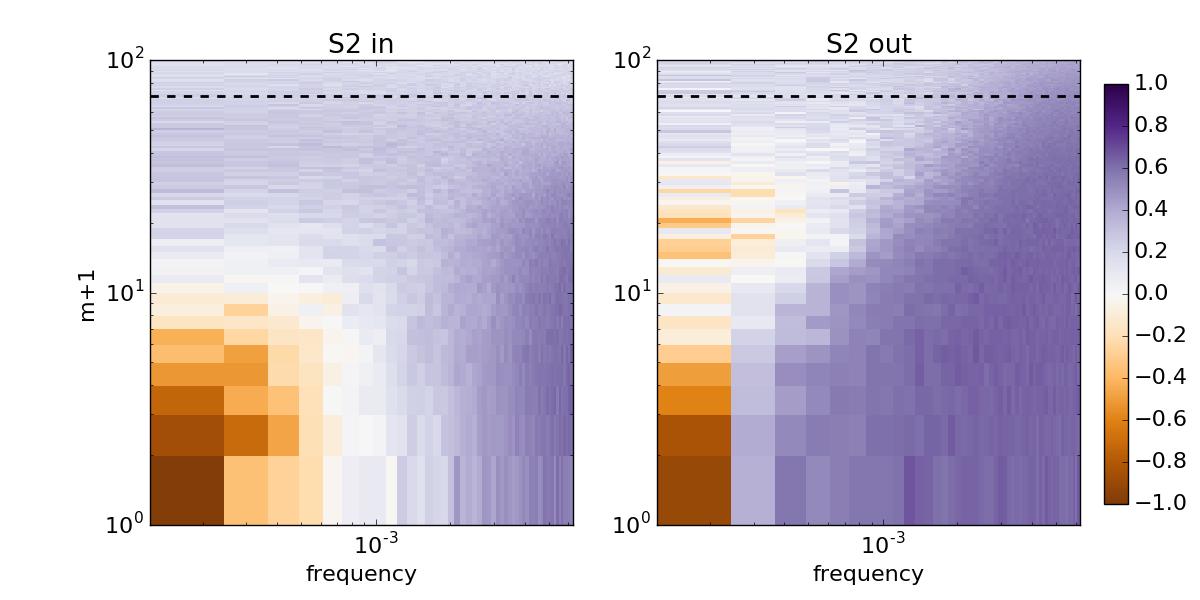}
\caption{Deviations from MAC balance in the bulk of simulation 
S2, as a function of azimuthal order $m$ and frequency, inside (left) and outside (right) the tangent cylinder.
The color maps show the quantity $(M-A)/C$ where $M$, $A$ and $C$ are respectively the rms of the Lorentz, Buoyancy and Coriolis terms in the vorticity equation.
All values are between -1 and 1, except for the mean flow (lower left corner).
The dashed horizontal line marks the unstable mode at onset of convection.
Frequency is in units of $\Omega$.
}
\label{fig:mac_S2}
\end{center}
\end{figure}

With our spatio-temporal decomposition, we are able to refine the main MAC balance of the two interior regions.
Figure \ref{fig:mac_S2} shows the relative importance of the Lorentz and Buoyancy terms.
Specifically, the Coriolis term is mostly balanced by the Lorentz term in purple regions, and by the buoyancy term in orange regions.

Inside the tangent cylinder, the Lorentz term is somewhat sub-dominant at large scales, where a balance between Coriolis and buoyancy terms describes well the largest scales and their slow evolution.
Through alignment of velocity and magnetic field, the Lorentz force shapes the flow while keeping itself weak.
At shorter time- and length-scales, the Lorentz term does eventually overcome buoyancy.

Outside the tangent cylinder, although the very weak mean flow is also controlled by a balance between Coriolis and buoyancy terms, the Lorentz force is more important.
In fact, buoyancy plays a primary role at intermediate length-scales only (around $m=20$).
Note that this azimuthal wavenumber is much smaller than that of the onset of convection ($m_c=67$).
A magnetostrophic balance, involving mostly Coriolis and Lorentz terms, controls the rapid dynamics of the flow outside the tangent cylinder.

\subsubsection{Flow invariance along the rotation axis}

In order to quantify the $z$-invariance of the flow, we introduce
\begin{equation}
G_m(\omega,s_0,s_1) = \frac{\int_{s_0}^{s_1} \langle\mathbf{U}_m(\omega,s,z)\rangle_z^2  H(s)\,s \mathrm{d} s }
					{\int_{s_0}^{s_1}\langle\mathbf{U}_m^2(\omega,s,z)\rangle_z  H(s)\,s \mathrm{d} s }
\end{equation}
where $\langle\mathbf{U}_m(\omega,s,z)\rangle_z$ is the $z$-average of $\mathbf{U}_m(\omega,s,z)$, the order $m$ Fourier coefficient of the velocity field at frequency $\omega$, cylindrical radius $s$ and distance $z$ above the equatorial plane.
$H(s)$ is the height of the cylinder of radius $s$ embedded in the sphere.
$G_m(s_0,s_1)$ measures the energy of the $z$-averaged flow relative to the energy of the total flow, and quantifies the geostrophy of the flow; for a geostrophic flow (i.e. independent of $z$), $G_m = 1$.
Note also that a quasi-geostrophic flow, which has $U_z$ depending (linearly) on $z$, will have $G_m < 1$.

\begin{figure}
\begin{center}
\includegraphics[width=0.7\textwidth]{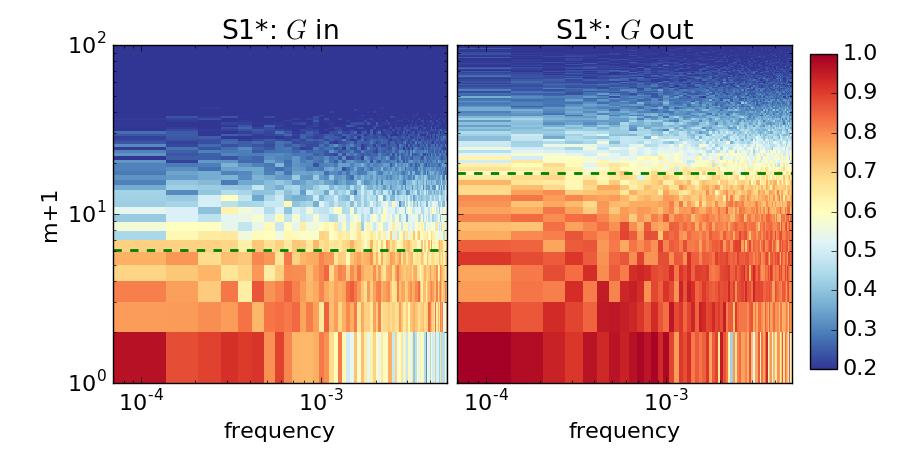}
\includegraphics[width=0.7\textwidth]{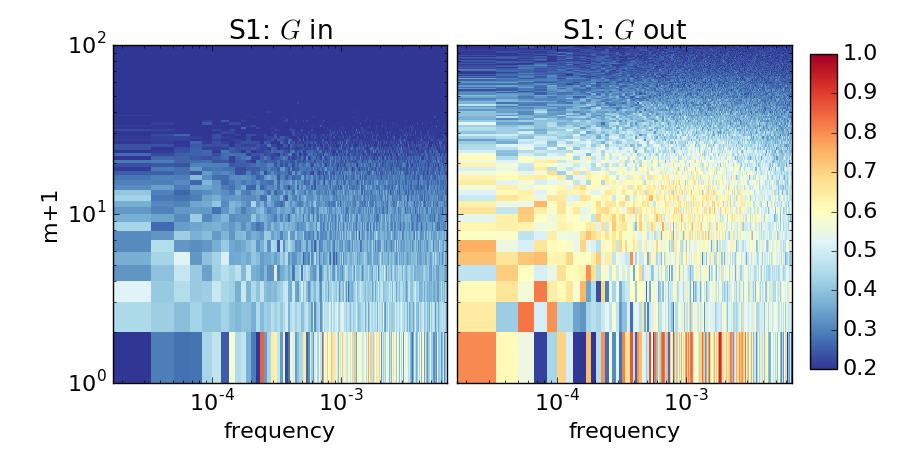}
\includegraphics[width=0.7\textwidth]{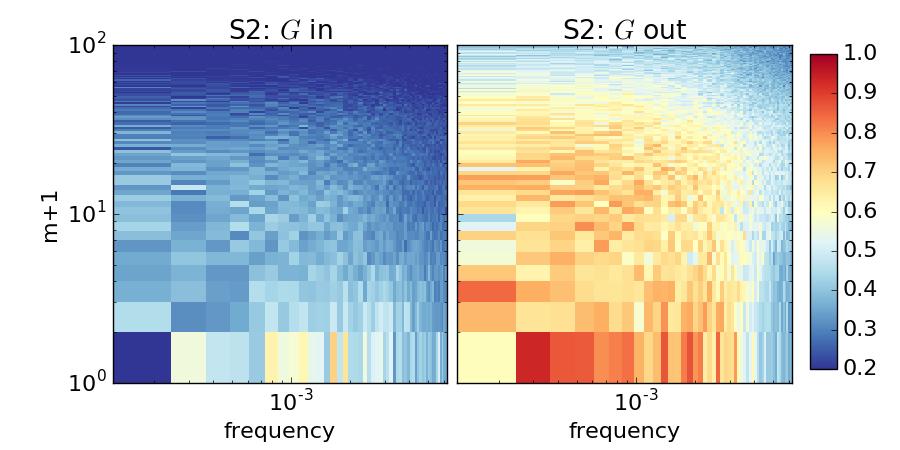}
\caption{Colormaps of $G$, measuring the geostrophy of the flow, as a function of azimuthal order $m$ and frequency, inside (left) and outside (right) the tangent cylinder.
Top: simulation S1* ($E=10^{-6}$, no magnetic field).
Middle: simulation S1 ($E=10^{-6}$, $\Lambda=8.2$, $Le=6.4 \times 10^{-3}$).
Bottom: simulation S2 ($E=10^{-7}$, $\Lambda=2.5$, $Le=1.6 \times 10^{-3}$).
Frequency is in units of $\Omega$.
$G=1$ means a geostrophic flow, $G=0$ is for a completely ageostrophic flow.
We find that for $G>0.6$ the flow already looks fairly columnar while for $G>0.8$ the flow displays sharp and straight columns.
The dashed line indicates the wavenumber $m$ for which $Ro_{i,o}(m) = Ro\,m/r_{i,o} = 0.1$ in S1*.
}
\label{fig:qgness}
\end{center}
\end{figure}

Figure \ref{fig:qgness} represents $G_m$ within the tangent cylinder and outside the tangent cylinder.
First we see that the large-scale flow in S1* have large values of $G_m$, indicating a high $z$-invariance of the flow, even within the tangent cylinder.
This invariance is lost at smaller scales, when the local Rossby number $Ro(m) = Ro \, m/\delta > 0.1$ -- where $\delta = r_o$ for the outer region and $r_i$ for the inner region.
When influenced by a strong magnetic field, the $z$-invariance is completely lost inside the tangent cylinder, while it is only lowered outside, all the more as the Lehnert number is decreased (compare results for S1 and S2, and see also Fig. \ref{fig:Uavg_zavg}), except for the axisymmetric mean flow.
The quasi-geostrophic description \citep[where the flow dynamics is treated as invariant along the rotation axis, e.g.][]{schaeffer2006,gillet2011,labbe2015} is thus relevant outside the tangent cylinder, even for time-scales which would correspond to several decades or even centuries (when rescaled using Alfvén or advective time-scales).

\subsection{Torsional Alfvén waves}

\begin{figure}
\begin{center}
\includegraphics[width=0.99\textwidth]{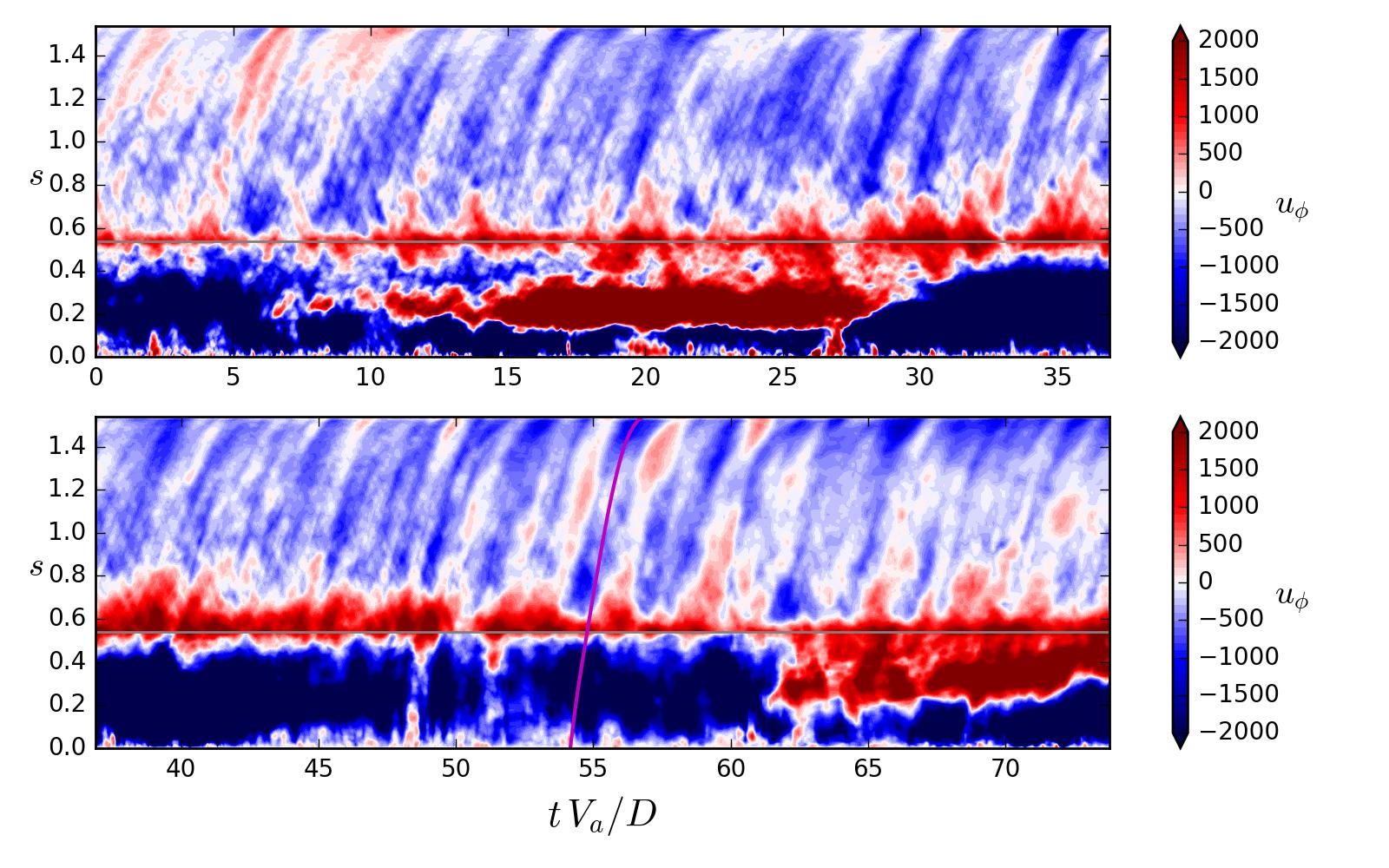}
\caption{space-time diagram of $z$-averaged zonal flow $u_\phi$ showing torsional wave propagation outside the tangent cylinder in the simulation S2 ($E=10^{-7}$, $Pm=0.1$).
Inside the tangent cylinder (marked by the horizontal grey line at $s=r_i$), the flow is averaged in the Northern hemisphere only.
Bottom panel continues the top one.
The magenta curve is the signature of a propagation at the expected torsional Alfvén wave speed.
}
\label{fig:taw_S2}
\end{center}
\end{figure}

The high $G$ values obtained for zonal flow at various frequencies together with a magnetic energy much larger than the kinetic energy suggest the presence of torsional waves.
Figure \ref{fig:taw_S2} shows a space-time diagram of $z$-averaged zonal flow.
Outside the tangent cylinder, the propagation of waves from the tangent cylinder ($s=0.54$) towards the equator ($s=1.54$) is striking.
There are also a few events that propagate in the other direction, which seem to be spawned in the fluid interior.
Similar waves are seen in S1, although the signal is less clear (not shown). In S0, however, no such wave pattern appears.
Interestingly, because the mean zonal flow is weak in our S1 and S2 simulations, we observe the torsional wave propagation outside the tangent cylinder without even removing the mean or employing any filtering.
This contrasts with previous studies, where the mean flow was subtracted or filtered out \citep{wicht2010, teed2014}.
We also note, that the frequency spectra of axisymmetric ($m=0$) motions do not display significant peaks that would correspond to torsional modes.
This absence of distinct peaks was also noted by \citet{gillet2015} for core flows inverted from geomagnetic data.
However, the measure of geostrophy $G_m$ peaks at several frequencies that are compatible with such modes (see Fig. \ref{fig:qgness} bottom right).

The torsional waves seem to originate mostly near the tangent cylinder ($s=0.54$).
Refering to figure \ref{fig:fb2d_S2}, we can see that for $m=0$, the Lorentz force dominates the region close to the inner-core (named \textit{bli}), whereas it stays weak in the vicinity of the tangent cylinder away from the boundary (region named \textit{tc}).
This hints to the Lorentz force close to the inner-core as the main source for torsional waves, which is in line with recent findings \citep{teed2015}.

Inside the tangent cylinder, the signal is dominated by the variability of the strong twisted polar vortex system, and no waves are seen there.
Note that with a calm TC, torsional waves are also expected there \citep[as observed by][]{teed2015}.
Reflection of the waves at the equator depends on both the magnetic Prandtl number $Pm$ \citep{schaeffer2012} and the mantle conductivity \citep{schaeffer2016b}.
For an insulating mantle, only weak reflection is expected even for $Pm$ as low as $0.1$.
Indeed no obvious reflection is observed in our simulations.

\begin{figure}
\begin{center}
\begin{minipage}{0.6\textwidth}
\includegraphics[width=1\textwidth]{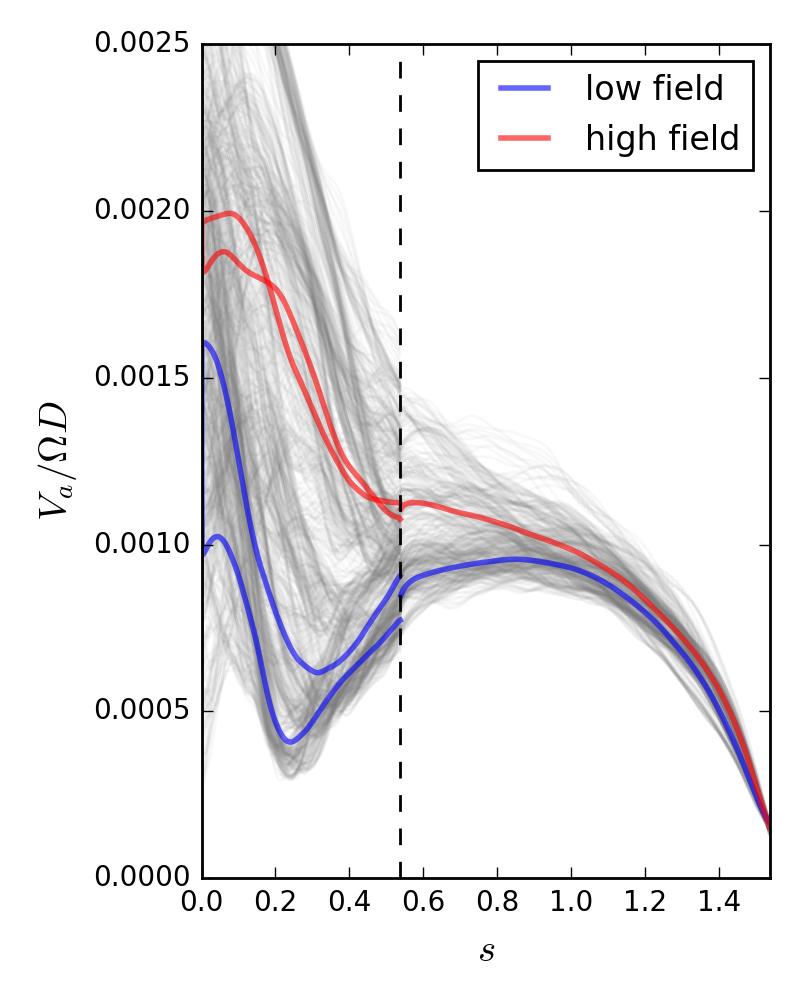}
\end{minipage}
\begin{minipage}{0.39\textwidth}
\includegraphics[width=1\textwidth]{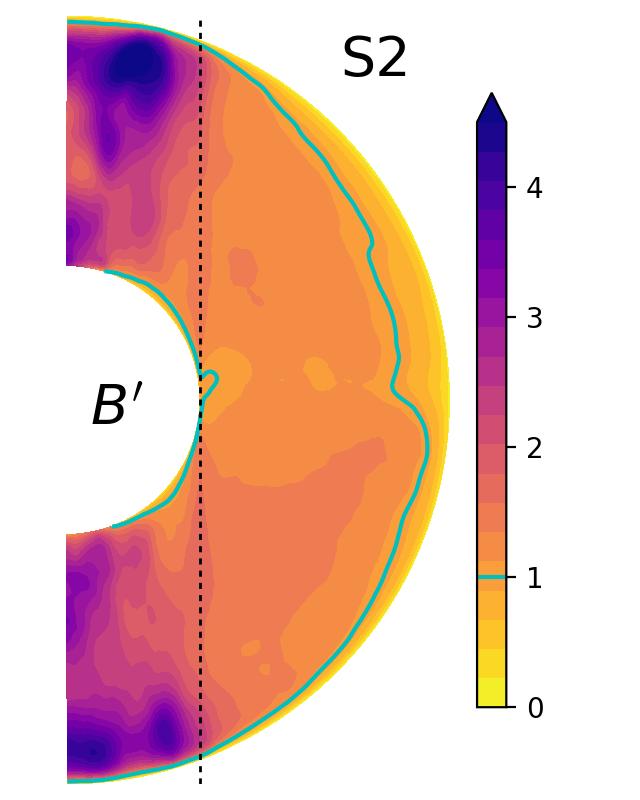}
\end{minipage}
\caption{\textit{Left:} rms value of the cylindrical radial magnetic field (averaged over $z$ and $\phi$) in S2, which is proportional to the torsional Alfvén wave propagation speed $V_a$.
The blue and red thick curves show the time-average on the first half of the time-serie (low field) and the second half (high field).
\textit{Right:} magnetic field fluctuation levels averaged in longitude and time in S2 (Elsasser units).
}
\label{fig:Bs2_S2}
\end{center}
\end{figure}

Figure \ref{fig:Bs2_S2} (left) represents the torsional Alfvén wave speed 
\begin{equation}
V_a(s,t) = \sqrt{\frac{ \int_z \int_\phi B_s^2 dz\,d\phi} {\int_z \int_\phi dz\,d\phi  }} .
\end{equation}
It is interesting to notice that the speed decreases towards the equator, as in \cite{gillet2011}, where it reaches a finite value about 6 times smaller than at mid-depth.
The fluctuations around the mean are large inside the tangent cylinder and decrease outwards.
This is consistent with the overall fluctuation level of $B$ shown in figure \ref{fig:Bs2_S2} (right), dominated by the inner region.

\subsection{Taylor constraint}		\label{sec:taylor}

We have shown that the most important terms in the vorticity equation are the Coriolis, Lorentz and Buoyancy terms, the other terms being much smaller.
When viscosity and inertia are completely removed, \cite{taylor1963} showed that an equilibrium exists where the net torque on each geostrophic cylinder is zero.
This is known as the Taylor constraint and solutions obeying this constraint are said to be in a Taylor state.
Torsional waves are triggered by deviations from this equilibrium.
Following \citet{wicht2010}, we quantify to what extent this constraint is enforced using
\begin{equation}
 \mathcal{T}(s,t) = \frac{ \int_z \int_\phi (j \times b).e_\phi } {\int_z \left| \int_\phi (j \times b).e_\phi \right| }
\end{equation}
which is zero if the force circulation on the geostrophic cylinder of radius $s$ is zero, and 1 if the Lorentz force keeps the same sign at various heights.

\begin{figure}
\begin{center}
\includegraphics[width=0.99\textwidth]{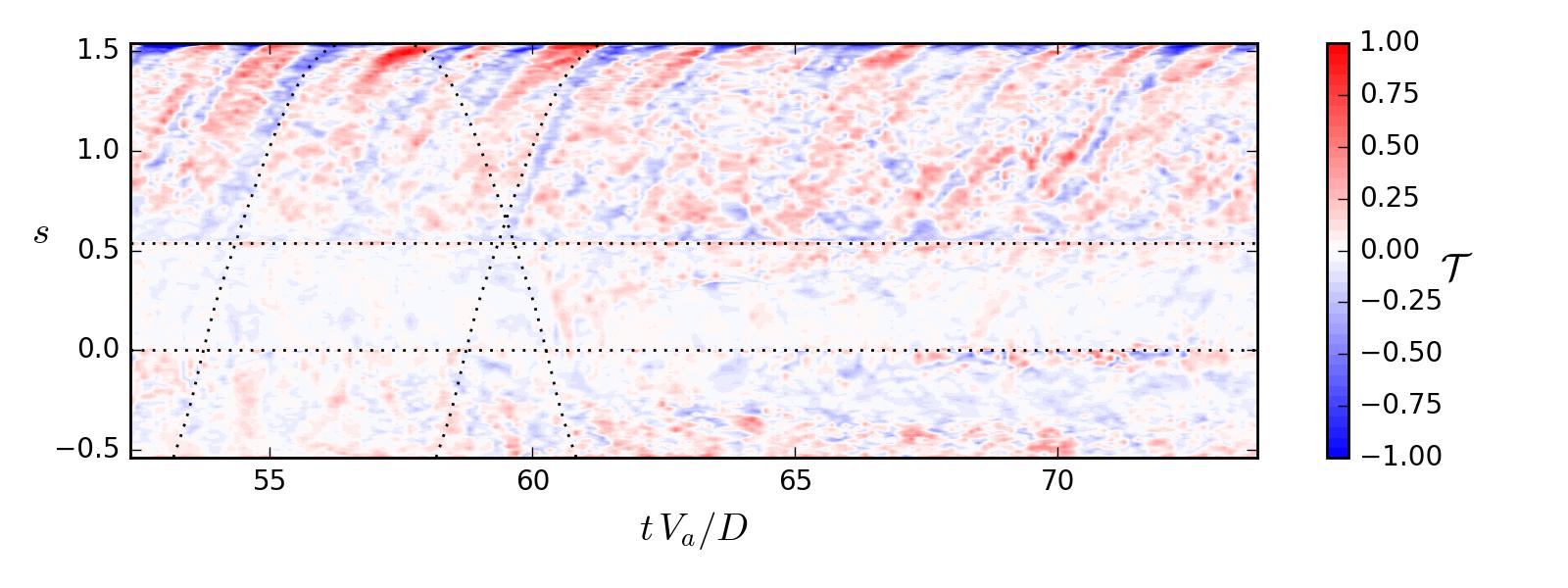}
\includegraphics[width=0.99\textwidth]{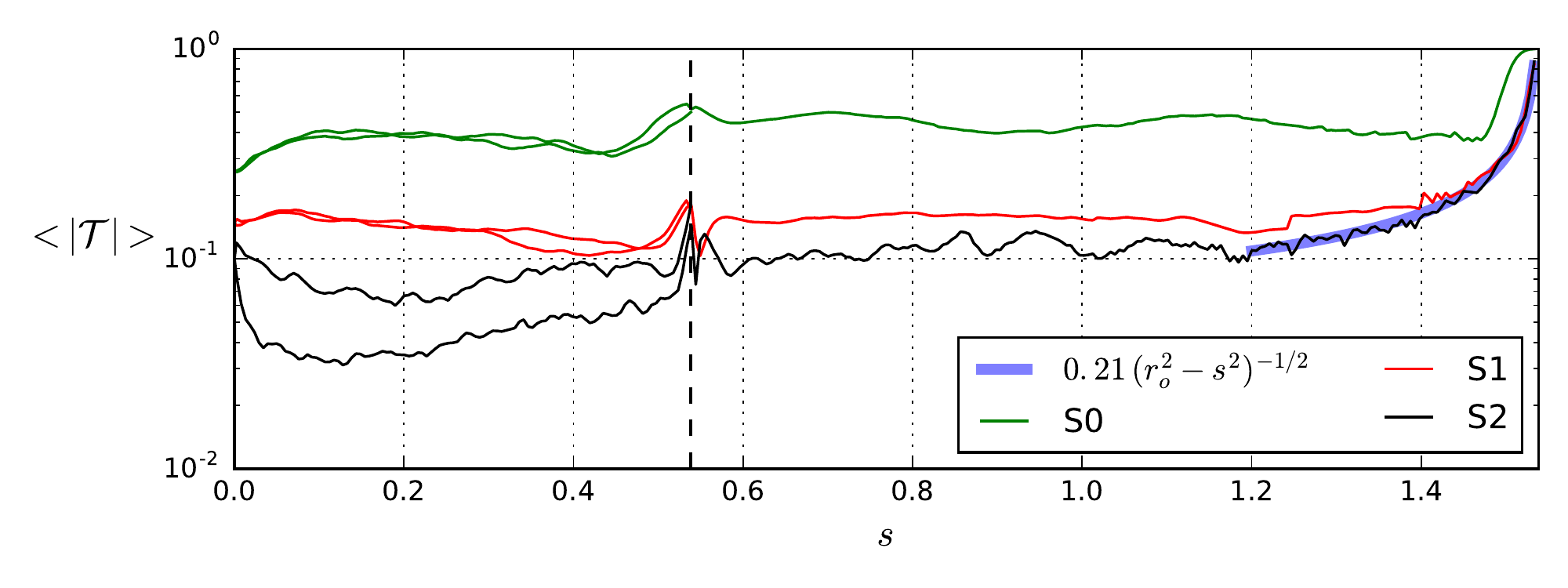}
\includegraphics[width=0.99\textwidth]{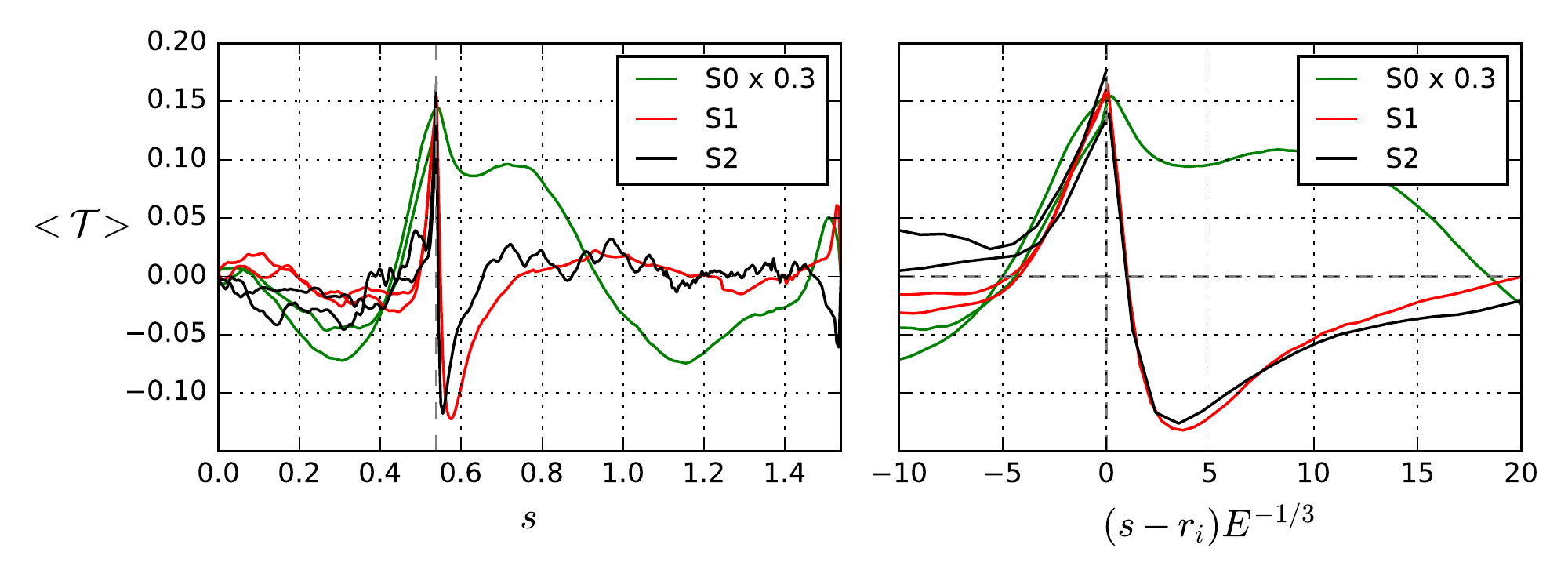}

\caption{Distance to the Taylor constraint as measured by $\mathcal{T}$.
Top: space-time diagram of $\mathcal{T}(s,z)$, in the end of simulation S2 where the signature of torsional Alfvén wave is dominant in the outer region $s>r_i$.
Within the tangent cylinder ($|s|<r_i$), the values for $\mathcal{T}$ in the northern (southern) hemisphere are given for $s>0$ ($s<0$).
Middle: time-average of the absolute value $\langle|\mathcal{T}|\rangle(s)$ for simulations S0 ($E=10^{-5}$), S1 ($E=10^{-6}$), and S2 ($E=10^{-7}$).
Bottom: time-average $\langle\mathcal{T}\rangle(s)$ (without the absolute value).
The two lines within the tangent cylinder $(s<r_i)$ are the separate values for the north and south hemispheres.
}
\label{fig:taylor}
\end{center}
\end{figure}

Figure \ref{fig:taylor} shows different views of this quantity $\mathcal{T}$.
The space-time diagram is dominated by torsional waves outside the tangent cylinder, while inside the tangent cylinder $\mathcal{T}$ is smaller.
The time-average of the absolute value of $\mathcal{T}$ shows this trend as well, and also shows that $\langle|\mathcal{T}|\rangle$ decreases significantly from simulation S1 to S2.
This may be linked to lower viscosity and stronger magnetic fields in S2.
However, the maximum close to the equator, which scales as $1/h(s)$ (where $h(s) = \sqrt{r_o^2 - s^2}$ is the height of the geostrophic cylinder) does not seem to change between S1 and S2.
It is in fact not surprising that $\mathcal{T}$ reaches 1 near the equator.
There, the height of geostrophic cylinders becomes too small for the magnetic torque to have sufficient variations in $z$ to cancel.
Hence, the measure of $\mathcal{T}$ is spoiled by geometric effects close to the equator, reaching values close to 1 independently of viscosity or inertia.
Fortunately, when averaging over the volume, it has less effect.
We find $\mathcal{T}_{S1}=0.16$ and $\mathcal{T}_{S2}=0.12$.
These numbers are significantly larger and decrease slower than those reported by \cite{aubert2017} for stress-free boundary conditions.
Considering that the Reynolds stress and viscosity in the bulk are negligible in S2, but that viscous forces balance magnetic forces near the outer boundary (see Fig. \ref{fig:fb2d_S2} \textit{out} and \textit{blo} regions for $m=0$), it suggests that viscous torques in the boundary layer (Ekman friction) dominates the torque balance on geostrophic cylinders.

The quantity $\langle|\mathcal{T}|\rangle$ includes contribution from torsional waves and other time-dependent fluctuations, and as such is not representative of the long-term equilibrium.
We checked that the time-averages of $\mathcal{T}$ (sign included) for S1 and S2 do not tend to zero but rather to very similar profiles when averaged over various time spans.
These averages $\langle\mathcal{T}\rangle$ are shown for S0, S1 and S2 in Figure \ref{fig:taylor} (bottom).
The amplitude is significantly smaller than that of $\langle|\mathcal{T}|\rangle$, and of the same level inside and outside the tangent cylinder.
The prominent peaks attached to the tangent cylinder scale as $E^{1/3}$ (or $E^{1/4}$, which are difficult to distinguish with one decade of $E$), pointing to an important role of viscosity there, compatible with a Stewartson layer.

\section{Discussion}

We have reached the $E=10^{-7}$ milestone for a geodynamo simulation with vigorous convection.
Our simulations do not significantly deviate from the scaling laws based on convective power input proposed by \citet{christensen2006}.
However, the low viscosity associated with detailed analyses including force balances at different time- and length-scales bring a renewed picture of planetary dynamos.

\begin{figure}
\begin{center}
\includegraphics[width=0.9\textwidth]{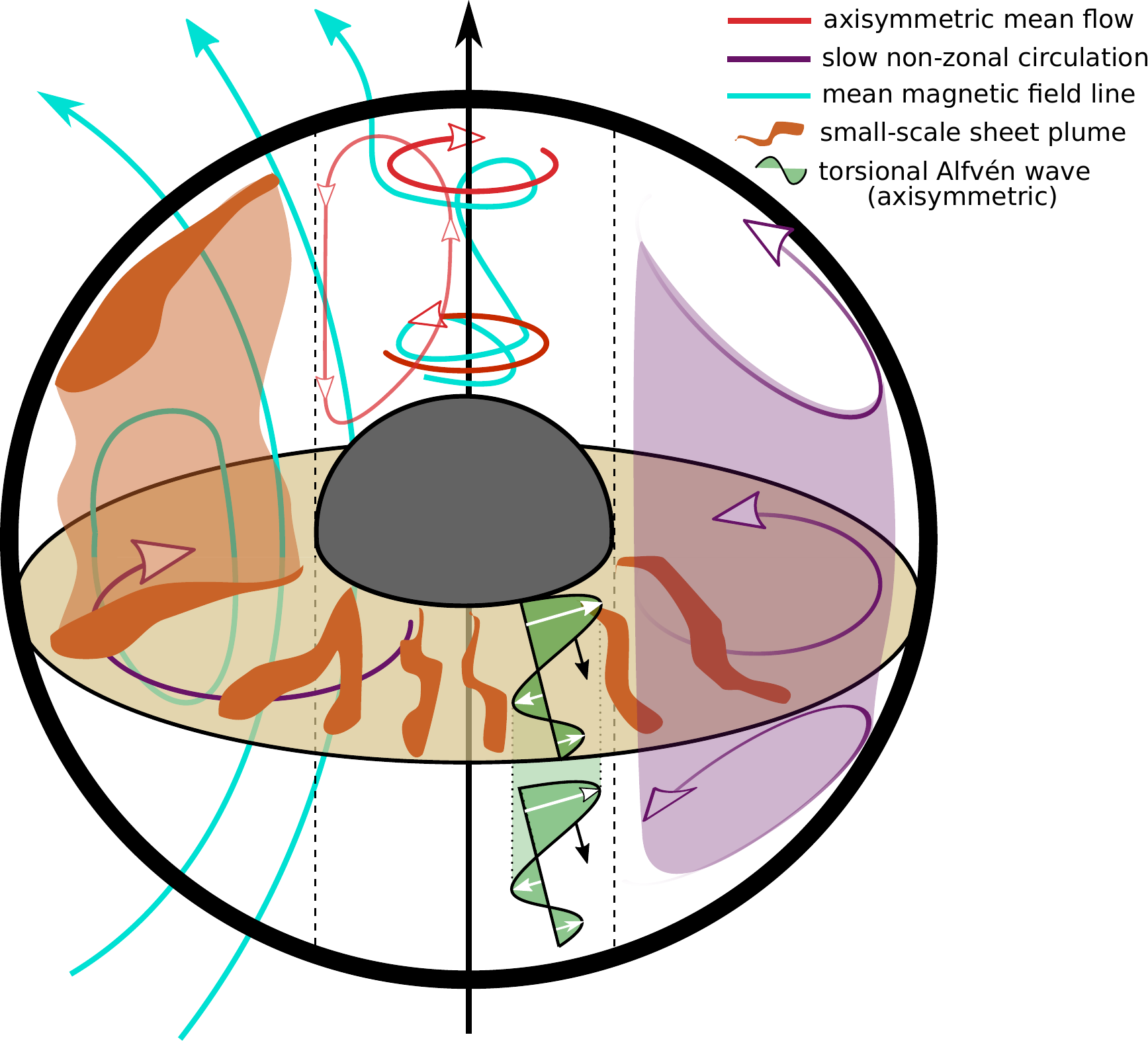}
\caption{Sketch of the internal dynamics in our lowest viscosity simulation S2 ($E=10^{-7}$, $Pm=0.1$, $Ra/Ra_c=6316$).}
\label{fig:sketch}
\end{center}
\end{figure}

The internal dynamics observed in the less viscous simulation S2 is summarized in Figure \ref{fig:sketch}.
Importantly, global scale, columnar, non-zonal circulations emerge on intermediate time-scales and are discussed in \S\ref{sec:gyre_drift}.
The zonal mean flow is strongly suppressed outside the tangent cylinder (TC), as seen in Fig. \ref{fig:avg}, where it is replaced by torsional Alfvén waves (Fig. \ref{fig:taw_S2}) triggered by Lorentz torque fluctuations at the inner-core boundary (Fig. \ref{fig:fb2d_S2}).
Inside the TC, the zonal flow is enhanced by the magnetic field and takes the form of a polar twisted vortex (Fig. \ref{fig:avg}), with important fluctuations in time (Fig. \ref{fig:ub_spec2d_S2}).
Note also that the flow is by no means invariant along the rotation axis inside the TC, contrarily to the non-magnetic case.
Torsional Alfvén waves are not visible in this very active inner region.

\subsection{Inhomogeneities (in space and in time)}

Our simulations point to strong spatial inhomogeneities (see Fig. \ref{fig:3D}) and to large temporal fluctuations of convective spherical dynamos  (see Fig. \ref{fig:nrj}).
The tangent cylinder (TC) separates regions with different dynamics. The inside of the TC is characterized by strong zonal flows and toroidal fields (see Fig. \ref{fig:avg}).
Albeit occupying a small portion of the fluid volume, it bears a significant fraction of the total kinetic (about $1/10$) and magnetic energies (about $1/4$).

\begin{figure}
\begin{center}
\includegraphics[width=0.99\linewidth]{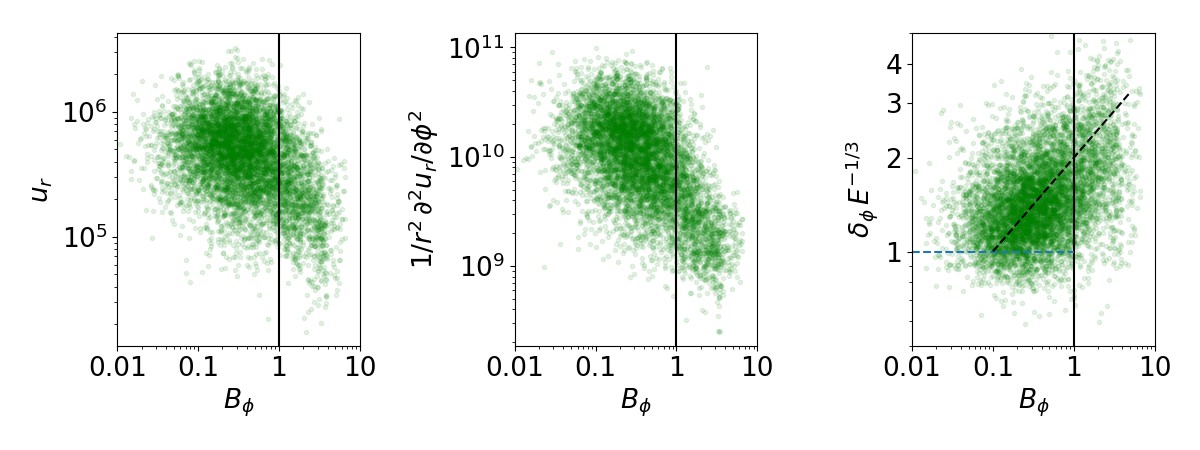}
\caption{Distribution of the amplitude of some local quantities in S2 as a function of the local azimuthal magnetic field $B_\phi$ (in Elsasser units), in the equatorial plane at mid-depth ($0.87<r<1.21$) and from 20 snapshots. For the radial velocity $u_r$ (left), its second $\phi$ derivative $1/r^2 \partial_{\phi\phi} u_r$ (middle) and the azimuthal length-scale $\delta_\phi = \sqrt{r^2 u_r/\partial_{\phi\phi} u_r}$ (left) obvious trends can be seen with the local magnetic field. Note that all quantities are smoothed over 20 azimuthal points to reduce the scatter.
The dashed line is $2 {B_\phi}^{0.3}$ to guide the eye. }
\label{fig:bscalecorrel}
\end{center}
\end{figure}

Our study also points to large heterogeneities outside the tangent cylinder, where regions of intense magnetic fields are observed next to regions of almost zero field (see Fig. \ref{fig:equat_S2} and \ref{fig:3D}).
This is linked to the magnetic field outside the tangent cylinder being mostly non-zonal (70\% of the magnetic energy)
in contrast with the field at the core surface, which is mostly an axisymmetric dipole (Fig. \ref{fig:Br}).
The heterogeneous magnetic field strength translates into heterogeneous length-scales for the convection.
To quantify this effect, we compare some local quantities associated with the radial flow in the equatorial plane to the local magnetic field.
We find that the most convincing correlations occur with the azimuthal field $B_\phi$, the component mostly perpendicular to the plume sheets.
In Fig.~\ref{fig:bscalecorrel}, it is clear that the amplitude of the radial flow decreases by a factor of about 10 where the azimuthal field is strong, hence suppressing convection (Fig.~\ref{fig:bscalecorrel} left).
Meanwhile, the length-scale $\delta_\phi$ increases up to 3 times from the viscous scale $E^{1/3}$ (Fig.~\ref{fig:bscalecorrel} right).
We therefore conclude that the primary effect of the magnetic field outside the TC is to damp convection.
As a secondary effect, the length-scale increases.
Note that in S1 these correlations are less spectacular (not shown).


Lastly, the fields are also heterogeneous in time.
This aspect has been largely overlooked as databases for previous calculations often record time-averaged values.
These large temporal fluctuations are even more intriguing as they are completely absent from our non-magnetic convection case (compare Fig. \ref{fig:u_spec2d_S1nob} with Fig. \ref{fig:ub_spec2d_S1}; see also Fig. \ref{fig:nrj}).
It appears that the dynamo regime is accompanied with global long-term correlation, otherwise absent from the hydrodynamics.


\subsection{Comparison with the state of the art}
At this stage, it is worthwhile to replace our findings in the context of the recent studies conducted
by \citet{yadav2016b}, \citet{sheyko2016} and \citet{aubert2017}, whose properties are recalled in table~\ref{tab:simus}. 

\subsubsection{Force balance}
\cite{yadav2016b} 
performed a systematic parameter survey down to an Ekman number equal to $10^{-6}$ and
demonstrated convincingly that in their dynamo simulations, the first order force balance 
(once the zeroth-order geostrophic balance had been removed) 
took place between the Lorentz force, the buoyancy force and the part of the Coriolis force that is
uncompensated by the pressure gradient. \cite{yadav2016b} stressed that, in the bulk of the spherical shell, 
inertial forces and viscous forces represented less than $5\%$ of these 3 dominant actors of the dynamics
(outside the boundary layers).  
This statement was based on an integrated analysis, that is on the average rms values of the various forces at work in the simulations.
Our simulation S2 refines this finding at more extreme conditions.
Not surprisingly, our spatial-temporal analysis of the vorticity equation reveals that this 3-term balance occurs in the bulk of the domain (`in', `tc', and `out' regions in Figure~\ref{fig:regions}), while the dynamics of the boundary layers is primarily governed by viscous and magnetic forces.

Inspection of the scale-dependent balance (space-time spectra of Fig.~\ref{fig:fb2d_S2}) shows that we must distinguish the regions inside and outside the tangent cylinder (TC).
Inside the TC, at large-scale (wavenumbers $m \lesssim 10$) and low frequency (representative of the most energetic flow, see Fig.~\ref{fig:ub_spec2d_S2}) the bulk 3-term MAC balance is mostly a 2-term balance between buoyancy and the uncompensated Coriolis force, representative of the thermal wind balance discussed above.
At smaller scale or higher frequency, magnetic effects overcome buoyancy.
To be precise, since we are dealing with the vorticity equation in our analysis, the curl of the Lorentz force becomes of comparable magnitude 
as the uncompensated Coriolis and larger than the buoyancy force (Fig.~\ref{fig:mac_S2}).
Outside the TC, the picture is different. For the energetic small-scale convection plumes ($20 \lesssim m \lesssim 70$, see Fig.~\ref{fig:ub_spec2d_S2}), a MAC balance is observed where Lorentz and buoyancy terms combine to balance the Coriolis term (Fig.~\ref{fig:mac_S2}).
However, the Lorentz force does not contribute much to the large-scale, low frequency force budget, which is again governed by a thermal wind balance.
We also remark that outside TC, inertia (advection) is not completely negligible.
In both regions, at frequencies higher than the Alfvén frequency (about $2 \times 10^{-3}$ for S2), the buoyancy becomes insignificant and the Lorentz forces alone balance the Coriolis terms at all length scales.

\begin{figure}
\begin{center}
\includegraphics[width=1\textwidth]{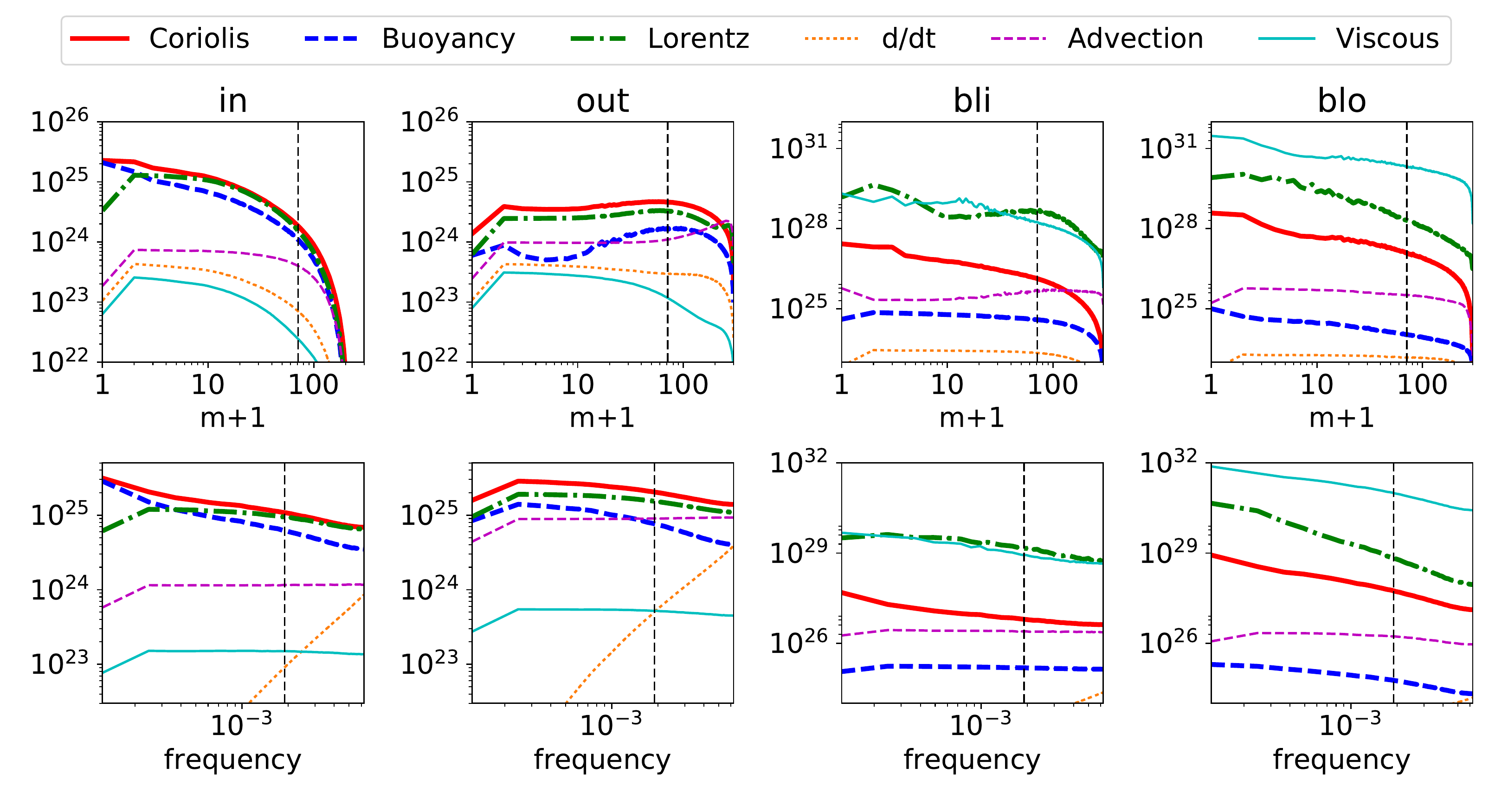}
\caption{$m$ spectra (top) and frequency spectra (bottom) of various terms in the vorticity equation of S2, in the regions defined in fig.~\ref{fig:regions} (except \textit{tc} because it is simply an intermediate between \textit{in} and \textit{out}).
This stacks the data shown in fig.~\ref{fig:fb2d_S2} by summing over frequencies (top) or over $m$ (bottom).
The vertical dashed line is the critical wavenumber at onset of convection $m_c=67$ or the Alfvén frequency ($Le=1.9 \times 10^{-3}$).
}
\label{fig:fb1d_S2}
\end{center}
\end{figure}

The analysis of \cite{aubert2017} separates neither the regions inside and outside the TC, nor the contributions at various time-scales.
Instead their decomposition in harmonic degree $\ell$ mixes contributions from the two regions.
To ease comparison with their work, we have transformed our dynamical balance analysis into a more global one to show averaged spatial or temporal dependence in Fig.~\ref{fig:fb1d_S2}.
The global force balance analysis of \cite[their Fig.~2b]{aubert2017} at $E=3\times10^{-6}$ shows that buoyancy does not enter the main balance at small scales ($\ell > 10$), a behavior not seen in our analysis in azimuthal wavenumber $m$, but maybe related to the frequency dependence in S2 (Fig.~\ref{fig:fb1d_S2} bottom).
In this integrated picture, the three main terms (Coriolis, Buoyancy and Lorentz) clearly dominate for $1 \leq m \lesssim 70$, matching the MAC balance paradigm inside the TC.
However, outside the TC, the stacking of Fig.~\ref{fig:fb1d_S2} brings the terms closer together, suggesting Coriolis and Lorentz dominate, with significant contribution of both buoyancy and advection.
It is thus important to separate time and space contributions to uncover the finer balance described in the previous paragraph.

\subsubsection{Flow length-scale}
In previous studies, a characteristic length-scale of convection is often defined by an average over the whole domain.
Figure \ref{fig:equat_S2} and \ref{fig:3D} shows the wide variety of length-scales present in simulation S2.
There are tiny plumes that form at the inner-core boundary (with measured width close to the $E^{1/3}$ viscous length-scale) and progressively widen while traveling across the shell.
These small plumes coexist with much larger ones (Fig.~\ref{fig:equat_S2}), and even with global-scale eddies (Fig. \ref{fig:Uavg_zavg}).
This is most likely due to the heterogeneous magnetic field, with large regions of low field next to regions of stronger field (see also Fig.~\ref{fig:bscalecorrel}).
Hence, even though the small viscous length-scale $E^{1/3}$ is present in our simulations, the kinetic energy spectrum peaks at much larger scales (Fig. \ref{fig:spec_avg}).
In fact, the kinetic energy spectra peak at $\ell \sim 10$ for all our dynamo simulations, much like those of \cite{aubert2017}.
A similar peak at $m \gtrsim 10$ appears in the spatial-temporal kinetic energy spectra outside the TC (Fig.~\ref{fig:ub_spec2d_S2} and \ref{fig:ub_spec2d_S1}).
Clearly, in this inhomogeneous, anisotropic system, reducing the convective motion to a single length-scale would be misleading.

\cite{yadav2016b} pointed out the fact that when decreasing the Ekman number while maintaining a significant level of supercriticality (as large as allowed by computing ressources), simulations of dynamo action driven by an imposed temperature drop across the shell  were characterized by large-scale flows in the bulk of the shell.
They ascribed this property to the emergence of the MAC balance (hence a stronger feedback of the Lorentz force on the flow). This result is in contrast with the previous work of \cite{sakuraba2009}, who found (at lower levels of supercriticality) that the emergence of large scales in the bulk of the domain  was only possible if fixed-flux conditions were applied.
Here we also rely on fixed-flux conditions; therefore, we can not entirely corroborate the conclusions of \cite{yadav2016b}.
That being said, our spatio-temporal  analysis clearly demonstrates that the flow becomes weaker and larger scale when and where the magnetic field is strong (recall Fig.~\ref{fig:equat_S2} and \ref{fig:bscalecorrel}).

\subsubsection{Reversals}
The simulations of \cite{sheyko2016} are also in the rapidly rotating regime, with nominal values of the Ekman number as small as $2.4\times10^{-6}$ when adopting our definition of $E$.
The ratio of kinetic to magnetic energy in these calculations is always larger than $1$: it varies between $3$ and $30$ depending on the parameters.
Accordingly, the kinetic energy spectrum is always above the  magnetic spectrum (their Fig.~4a), and no sign of scale separation between the velocity and magnetic fields is to be seen, despite the small value of the magnetic Prandtl number ($0.04-0.05$).
Because the magnetic energy is small, these dynamos can operate in a reversing mode driven by Parker waves \citep[see also][]{schaeffer2006,busse2006parameter,tobias2013,schrinner2014topology}.
This cyclic behavior is akin to that of the solar dynamo, whose polarity changes every $11$~years. 
The field being weak, the emergence of a large scale flow in the bulk of the domain, as highlighted here when the field is strong and by \cite{yadav2016b}, does not seem to be happening.
In addition, these simulations do not exhibit a strong dynamical contrast between the interior of the tangent cylinder and its exterior (our `in' and 'out' regions).
Furthermore, we emphasize that, to our knowledge, no simulation with strong field (i.e. magnetic energy larger than kinetic energy) has ever exhibited polarity reversals.
It is therefore questionable whether existing reversing simulations are relying on the same mechanism as the one governing the Earth's core dynamo.
Finally, we note that it would be interesting to study the role of the highly active polar regions for the dynamo action and for polarity reversals.


\subsection{From numerical simulations to Earth's core regime}

Present-day numerical simulations, including the ones presented in this article, remain far from matching the physical parameters expected for the Earth core.
However, some encouraging comparisons can be made that also suggest simulations can shed some light on the Earth's core dynamics.

\subsubsection{Invariance along the rotation axis}
Assuming $z$-invariance of motions within the core has proven to be a fecund way of analyzing the secular variation of the Earth's magnetic field \citep[e.g.][]{gillet2010,gillet2011}.
If this is characteristic of the rapidly-rotating regime in which the Earth's core operates, it is important to see if simulations reproduce this trend.
Figure \ref{fig:qgness} shows that the magnetic field reduces the $z$-invariance of the flow.
However, outside of the tangent cylinder, the most energetic motions (compare Fig.~\ref{fig:qgness} with Fig.~\ref{fig:ub_spec2d_S1} and \ref{fig:ub_spec2d_S2}) are still mostly $z$-invariant.
Shortest time- and length-scales, which are also less energetic, lose their $z$-invariance, as expected \citep{nataf2015}.
We note that the flow in S2 is more $z$-invariant than in S1, due to a lower influence of the magnetic field as measured by both the Elsasser and Lehnert numbers.
Because the Earth's core has a larger Elsasser but a smaller Lehnert number than S2, it is difficult to conclude whether the Earth's core is globally more or less $z$-invariant than S2, but our simulations suggest that the large scale, slow motions can be well approximated by columnar flows outside the TC.

\subsubsection{Global circulation and westward drift}	\label{sec:gyre_drift}

A prominent feature of the geomagnetic field is the westward drift of the magnetic flux patches, especially near the equator and in the Atlantic hemisphere \citep[e.g.][]{finlay2003}.
This westward drift has been associated with a large-scale, $z$-invariant, eccentric gyre \citep[e.g.][]{pais2008,pais2015,gillet2015}.

Similarly, a long-term, large-scale, $z$-invariant circulation forms spontaneously in S2 (Fig.~\ref{fig:Uavg_zavg} and \ref{fig:sketch}) and is responsible for the westward drift of the magnetic field in one hemisphere (see Fig.~\ref{fig:wdrift}).
We emphasize that our setup and its boundary conditions have all a strict spherical symmetry (i.e. no heterogeneous thermal boundary conditions at the mantle or the inner-core).
In contrast, the Coupled Earth simulations initiated by \citet{aubert2013CE} produce these features using a heterogeneous heat flux at the boundaries together with a gravitational coupling between inner-core and mantle that must exceed the direct coupling at the CMB \citep[see detailed analysis by][]{pichon2016}.
We remark however that even though the intense jet localized near the tangent cylinder in S1 and S2 (see Figure \ref{fig:Uavg_zavg}) looks similar to the ones inferred from geomagnetic data \citep[e.g.][]{pais2008,schaeffer2011,pais2015,gillet2015,livermore2016}, it flows in the opposite direction.
Nevertheless, it is interesting that a simple homogeneous model like S2 can reproduce features like the westward drift and a non-zonal global circulation.

\subsubsection{Influence of the magnetic field}		\label{sec:infmag}
Our simulations indicate that the magnetic over kinetic energy ratio increases as we get closer to Earth's parameters, in agreement with the scaling laws of \citet{christensen2006}.
This is reassuring, but we get a maximum ratio of about 12 while it approaches $10^4$ in the core.

Although the force balance may suggest a mild effect of the magnetic field on the mean flow, this is clearly not the case.
As shown in figure \ref{fig:nrj}, when the magnetic field is turned off (everything else being kept the same), the kinetic energy quickly increases by a factor 10.
One also observes that the non-magnetic convection in S1* is more evenly distributed in the whole shell, with the formation of a strong zonal jet outside the tangent cylinder, instead of a strong twisted polar vortex in the dynamo case (see figures \ref{fig:avg}, \ref{fig:equat_S1}).
In the presence of magnetic fields, the flows are organized to limit induction, by aligning magnetic and velocity field wherever possible.
This effect can be understood as a generalization of Ferraro's law of isorotation \citep{aubert2005}.
In this way, the strong magnetic field obtained in S1 and S2 completely changes the flow.
Not only does it completely suppress the geostrophic jets, but the convective plumes extend further into the shell (because they are not blocked by the zonal jets anymore).
The plumes are also of larger scale in the bulk, although they often emerge at a very small scale from the ICB, even smaller than without the magnetic field.

Inside the TC, the mean magnetic field intensity reaches a value that corresponds to the Elsasser number $\Lambda > 1$ (Fig.~\ref{fig:avg}), and the mean field and flow are largely aligned, limiting the amount of induction.
Of course, substantial induction is still occurring to sustain the strong field there (locally up to $\Lambda \gtrsim 20$, see Fig.~\ref{fig:3D}), but much less than if the fields were unaligned.
Indeed, focusing on the forces acting on the mean flow that are symmetric about the rotation axis, we find a dominant balance between the Coriolis and buoyancy forces (see Fig. \ref{fig:fb2d_S2} and \ref{fig:mac_S2}).
This indicates that the zonal flows within the TC are very likely thermal winds driven by the density contrast between a light fluid inside the TC and a heavier one outside, which are nonetheless 'shaped' by the strong magnetic field.
The polar vortices inferred within the Earth's core \citep{olson1999} suggest that this regime found inside the TC in our dynamo simulations is indeed realistic.

\subsubsection{Density segregation}

It is tantalizing to relate the massive density contrast depicted in the codensity snapshot of Fig.~\ref{fig:3D} as well as in the mean codensity maps of Fig.~\ref{fig:avg} to the seismic velocity anomalies detected in the Earth's core \citep[e.g.][]{souriau1990} and usually attributed to inner core anisotropy.
In our simulations, the amplitude of the zonal flow and density anomaly contrast scale well according to the thermal wind balance on the global scale
\begin{equation}
2\Omega U \sim \frac{\delta\rho}{\rho} g.  \label{eq:thwind}
\end{equation}
When applied to the Earth's core ($\Omega \simeq 7 \times 10^{-5}$~s$^{-1}$, $U \simeq 30$~km/yr, $g \simeq 10$~m/s) it yields $\delta\rho/\rho \simeq 10^{-7}$, which is orders of magnitude too small to be detectable by seismology as such \citep[see also][]{stevenson1987,aurnou2003}.
Therefore, one should not expect any seismic signature unless seismic velocities vary significantly with chemical composition without much density change.
If there is such an effect, density anomalies  would be related with a chemical segregation within the Earth's core akin to the density separation seen in our simulations.

Interestingly, we observe that the establishment of a strong lateral density gradient requires intense magnetic fields, even though the latter do not directly enter the thermal wind balance (see \S\ref{sec:infmag}).
Furthermore, we remark that a stably stratified region appears within the TC, but only in the dynamo simulations (Fig.~\ref{fig:avg} and \ref{fig:Cprof}).
Analysis of the zonal force balance within the TC shows that the Lorentz force slightly helps the Coriolis force to balance the buoyancy force.
It may thus be worth studying further the impact of magnetic forces on density separation and how they affect the 'thermal' wind balance (eq. \ref{eq:thwind}) in planetary cores.

\subsubsection{The role of viscosity}

Despite reaching very low viscosities, our simulations are not devoid of viscous effects.
They are of course important in the dynamics of the boundary layers, as seen in Fig.~\ref{fig:fb2d_S2}.
This may result in torques that contribute to deviations from the Taylor state (see \S\ref{sec:taylor}).
But viscosity also appears to control two other aspects.
First, the region of transition at the tangent cylinder seems to be controlled by viscosity, scaling with the Ekman number as a Stewartson layer (Fig.~\ref{fig:taylor}).
Second, where the magnetic field is weak, convective plumes occur at the viscous scale $E^{1/3}$ (see also Fig.~\ref{fig:bscalecorrel}).
Whether these features are significant for the Earth's core are open questions.

\subsection{Perspectives and future work.}

Our direct simulations may serve as benchmarks for developing large eddy simulations schemes \citep[see e.g.][]{buffett2003, nataf2015, aubert2017} and for validating asymptotic models \citep[e.g.][]{calkins2016}, which are both needed to further progress towards realistic dynamos.
Several terabytes of data will be held available for a few years to allow further analysis.
It would be also instructive to characterize the turbulence and the dynamo mechanism.
\citet{calkins2016} have recently emphasized the significance of the ordinary Prandtl number in the selection of the dynamo mechanism in a rotating layer.
We also have to investigate whether our results hold for $Pr \neq 1$.

Because of the apparently important role played by the tangent cylinder (TC), it would be of great interest to revisit the role of a conducting inner-core in this parameter range.
Indeed, it might enable the strong toroidal field to reach the boundary and alter the scale of the plumes originating there.
More importantly, if the dynamics within the TC influence significantly the magnetic field generation, we should compare dynamos with and without inner-core, the latter having relevance for the early Earth or other planets.
It may also be interesting to isolate the region within the tangent cylinder to study its own dynamics decoupled from the outer region.

\subsubsection*{Acknowledgements}

{
\footnotesize
The XSHELLS, SHTns and SINGE codes used in this paper are freely available at \url{https://bitbucket.org/nschaeff}.
We thank Julien Aubert and the geodynamo team for fruitful discussions and comments.
We also thank two anonymous referees for their constructive comments that helped improve this paper.
We acknowledge PRACE and GENCI for awarding us access to resource Curie based in France at TGCC (grant pa1413 and t2014047258); GENCI awarded further access to Occigen (CINES) and Turing (IDRIS) under grant x2015047382 and x2016047382.
Part of the computations were also performed on the Froggy platform of CIMENT (\texttt{https://ciment.ujf-grenoble.fr}), supported by the Rh\^ one-Alpes region (CPER07\_13 CIRA), OSUG@2020 LabEx (ANR10 LABX56) and Equip@Meso (ANR10 EQPX-29-01).
Heavy data post-processing was partly performed on the S-CAPAD platform, IPGP, France.
This work was funded by the French {\it Agence Nationale de la Recherche} under grants ANR-11-BS56-011 (AVSGeomag), ANR-13-BS06-0010 (TuDy) and ANR-14-CE33-0012 (MagLune).
ISTerre is part of Labex OSUG@2020 (ANR10 LABX56).
All figures were produced using matplotlib (\url{http://matplotlib.org/}) except Fig. \ref{fig:3D} rendered with paraview (\url{http://www.paraview.org/}) and Fig. \ref{fig:sketch}, drawn with inkscape (\url{https://inkscape.org/}).
}

\appendix

\section{Codensity profiles}	\label{sec:codensity}

\begin{figure}
\begin{center}
\includegraphics[width=0.6\textwidth]{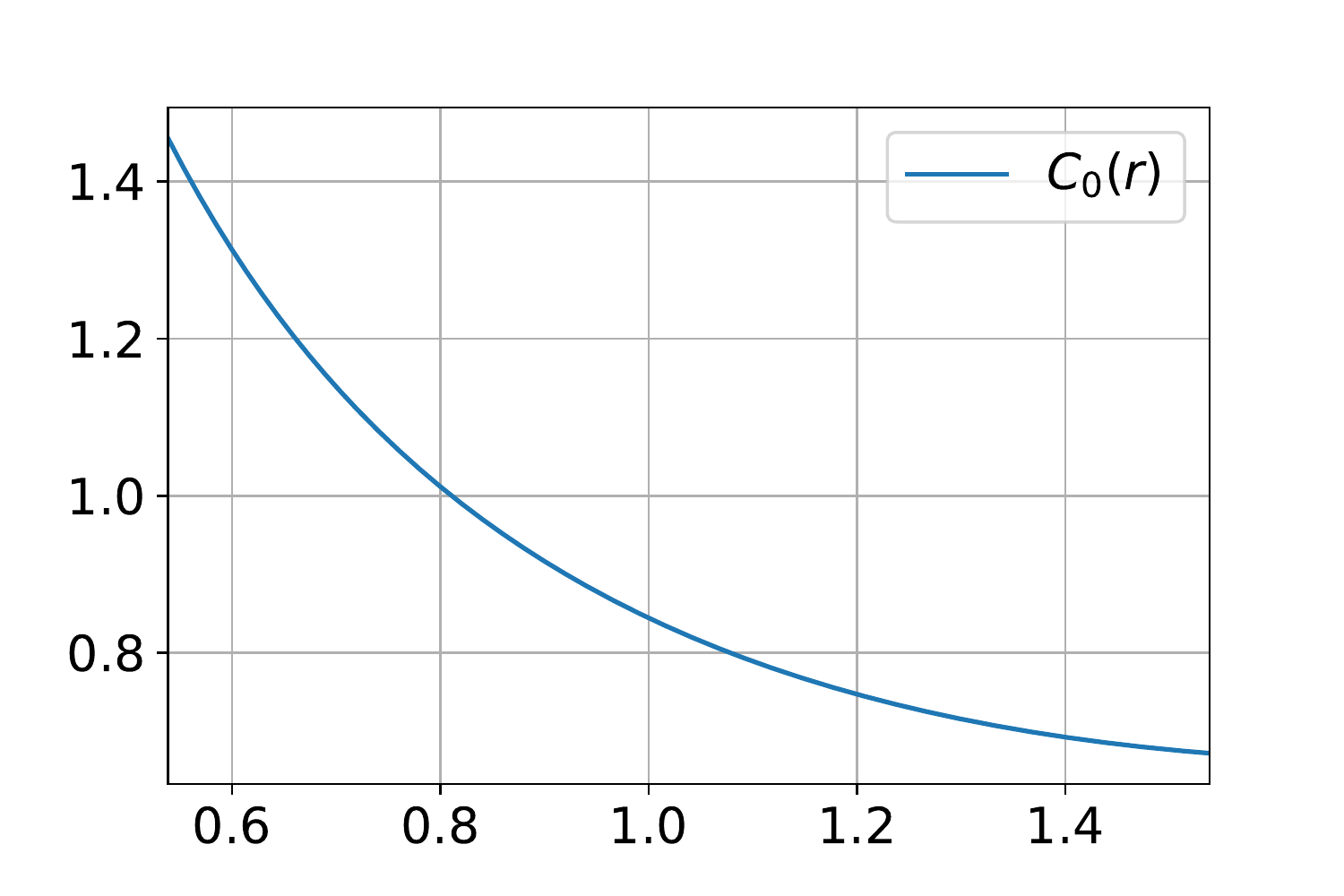}
\caption{Conductive codensity profile $C_0(r)$ given by equation \ref{eq:C0}.
}
\label{fig:C0}
\end{center}
\end{figure}

\begin{figure}
\begin{center}
\includegraphics[width=0.99\textwidth]{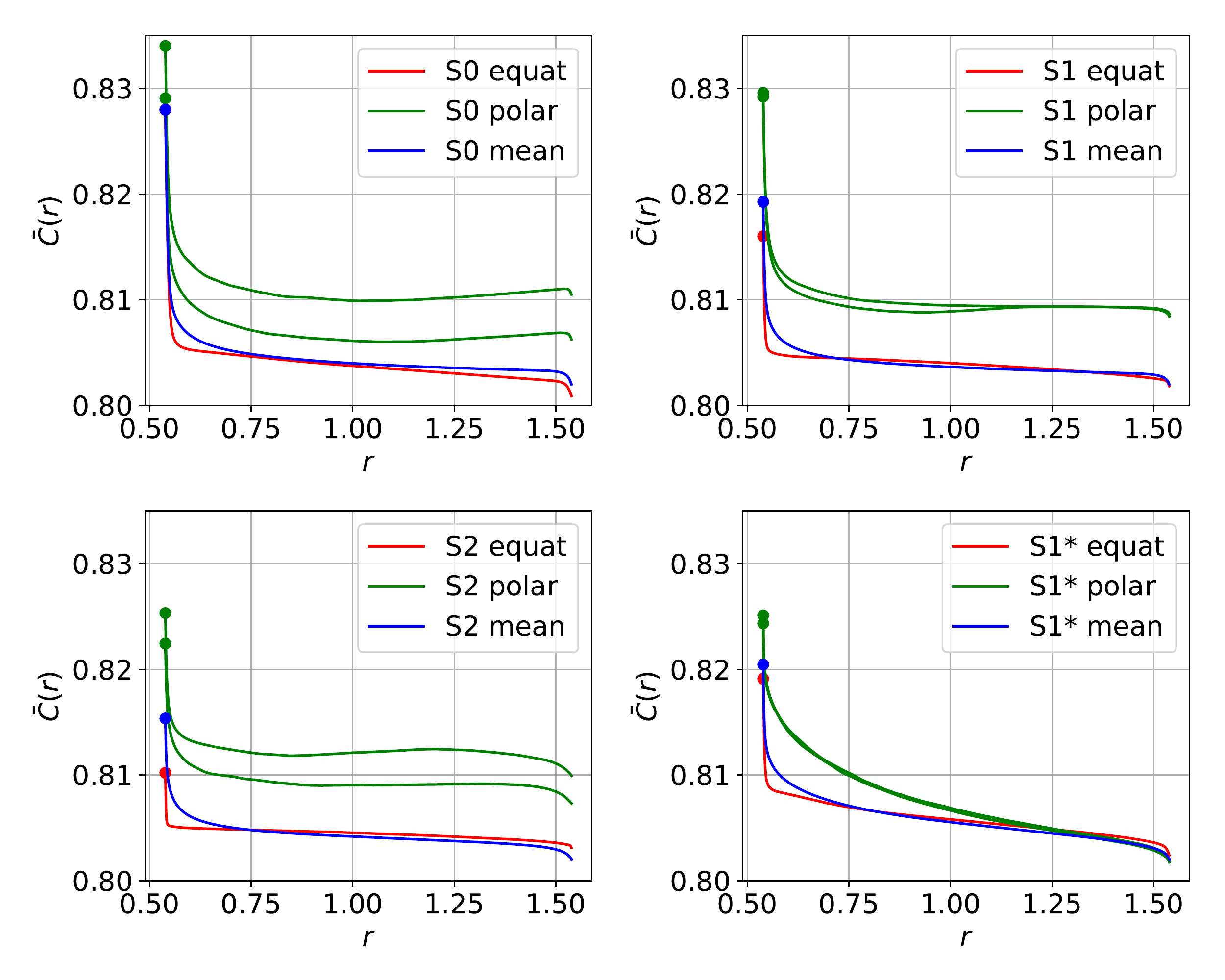}
\caption{Time-averaged total codensity profiles ($C+C_0$) in our simulations.
The two polar profiles (north and south, in green) are represented together with the mean equatorial profile (averaged in longitude, red) and the global mean profile (averaged on spherical shells, blue).
}
\label{fig:Cprof}
\end{center}
\end{figure}

In equation \ref{eq:codensity}, the conductive codensity profile $C_0(r)$ is obtained using the thermochemical model of \citet{aubert2009}, 
with a fraction $f_i = 0.75$ of buoyancy due to light element release at the inner-core boundary and the remaining $0.25$ due to internal heating or equivalently secular cooling:
\begin{equation}
C_0(r) =  c_i \frac{r^2}{2} + c_o \frac{1}{r},   \label{eq:C0}
\end{equation}
with
\begin{equation}
c_i = \frac{(2f_i-1)}{r_o^3 - r_i^3} \simeq 0.143, \quad \quad c_o = \frac{f_i r_o^3 - (1-f_i) r_i^3}{r_o^3 - r_i^3} \simeq 0.772.
\end{equation}
This profile $C_0$ is represented in figure \ref{fig:C0}, and has a gradient at the outer boundary $\beta = -\partial_r C_0|_{r_o} = -c_i r_o + c_o/r^2 \simeq 0.1056$.


Time-averaged, total codensity profiles ($C+C_0$) obtained in our simulations are represented in figure \ref{fig:Cprof}.
The codensity difference across the shell is smallest in the equatorial plane, implying efficient convection takes place there.

%
%


\footnotesize
\setlength{\bibsep}{1ex plus 0.3ex}
\bibliographystyle{elsarticle-harv}

\end{document}